\documentclass[graybox]{svmult}
 
\usepackage{mathptmx}       % selects Times Roman as basic font
\usepackage{helvet}         % selects Helvetica as sans-serif font
\usepackage{courier}        % selects Courier as typewriter font
\usepackage{type1cm}        % activate if the above 3 fonts are
\usepackage{makeidx}         % allows index generation
\usepackage{graphicx}        % standard LaTeX graphics tool
\usepackage{multicol}        % used for the two-column index

\usepackage{eucal}
\usepackage{bm}
\usepackage{cite}
\usepackage{color}

\def\vc#1{\mbox{\vec #1}}
 
\makeindex             % used for the subject index
\begin{document}
 
%\title*{Alpha-Particle Condensation in Nuclei}
\title*{Nuclear Alpha-Particle Condensates}
\author{T.~Yamada, Y.~Funaki, H.~Horiuchi, G.~R\"opke, P.~Schuck, and A.~Tohsaki}
\institute{T.~Yamada \at Laboratory of Physics, Kanto Gakuin University, Yokohama 236-8501, Japan, \at \email{yamada@kanto-gakuin.ac.jp}
\and Y.~Funaki \at Institute of Physics, University of Tsukuba, Tsukuba 305-8571, Japan
\and H.~Horiuchi \at Research Center for Nuclear Physics (RCNP), Osaka University, Osaka 567-0047, Japan, and \at International Institute for Advanced Studies, Kizugawa 619-0225, Japan,
\and G.~R\"opke \at Institut f\"ur Physik, Universit\"at Rostock, D-18051 Rostock, Germany
\and P.~Schuck \at Institut de Physique Nucl\'eaire, CNRS, UMR 8608, Orsay, F-91406, France, and \at Universit\'e Paris-Sud, Orsay, F-91505, France, and \at Labratoire de Physique et Mod\'elisation des Milieux Condens\'es, CNRS et Universit\'e Joseph Fourier, 25 Av. des Martyrs, BP 166, F-38042 Grenoble Cedex 9, France
\and A.~Tohsaki \at Research Center for Nuclear Physics (RCNP), Osaka University, Osaka 567-0047, Japan
}
\maketitle
 
\abstract{
The $\alpha$-particle condensate in nuclei is a novel state described by a product state of $\alpha$'s, all with their c.o.m. in the lowest $0S$ orbit. We demonstrate that a typical $\alpha$-particle condensate is the Hoyle state ($E_{x}=7.65$ MeV, $0^+_2$ state in $^{12}$C), which plays a crucial role for the synthesis of $^{12}$C in the universe. The influence of antisymmentrization in the Hoyle state on the bosonic character of the $\alpha$ particle is discussed in detail. It is shown to be weak. The bosonic aspects in the Hoyle state, therefore, are predominant. It is conjectured that $\alpha$-particle condensate states also exist in heavier $n\alpha$ nuclei, like $^{16}$O, $^{20}$Ne, etc. For instance the $0^+_6$ state of $^{16}$O at $E_{x}=15.1$ MeV is identified from a theoretical analysis as being a strong candidate of a $4\alpha$ condensate. The calculated small width ($34$ keV) of $0^+_6$, consistent with data, lends credit to the existence of heavier Hoyle-analogue states.  In non-self-conjugated nuclei such as $^{11}$B and $^{13}$C, we discuss candidates for the product states of clusters, composed of $\alpha$'s, triton's, and neutrons etc. The relationship of $\alpha$-particle condensation in finite nuclei to quartetting in symmetric nuclear matter is investigated with the help of an in-medium modified four-nucleon equation. A nonlinear order parameter equation for quartet condensation is derived and solved for $\alpha$ particle condensation in infinite nuclear matter. The strong qualitative difference with the pairing case is pointed out.
}
 
%%%%%%%%%%%%%%%%%%%%%%%%%%%%%%%%%%%%%%%%%%%
\section{Introduction}\label{sec:1}
%%%%%%%%%%%%%%%%%%%%%%%%%%%%%%%%%%%%%%%%%%%
 
Cluster as well as mean-field pictures are crucial to understand the structure of light nuclei~\cite{wildermuth77,ptp_supple_68}. It is well known that many states in light nuclei as well as neutron rich nuclei~\cite{oertzen06} and hypernuclei~\cite{hiyama09} have cluster structures. Recently, it was found that certain states in self-conjugate nuclei around the $\alpha$-particle disintegration threshold can be described dominantly as product states of $\alpha$ particles, all in the lowest $0S$ orbit. They are called "$\alpha$-particle condensate states``. Considerable theoretical and experimental work has been devoted to this since this idea was first put forward in 2001\cite{thsr}.
 
The ground state of $^{8}$Be has a pronounced $\alpha$-cluster structure~\cite{hiura72,qmc}. Its average density in the $0^{+}$ ground state is, therefore, very low, only about a third of usual nuclear saturation density. The two $\alpha$ particles are held together only by the Coulomb barrier and $^8$Be is, therefore unstable but with a very long life time ($10^{-17}$~s). No other atomic nucleus is known to have such a structure in its ground state. However, it is demonstrated with a purely microscopic approach that, e.g. $^{12}$C also has such a structure but as an excited state~\cite{uegaki,kamimura,funaki03,chernykh07}: the famous ''Hoyle`` state~\cite{hoyle,fowler}, i.e.~the $0^{+}_2$ state at 7.65 MeV~\cite{ajzenberg86}. It is formed by three almost independent $\alpha$ particles, only held together by the Coulomb barrier. It is located about 300 keV above the disintegration threshold into $3\alpha$ particles and has a similar life time as $^8$Be, i.e. also very long. A new-type of antisymmetrized $\alpha$-particle product state wave function, or THSR $\alpha$-cluster wave function proposed by Tohsaki, Horiuchi, Schuck, and R\"opke~\cite{thsr,schuck07,nupecc,funaki09,brenner} describes well the structure of the Hoyle state. The THSR wave function is analogous to the (number-projected) BCS wave function~\cite{Ring_Schuck}, replacing, however, Cooper pairs by $\alpha$ particles (quartets).  The $3\alpha$ particles, to good approximation, can be viewed to move in their own bosonic mean field where they occupy the lowest $0S$ level. We, therefore, talk about an alpha particle condensate. A more accurate theory reveals that there exist residual correlations, mostly of the Pauli type, among the alpha particles and, in reality, their occupation of the $0S$ level is reduced but still amounts to over $70~\%$~\cite{matsumura04,yamada05,funaki10}. This number is typical for nuclear mean field approaches. The theory~\cite{thsr,funaki03} reproduces almost all measured data of the Hoyle state, as for instance the inelastic form factor from $(e,e')$, very accurately. It is predicted that the Hoyle state has about triple to quadruple volume compared with the one of the $^{12}$C ground state. Excitations of one alpha out of the condensate into $0D$ and $1S$ states of the mean field can be formed and the $2^{+}_{2}$~\cite{funaki05,funaki06,yamada05} and  $0^{+}_{3}$~\cite{kurokawa05} states in $^{12}$C are reproduced in this way (the latter, so far only tentatively). This triplet of states are precisely the ones which, even with the most modern no-core shell model codes~\cite{nocore,navratil09}, cannot be reproduced at all. The new interpretation of the Hoyle state as an $\alpha$ condensate has stimulated a lot of theoretical and experimental works on $\alpha$-particle condensation phenomena in light nuclei~\cite{funaki05,funaki06,kurokawa05,itoh04,freer05,freer09,kokalova05,ohkubo,takashina,enyo,wakasa,Bordeier,khoa11}.
 
The establishment of the novel aspects of the Hoyle state incited us to conjecture $4\alpha$ condensation in $^{16}$O. The theoretical calculation~\cite{funaki08} of the OCM (orthogonality condition model) type\cite{saito68} succeeded in describing the structure of the first six $0^+$ states up to about 16 MeV, including the ground state with its closed-shell structure, and showed that the $0^{+}_6$ state at $15.1$ MeV around the $4\alpha$ threshold is a strong candidate
for a $4\alpha$-particle condensate, having a large $\alpha$ condensate fraction of $60~\%$. Similar gas-like states of $\alpha$ clusters have been predicted around their $\alpha$ cluster disintegration thresholds in self-conjugate $A=4n$ nuclei with the THSR wave function~\cite{thsr,tohsaki_nara} and the Gross-Pitaevskii-equation approach~\cite{yamada04}. Besides the $4n$ nuclei, one can also expect  cluster-gas states composed of alpha and triton clusters (including valence neutrons, etc.) around their cluster disintegration thresholds in $A \not= 4n$ nuclei, in which all clusters are in their respective $0S$ orbits, similar to the Hoyle state with its $(0S_{\alpha})^3$ configuration. The states, thus, can be called "Hoyle-analogues`` in non-self-conjugated nuclei. It is an intriguing subject to investigate whether or not Hoyle-analogue states exist in $A \not= 4n$ nuclei, for example, $^{11}$B, composed of $2\alpha$ and a $t$ cluster~\cite{kawabata07,enyo07,yamada10} or $^{13}$C, composed of $3\alpha$ and $1n$~\cite{kawabata08,yamada08_IJMPE,yoshida09}. The $2\alpha+t$ ($3\alpha+n$) OCM~\cite{yamada10,yamada08_IJMPE} calculation indicates that the $1/2^{+}_{2}$ ($1/2^{+}_3$) state at $E_x=11.95$ (12.14) MeV just above the $2\alpha+t$ ($3\alpha+n$) threshold is a candidate for the Hoyle-analogue.
 
It has been pointed out that in homogeneous nuclear matter and asymmetric matter $\alpha$ condensation is a possible phase~\cite{roepke98,beyer00,slr09,sogo10_quartet,sogo10} at low densities. Therefore, the above mentioned $\alpha$-particle product states in finite nuclei is related to Bose-Einstein
condensation (BEC) of $\alpha$ particles in infinite matter. The infinite matter study used a four particle (quartet) generalization of the well known Thouless criterion for the onset of pairing as a function of density and temperature. The particular finding in the four nucleon case was that $\alpha$-particle condensation can only occur at very low densities where the quartets do not overlap appreciably. This result is consistent with the structure of the Hoyle state as well as the $0^+_6$ state of $^{16}$O, in which the average density is about one third or one fourth of the saturation density. It is interesting to note that the low density condition for quartetting was in the meanwhile confirmed in Ref.~\cite{quartet} with a theoretical study in cold atom physics.
 
%%%%%%%%%%%%%%%%%%%%%%%%%%%
\begin{figure}[t]
\begin{center}
\includegraphics[width=0.9\hsize]{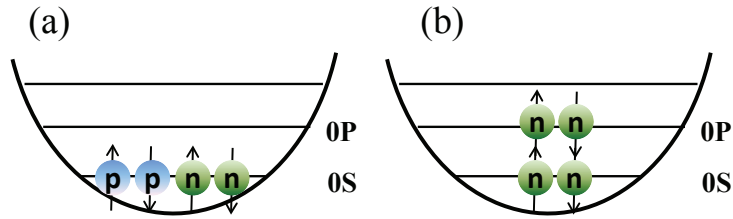}
\caption{(Color online) Sketch of (a)~$\alpha$-particle configuration with the two protons and two neutrons occupying the lowest $0S$ level in the mean field potential of harmonic oscillator shape, and (b)~the energetically lowest configuration in the case of four neutrons with two neutrons in the $0S$ orbit and the other two in the $0P$ orbit.} \label{fig:sketch}
\end{center}
\end{figure}
%%%%%%%%%%%%%%%%%%%%%%%%%%%
 
At this point it may be worthwhile to remark that nuclear physics is predestinated for cluster physics. This stems from the fact that in nuclear physics there are four different fermions (proton-neutron, spin up-down), all attracting one another with about equal strength. Such a situation is very rare in interacting fermion systems. Most of the time there are only two species of fermions, as e.g. electrons, spin up-down. However, four different fermions are needed to form a quartet. This is easily understood in a mean field picture where the four nucleons can be put into the lowest $0S$ level of a harmonic potential, whereas were there only neutrons two of four neutrons would have to be put into the $p$-orbit which is energetically very penalizing, see Fig.~\ref{fig:sketch}. This is the reason why there is no bound state of four neutrons, while the $\alpha$-particle is very strongly bound. However, recently experiments in cold atom physics try to trap more than one species of fermions~\cite{olk08} which then also may open up interesting cluster physics in that field.
 
The purpose of this lecture is to demonstrate the novel aspects of nuclear $\alpha$-particle condensates, in particular, emphasizing the structure study of $^{12}$C and $^{16}$O with the THSR wave function and the OCM approach.
 
The paper is organized as follows. In Sec.~\ref{sec:2} we first review briefly the RGM framework to describe $n\alpha$ nuclear states~\cite{wheeler37,ptp_supple_62}, which is basic for the THSR wave function and OCM. Then, we formulate the THSR wave function and OCM. Before discussing the Hoyle state, we study the structure of $^{8}$Be with the THSR wave function, and discuss the difference between the THSR-type wave function and Brink-type wave function~\cite{brink,margenau} in Sec.~\ref{sec:3}. The latter type of wave function is based on a geometrical, crystal-like viewpoint of the cluster structure. Section~\ref{sec:4} is dedicated to a discussion of the structure of the Hoyle state, studying the antisymmetrization effect among the $3\alpha$ clusters, occupation probability and momentum distribution of $\alpha$ particles, and the de Broglie wave length, etc. Then, we discuss the Hoyle-analogue states in $^{16}$O with the $4\alpha$ OCM and THSR wave function, together with $^{11}$B and $^{13}$C. The Gross-Pitaevskii-equation approach is devoted to investigate $\alpha$-particle condensation in heavier $4n$ nuclei. In Sec.~\ref{sec:5}, we focus on the $\alpha$-particle condensation in nuclear matter and its relation with that in finite nuclei. The density dependence of the $\alpha$ condensation fraction is discussed and a 'gap' equation for the $\alpha$ particle order parameter is established and solved. The strong qualitative difference with the pairing case is discussed. Finally, in Sec.~\ref{sec:6} we present the summary and conclusions.
 
\newpage
%%%%%%%%%%%%%%%%%%%%%%%%%%%%%%%%%%%%%%%%%%%%%%%%%%%%%%%%%%%%%%%%%
\section{Formulation of alpha-condensation \\ --- THSR wave function and OCM approach ---}\label{sec:2}
%%%%%%%%%%%%%%%%%%%%%%%%%%%%%%%%%%%%%%%%%%%%%%%%%%%%%%%%%%%%%%%%%

%%%%%%%%%%%%%%%%%%%%%%%%%%%%%%%%%%%%%%%%%%%
\subsection{Resonating Group Method (RGM)}\label{subsec:2-1}
%%%%%%%%%%%%%%%%%%%%%%%%%%%%%%%%%%%%%%%%%%%
 
The microscopic $n\alpha$ wave function $\Psi_{n\alpha}$ incorporating $\alpha$-cluster substructures can in general be expressed in the following RGM form~\cite{wheeler37,ptp_supple_62}:
\begin{eqnarray}
&&\Psi_{n\alpha}=\mathcal{A}\left\{\chi\left(\bm{\xi}\right)\prod_{i=1}^{n}\phi_{\alpha_i}\right\}
      = \int {d\vc{a}} {\Psi_{n\alpha}(\vc{a})} {\chi (\vc{a})}, \label{total_wf_rgm} \\
&&\Psi_{n\alpha}(\vc{a})\equiv \mathcal{A}\left\{\prod_{j=1}^{n-1}\delta(\bm{\xi}_j-\vc{a}_j)\prod_{i=1}^{n}\phi_{\alpha_i}\right\},  \label{total_wf_rgm_2}
\end{eqnarray}
with $\mathcal{A}$ the antisymmetrizer of $4n$ nucleons. The intrinsic wave function of the $i$-th $\alpha$ cluster, $\phi_{\alpha_i}$, is taken as a Gaussian (with size parameter $b$),
\begin{equation}
\phi_{\alpha_i} \propto \exp\Big[-\sum_{1\leq k<l \leq4}({\vc r}_{i,k} - {\vc r}_{i,l})^2/(8b^2)\Big],\label{eq:2}
\end{equation}
representing the intrinsic spatial part of the $(0s)^4$ shell-model configuration, where $\{\vc{r}_{i,1},\cdots,\vc{r}_{i,4}\}$ denote the coordinates of the four nucleons in  the $i$-th cluster. The spin-isospin part in Eq.~(\ref{eq:2}) is not explicitly written out but supposed to be of scalar-isoscalar form. We will not mention it henceforth. The wave function $\chi$ for the c.o.m. motion of the $\alpha$'s is chosen translationally invariant and depends only on the corresponding Jacobi coordinates $\bm{\xi}=\{\bm{\xi}_1,\bm{\xi}_2,\cdots,\bm{\xi}_{n-1}\}$. The function $\Psi_{n\alpha}(\vc{a})$ in Eq.~(\ref{total_wf_rgm_2}) describes the $\alpha$-cluster state located at the relative positions specified by a set of the Jacobi parameter coordinates $\vc{a}=\{\vc{a}_1, \vc{a}_2, \cdots, \vc{a}_{n-1}\}$.
 
The internal part of the Hamiltonian for the relevant $A=4n$ nucleus is composed of kinetic energy $-\frac{\hbar^2}{2M}\nabla_i^2$, with nucleon mass $M$, the Coulomb force $(V_{ij}^C)$, the effective two-nucleon $(V_{ij}^{(2)})$ and three-nucleon $(V_{ijk}^{(3)})$ interactions:
\begin{equation}
H=-\sum_{i=1}^{4n}\frac{\hbar^2}{2M}\nabla_i^2 - T_{\rm G} +\sum_{i<j}^{4n}{V}_{ij}^C + \sum_{i<j}^{4n}{V}_{ij}^{(2)} + \sum_{i<j<k}^{4n}{V}_{ijk}^{(3)}, \label{eq:H_for_THSR}
\end{equation}
where the c.o.m. kinetic energy of the total system $T_{\rm G}$ is subtracted.
 
The Schr\"odinger equation for the fermionic $n\alpha$ system is
\begin{eqnarray}
H\Psi_{n\alpha}=E\Psi_{n\alpha}. \label{schrodinger_eq}
\end{eqnarray}
Substituting the total wave function of Eq.~(\ref{total_wf_rgm_2}) into Eq.~(\ref{schrodinger_eq}), we obtain the equation of motion for the relative wave function $\chi$,
\begin{eqnarray}
\int d\vc{a}' \left\{ H(\vc{a}, \vc{a}') - E N(\vc{a}, \vc{a}') \right\} \chi (\vc{a}')=0,
  \hspace*{5mm}{\rm or}\hspace*{5mm}
  \left(\mathcal{H} - E\mathcal{N}\right)\chi = 0, \label{rgm_eq}
\end{eqnarray}
where the Hamiltonian and norm kernels, $H(\vc{a},\vc{a}')$ and $N(\vc{a},\vc{a}')$, are defined as
\begin{eqnarray}
\left\{ \begin{array}{c} H(\vc{a}, \vc{a}') \\ N(\vc{a}, \vc{a}') \end{array} \right\}
  = {\langle \Psi_{n\alpha}(\vc{a}) \mid \left\{ \begin{array}{c} H \\ 1 \end{array} \right\}
      \mid \Psi_{n\alpha}(\vc{a}')\rangle}. \label{eq:RGM_kernel}
\end{eqnarray}
Equation~(\ref{rgm_eq}) is called the RGM equation~\cite{ptp_supple_62}. One also can formulate the RGM framework for non $4n$ nuclei such as $^{11}$B and $^{13}$C with the microscopic $2\alpha+t$ and $3\alpha+n$ cluster model, respectively.
 
%%%%%%%%%%%%%%%%%%%%%%%%%%%%%%%%%%%%%%%%%%%
\subsection{THSR wave function}\label{subsec:2-2}
%%%%%%%%%%%%%%%%%%%%%%%%%%%%%%%%%%%%%%%%%%%
 
In the THSR description~\cite{thsr,funaki09}, the relative wave function $\chi$ in Eq.~(\ref{total_wf_rgm}) is expressed in the following $n\alpha$ condensation form,
\begin{eqnarray}
\chi_{n\alpha}^{\rm THSR}(B:\vc{R}_1,\vc{R}_2,\cdots,\vc{R}_n) &=& \prod_{i=1}^{n} \varphi_0(B:\vc{R}_i-\vc{X}_{\rm G}), \label{eq:thsr_xi}\\
\varphi_0(B:{\vc{R}}) &=& \exp(-2\vc{R}^2/B^2), \label{eq:thsr_0s_wf}
\end{eqnarray}
where $\vc{R}_i=(\vc{r}_{i,1}+\cdots+\vc{r}_{i,4})/4$ denotes the c.o.m. coordinate of the $i$-th $\alpha$ particle, $\vc{X}_{\rm G}=(\vc{R}_{1}+\cdots+\vc{R}_{n})/n$ is the total c.o.m. coordinate of the $n\alpha$ system and ${\varphi_0(B:\vc{R})}$ represents a Gaussian with a large width parameter $B$ which is of the nucleus' dimension. Usually, one uses Jacobi coordinates $\{\bm{\xi}_i\}$ splitting off the total c.o.m. part of the wave function. Then the THSR ansatz for $\chi$ in Eq.~(\ref{eq:thsr_xi}) is given by
\begin{equation}
\chi_{n\alpha}^{\rm THSR}(B:\vc{R}_1,\vc{R}_2,\cdots,\vc{R}_n) =\exp\Big(-2\sum_{i=1}^{n-1}\mu_i \frac{\bm{\xi}_{i}^2}{B^2} \Big), \label{eq:thsr_spherical}
\end{equation}
with $\mu_i=i/(i+1)$. A slight generalization of Eq.~(\ref{eq:thsr_spherical}) is possible, taking into account  nuclear deformation (see Sec.~\ref{sec:3}).  With Eqs.~(\ref{eq:thsr_xi}) and (\ref{total_wf_rgm}), one can write the THSR wave function in the following $n\alpha$ product form,
\begin{eqnarray}
{\Psi_{n\alpha}}~~~\rightarrow~~~{\langle \vc{r}_{1,1},\cdots,\vc{r}_{n,4}|{\rm THSR}\rangle} &=& \mathcal{A} [\psi_{\alpha_1}\psi_{\alpha_2}\cdots\psi_{\alpha_n}],  \label{eq:5}
\end{eqnarray}
where
\begin{eqnarray}
{|{\rm THSR}\rangle} &=& {|{\rm THSR} (B) \rangle} \equiv \mathcal{A}{|B\rangle},  \label{eq:6} \\
{\langle \vc{r}_{1,1},\cdots,\vc{r}_{n,4} | B \rangle} &=& \psi_{\alpha_1}\psi_{\alpha_2}\cdots\psi_{\alpha_n},  \label{eq:7}
\end{eqnarray}
where $\psi_{\alpha_i}=\varphi_0(B:\vc{R}_{i}-\vc{X}_{\rm G})\phi_{\alpha_i}$ and definitions of Eqs.~(\ref{eq:6}) and (\ref{eq:7}) will be useful later.
Equations (\ref{eq:5})$\sim$(\ref{eq:7}) show the analogy of the THSR wave function with the number-projected BCS wave functions for pairing
\begin{eqnarray}
{\langle \vc{r}_{1,1},\cdots,\vc{r}_{n,2}|{\rm BCS}\rangle} = \mathcal{A} [\phi_{\rm pair}(\vc{r}_{1,1},\vc{r}_{1,2})\phi_{\rm pair}(\vc{r}_{2,1},\vc{r}_{2,2})\cdots\phi_{\rm pair}(\vc{r}_{n,1},\vc{r}_{n,2})],  \label{eq:8}
\end{eqnarray}
where $\phi_{\rm pair}(\vc{r}_{i,1},\vc{r}_{i,2})$ denotes the Cooper pair wave function.
 
The product of $n$ identical $0S$ wave functions in Eq.~(\ref{eq:5}) reflects the boson condensate character. This feature is realized as long as the action of the antisymmetrizer in Eq.~(\ref{total_wf_rgm}) is sufficiently weak. On the other hand, in the limit where $B$ is taken to be $B=b$, the normalized THSR wave function is equivalent to an SU(3) shell model wave function with the lowest harmonic oscillator quanta~\cite{bayman58,yamada_ptp_08}; for example, in the case of the $2\alpha$, $3\alpha$ and $4\alpha$ systems, they respectively are given by
\begin{eqnarray}
&&\lim_{B \rightarrow b} N_{2\alpha}(B)\Psi_{2\alpha}(B) = {| (0s)^4(0p)^4; (\lambda\mu)=(4,0),J^\pi=0^+\rangle}, \label{eq:su(3)_8Be}\\
&&\lim_{B \rightarrow b} N_{3\alpha}(B)\Psi_{3\alpha}(B) = {| (0s)^4(0p)^{8}; (\lambda\mu)=(0,4),J^\pi=0^+\rangle}, \label{eq:su(3)_12C}\\
&&\lim_{B \rightarrow b} N_{4\alpha}(B)\Psi_{4\alpha}(B) = {| (0s)^4(0p)^{12}; (\lambda\mu)=(0,0),J^\pi=0^+\rangle}, \label{eq:su(3)_16O}
\end{eqnarray}
where the $N_{n\alpha}(B)$ are the normalization factors. The shell model wave functions in Eqs.~(\ref{eq:su(3)_8Be}), (\ref{eq:su(3)_12C}) and (\ref{eq:su(3)_16O}) are the dominant configurations of the ground-state wave functions of $^{8}$Be, $^{12}$C and $^{16}$O, respectively. In fact, the components of Eqs.~(\ref{eq:su(3)_12C}) and (\ref{eq:su(3)_16O}) in the ground states of $^{12}$C and $^{16}$O have weights over $60~\%$ and $90~\%$, respectively, because both of the states have shell-model-like compact structures. On the other hand, in the case of the ground state of $^8$Be, the component of Eq.~(\ref{eq:su(3)_8Be}) is as small as about $20~\%$ but still the largest, and the remaining components distribute monotonously in a lot of higher SU(3) configurations, when one expands the wave function of the $^{8}$Be ground state in terms of the SU(3) basis. This characteristic comes from a pronounced $2\alpha$ cluster structure in the ground state (see Sec.~\ref{sec:3}).

The wave functions of the quantum states in $A=4n$ nucleus can be expanded using the $n\alpha$ THSR wave function, like
\begin{equation}
\Psi_k = \sum_m f_k (B^{(m)}) \Psi_{n\alpha}(B^{(m)}), \label{eq:hwwf}
\end{equation}
where $\Psi_{n\alpha}(B^{(m)})$ is the $n\alpha$ THSR wave function which has the form of
\begin{equation}
\Psi_{n\alpha}(B^{(m)})= \mathcal{A}[ \chi_{n\alpha}^{\rm THSR}(B^{(m)};\vc{R}_1,\vc{R}_2,\cdots,\vc{R}_n) \phi_{\alpha_1}\phi_{\alpha_2}\cdots\phi_{\alpha_n}]. \label{eq:thsr}
\end{equation}
The discrete variational parameters $B^{(m)}$ represent the generator coordinate of  the Hill-Wheeler ansatz. The expansion coefficients $f_k (B^{(m)})$ and the corresponding eigenenergy $E_k$ for the $k$-th eigenstate are obtained by solving the following Hill-Wheeler equation~\cite{hill53,griffin57},
\begin{eqnarray}
\sum_{m^\prime} \left\langle \Psi_{n\alpha}(B^{(m)}) \Big| H-E_k  \Big| \Psi_{n\alpha}(B^{(m^\prime)}) \right\rangle f_k (B^{(m^\prime)}) =0.  \label{eq:hw}
\end{eqnarray}
This equation has the same structure as  the RGM equation in Eq.~(\ref{rgm_eq}) but reducing it to a one parameter equation. The superposition of THSR wave functions fine tunes the results but a single THSR wave function with an optimized $B$-value already yields excellent results as will be demonstrated below.
 
%%%%%%%%%%%%%%%%%%%%%%%%%%%%%%%%%%%%%%%%%%%
\subsection{$n\alpha$ boson wave function and OCM}\label{subsec:2-3}
%%%%%%%%%%%%%%%%%%%%%%%%%%%%%%%%%%%%%%%%%%%
 
In order to study the bosonic properties of the $n\alpha$ system, one needs to map the microscopic (fermionic) $n\alpha$ cluster model wave function $\Psi_{n\alpha}$ in Eq.~(\ref{total_wf_rgm}) onto an $n\alpha$ boson wave function $\Phi_{n\alpha}^{(B)}$. The RGM framework given in Sec.~\ref{subsec:2-1} is useful and appropriate for the mapping. Taking into account the normalization of $\Psi_{n\alpha}$, $1 = {\langle \Psi_{n\alpha} | \Psi_{n\alpha} \rangle} = {\langle \chi(\bm{\xi}) | N(\bm{\xi},\bm{\xi}') | \chi(\bm{\xi}') \rangle}$, the $n\alpha$ bosonic wave function is provided in the following form~\cite{ptp_supple_62},
\begin{eqnarray}
\Phi_{n\alpha}^{(B)}(\bm{\xi}) \equiv \mathcal{N}^{1/2}\chi = \int d{\bm{\xi}}' {N}^{1/2}(\bm{\xi}, {\bm{\xi}}') \chi({\bm{\xi}}'),\label{boson_wf}
\end{eqnarray}
where $\chi$ represents the relative wave function with the set of Jacobi coordinates, $\bm{\xi}=\{\bm{\xi}_1,\bm{\xi}_2,\cdots,\bm{\xi}_{n-1}\}$, with respect to the c.o.m.~of $\alpha$ clusters. The square-root matrix $N^{1/2}(\bm{\xi},\bm{\xi}')$ is related to the norm kernel of the $n\alpha$ RGM wave function in Eq.~(\ref{eq:RGM_kernel}). It is noted that $\Phi_{n\alpha}^{(B)}$ depends only on the Jacobi coordinates $\bm{\xi}$, and all of the internal coordinates of $n\alpha$ particles are integrated out in $\Phi_{n\alpha}^{(B)}$.
 
From the RGM equation (\ref{rgm_eq}), the equation of motion for $\Phi_{n\alpha}^{(B)}(\bm{\xi})$ is obtained in the form
\begin{eqnarray}
\left({\mathcal{N}^{-1/2}}\mathcal{H}{\mathcal{N}^{-1/2}}-E\right)\Phi_{n\alpha}^{(B)}=0, \label{boson_eq}
\end{eqnarray}
where $\mathcal{H}$ denotes the Hamiltonian kernel defined in Eq.~(\ref{rgm_eq}). Then, one can interpret $\mathcal{N}^{-1/2}\mathcal{H}\mathcal{N}^{-1/2}$ as the nonlocal $n\alpha$ boson Hamiltonian. In Eq.~(22) care should be taken that before inversion all zero eigenvalues of the norm $\mathcal N$ are properly eliminated. The eigenfunctions belonging to the zero eigenvalues are the so-called Pauli forbidden states $u_F({\bf r})$ which satisfy the condition $\mathcal{N}u_F=\mathcal{A}\left\{u_F(\vc{\xi})\prod_{i=1}^{n}\phi_{\alpha_i}\right\}=0$.
 
The boson wave function has the following properties: 1) $\Phi_{n\alpha}^{(B)}$ is totally symmetric for any 2$\alpha$-particle exchange, 2) $\Phi_{n\alpha}^{(B)}$ satisfies the equation motion (\ref{boson_eq}), and 3) $\Phi_{n\alpha}^{(B)}$ is orthogonal to the Pauli forbidden states $u_F(\vc{r})$. In order to obtain the boson wave function $\Phi_{n\alpha}^{(B)}$, we need to solve the equation of motion of the bosons in Eq.~(\ref{boson_eq}). Solving the boson equation, however, is difficult in general even for the 3$\alpha$ case. Thus, it is requested to use more feasible frameworks for the study of the bosonic properties and the amount of $\alpha$ condensation for the $N\alpha$ system. One such framework is OCM (orthogonality condition model)~\cite{saito68}. The OCM scheme, which  is an approximation to  RGM, is known to describe nicely the structure of low-lying states in light nuclei~\cite{saito68,horiuchi74,Suz76,ptp_supple_68,fukatsu89,Kat92,yamada05,kurokawa05,funaki08}. The essential properties of the $n\alpha$ boson wave function $\Phi_{n\alpha}^{(B)}$, as mentioned above, can be taken into account in OCM in a simple manner. We will demonstrate this below.
 
In OCM, the $\alpha$ cluster is treated as a point-like particle. We approximate the nonlocal $n\alpha$ boson Hamiltonian  in Eq.~(\ref{boson_eq}) by an effective (local) one, that is $H^{{\rm (OCM)}}$,
\begin{eqnarray}
  && {\mathcal{N}^{-1/2}}\mathcal{H}{\mathcal{N}^{-1/2}} \sim H^{{\rm (OCM)}} \\
  && H^{{\rm (OCM)}}\equiv \sum_{i=1}^n T_i - T_{G}+\sum_{i<j=1}^{n} V_{2\alpha}^{\rm eff}(i,j)+\sum_{i<j<k=1}^{n}V_{3\alpha}^{\rm eff}(i,j,k),
      \label{hamiltonian_ocm}
\end{eqnarray}
where $T_i$ denotes the kinetic energy of the {\it i}-th $\alpha$ cluster, and the center-of-mass kinetic energy $T_{G}$ is subtracted from the Hamiltonian. The effective local 2$\alpha$ and 3$\alpha$ potentials are presented as $V_{2\alpha}^{\rm eff}$ (including the Coulomb potential) and $V_{3\alpha}^{\rm eff}$, respectively. Then, the equation of the relative motion of the $n\alpha$ particles with $H^{{\rm (OCM)}}$, called the OCM equation, is written as
\begin{eqnarray}
 &&\left\{ H^{{\rm (OCM)}}-E \right\}\Phi^{{\rm (OCM)}}_{n\alpha}=0, \label{ocm_eq} \\
 &&{\langle u_F \mid \Phi^{{\rm (OCM)}}_{n\alpha} \rangle}=0, \label{orthogonality_condition}
\end{eqnarray}
where $u_F$ denotes the Pauli-forbidden state of the $n\alpha$ system as mentioned above. In the case of $2\alpha$ system, the Pauli-forbidden states between the two $\alpha$-particles are $0S$, $0D$ and $1S$ states with the total oscillator quanta $Q$ less than $4$. It is pointed out that the Pauli-forbidden states in the $n\alpha$ system can be constructed from those of the $2\alpha$ system~\cite{horiuchi77}.
 
The bosonic property of the wave function $\Phi$ can be taken into account by symmetrizing the wave function with respect to any 2$\alpha$-particle exchange,
\begin{eqnarray}
\Phi^{{\rm (OCM)}}_{n\alpha}=\mathcal{S}\Phi^{{\rm (OCM)}}_{n\alpha}(1,2,\cdots,n),
\label{eq:ocm_wf}
\end{eqnarray}
where $\mathcal{S}$ denotes the symmetrization operator, $\mathcal{S}=(1/\sqrt{n!}) \sum_k \mathcal{P}_k$, where the sum runs over all permutations $\mathcal{P}$ of the $n$ $\alpha$-particles. It is noted that the completely collapsed state of the $n\alpha$ particles is forbidden within the present framework because of the Pauli-blocking effect in Eq.~(\ref{orthogonality_condition}).
 
The OCM equation (\ref{ocm_eq}) with the condition (\ref{orthogonality_condition}) is solved with the help of the Gaussian expansion method (GEM)~\cite{kamimura88,hiyama03}. Combining OCM and GEM provides a powerful tool to study the structure of light nuclei~\cite{yamada05,funaki08,yamada08_IJMPE,yamada10} as well as light hypernuclei~\cite{hiyama97,hiyama09}, because the Pauli-blocking effect among the clusters is properly taken into account and GEM covers an approximately complete model space~\cite{kamimura88,hiyama03}. It is also useful to apply Kukulin's method~\cite{kukulin} for removing the Pauli-forbidden states $u_F$'s from the wave function $\Phi^{{\rm (OCM)}}_{n\alpha}$. The present OCM-GEM framework, for example, in the case of $^{16}$O, can cover a model space large enough to describe the dilute $\alpha$ gas-like configuration, as well as $\alpha + {^{12}}$C cluster and shell-model-like ground state structures.
 
%%%%%%%%%%%%%%%%%%%%%%%%%%%%%%%%%%%%%%%%%%%%%%%%%%%%%
\subsection{Single $\alpha$-particle density matrix and occupation probabilities}\label{subsec:2-4}
%%%%%%%%%%%%%%%%%%%%%%%%%%%%%%%%%%%%%%%%%%%%%%%%%%%%%
 
A literal interpretation of an $\alpha$ condensate in a finite system is that all the $\alpha$ particles occupy the lowest $0S$-wave orbit of an $\alpha$ mean field potential. Due to residual interactions and the action of the Pauli principle, the occupation probabilities may spread out over several orbits, but a particular orbit  should be occupied with a significant probability if a state is called a condensate. The occupation probability can be calculated by solving the eigenvalue problem of a single $\alpha$-particle density matrix~\cite{matsumura04,yamada05,funaki10,suzuki_02,suzuki_08,yamada08_obdm,yamada09_obdm}.
 
The single $\alpha$-particle density matrix for the $n\alpha$ boson system can be defined with the use of the $n\alpha$ Boson wave function $\Phi_{n\alpha}^{(B)}(\bm{\xi})$ in Eq.~(\ref{boson_wf}) mapped from the translationally invariant normalized microscopic $n\alpha$ wave function in Eq.~(\ref{total_wf_rgm}),
\begin{eqnarray}
&&{\rho^{(1)}_{\rm int}(\vc{q}_1,{\vc{q}_1}')} = {\left(\frac{n}{n-1}\right)^{3}} \rho^{(1)}_{\rm int,J}(\bm{\xi}_1,{\bm{\xi}_1}'),\label{eq:one_particle_density_matrix}\\
&&{\rho^{(1)}_{\rm int,J}(\bm{\xi}_1,{\bm{\xi}_1}')} = {\int \prod_{i=2}^{n-1}{d\bm{\xi}_i} {{\Phi_{n\alpha}^{(B)}}^{*}(\bm{\xi}_1,\bm{\xi}_2,\cdots,\bm{\xi}_{n-1})} {{\Phi_{n\alpha}^{(B)}}({\bm{\xi}_1}',\bm{\xi}_2,\cdots,\bm{\xi}_{n-1})}},\hspace*{5mm}\label{eq:density_matrix_Jacobi}
\end{eqnarray}
where $\vc{q}_1=\frac{n-1}{n}\bm{\xi}_1=\vc{R}_1-\vc{X}_{\rm G}$ is the 1st particle coordinate ($\vc{R}_1$) with respect to the c.o.m coordinate of the system ($\vc{X}_{\rm G}$) and $\bm{\xi}_1$ denotes the relative coordinate between the 1st particle and the remaining $(n-1)$ ones. The factor in Eq.~(\ref{eq:one_particle_density_matrix}) is the Jacobian ${\partial \bm{\xi}_1}/{\partial \vc{q}_1}$. Since the wave function $\Phi_{n\alpha}^{(B)}(\bm{\xi})$ is totally symmetric with respect to particle permutation, the choice of the 1st particle is arbitrary. The definition (\ref{eq:density_matrix_Jacobi}) is called the Jacobi-type one-particle density matrix. The diagonal density matrix $\rho^{(1)}_{\rm int}(\vc{q},{\vc{q}})$ stands for the density distribution of $\alpha$ particles with respect to the c.o.m.~coordinate of the $n\alpha$ system. The eigenvalue problem of the density matrix $\rho^{(1)}_{\rm int}$,
\begin{eqnarray}
\int d\vc{q}' \rho^{(1)}_{\rm int}(\vc{q},{\vc{q}'}) \varphi(\vc{q}') = \lambda \varphi(\vc{q}),
\label{eq:eigen_eq_one_body_density}
\end{eqnarray}
gives the single $\alpha$-particle orbit $\varphi(\vc{q})$ and its occupation probability $\lambda$, where $\vc{q}$ is measured from the c.o.m. coordinate of the system. The spectrum of eigenvalues of the density matrix $\rho^{(1)}_{\rm int}$ gives information on the occupancy of the orbits of the system and it is obviously equal to that of $\rho^{(1)}_{\rm int,J}$. The occupation probability is labeled with the angular momentum $L$ and the quantum number of a positive integer $n_L$, like $L_{n_{L}}$. In this article, for a single-$\alpha$ orbit with an angular momentum $L$, we denote the largest occupation probability as $L_1$ ($n_L=1$), the second largest as $L_2$ ($n_L=2$), the third largest as $L_3$ ($n_3=3$), etc. Please notice that the positive number $n_L$ is different from the number of nodes for the radial part of the corresponding single-$\alpha$ orbit $\varphi(\vc{q})$ (for instance, in Fig.~\ref{fig:12C_alpha_orbits}, the single-$\alpha$ orbits labeled as $L=0$ and $n_L=1$ ($S_1$) have $2S$ and $0S$ nodal behaviors for the ground state ($0^+_1$) and the Hoyle state ($0^+_2$), respectively).
 
Let us remind that one should use the Jacobi coordinate system for the choice of the internal coordinates of the density matrix. If an internal coordinate system other than the Jacobi coordinate system is adopted (for example, that adopted by Pethick and Pitaevskii (PP)~\cite{pethick00}), an unphysical result is obtained even for  condensation of a finite number of ideal bosons in a harmonic trap, contrary to what PP expected~\cite{pita}. Two physically motivated criteria for the choice of the adequate coordinate system lead to a unique answer for the internal one-particle density matrix, i.e.~the Jacobi-type internal density matrix, while the PP-type one-body density matrix does not satisfy the criteria (see Refs.~\cite{yamada08_obdm,yamada09_obdm} for details).
 
In general even for the $3\alpha$ system, one encounters  numerical difficulties to obtain the boson wave function mapped from the microscopic $n\alpha$ wave function, $\Phi^{(B)}_{n\alpha} = \mathcal{N}^{1/2}\chi$ in Eq.~(\ref{boson_wf}), by solving the boson equation in Eq.~(\ref{boson_eq}), as mentioned above. Thus, it is hard in general to calculate the one-body density matrix for the $\alpha$ particle $\rho^{(1)}_{\rm int}$ in Eq.~(\ref{eq:one_particle_density_matrix}). To overcome this difficulty, the following two approximate ways have so far been proposed to evaluate the density matrix. One is, as for the boson wave function, to use the $n\alpha$ OCM wave function (\ref{eq:ocm_wf}) obtained by solving the OCM equations~(\ref{ocm_eq}) and (\ref{orthogonality_condition}), i.e.~$\Phi^{(B)}_{n\alpha} \simeq \Phi^{{\rm (OCM)}}_{n\alpha}$~\cite{yamada05,funaki08}. The application of this method was done for the $3\alpha$ and $4\alpha$ systems, the results of which are presented in Secs.~\ref{subsec:4-1} and \ref{subsec:4-2}.  The other is to make the following approximation for the boson wave function as proposed in Refs.~\cite{matsumura04,funaki10}, $\Phi^{(B)}_{n\alpha}(\bm{\xi}) \simeq \mathcal{N}\chi/{\sqrt{\langle \mathcal{N}\chi | \mathcal{N}\chi \rangle}}$. This method was used for the $3\alpha$ and $4\alpha$ THSR wave functions. The results are discussed in Sec.~\ref{subsec:4-2}. The two approximations give quantitative similar results for the occupation probabilities and for the single-$\alpha$ orbits in the $3\alpha$ and $4\alpha$ systems, which are obtained by solving the eigenvalue problem of Eq.~(\ref{eq:eigen_eq_one_body_density}).

%%%%%%%%%%%%%%%%%%%%%%%%%%%%%%%%%%%%%%%%%%%
\section{THSR wave function vs~Brink wave function for $^{8}$Be}\label{sec:3}
%%%%%%%%%%%%%%%%%%%%%%%%%%%%%%%%%%%%%%%%%%%
 
\begin{figure}[t]
\begin{center}
\includegraphics[width=0.75\hsize]{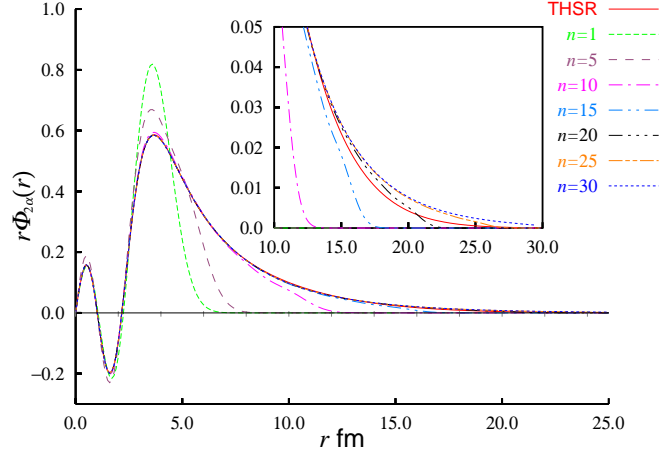}
\caption{(Color online) Comparison of THSR wave function with a single component ``Brink'' wave function with $D=3.45$~fm (denoted by $n=1$). The convergence rate with the superposition of several $(n)$ ``Brink'' wave functions is also shown. The line denoted by $n=30$ corresponds to the full RGM solution. The Volkov No.1 force is taken with Majorana parameter value $M=0.56$. Figure is taken from Ref.~\cite{funaki09}.} \label{fig:conv_rel_wf}
\end{center}
\end{figure}

Before discussing the Hoyle state, it is instructive to study $^8$Be in some detail because even this nucleus which is known to have intrinsically a two-alpha dumbbell structure~\cite{hiura72,qmc} can very well be described in the laboratory frame with the THSR wave function. Let us repeat Eq.~(\ref{total_wf_rgm}) for this particular case
\begin{equation}
\Psi_{2\alpha}=\mathcal{A}[\chi({\vc r}) \phi_{\alpha_1} \phi_{\alpha_2}], \label{eq:8Be_RGM}
\end{equation}
with the relative coordinate between the $2\alpha$ particles, $\vc{r} = \vc{R}_1-\vc{R}_2$. Note that Eq.~(\ref{eq:8Be_RGM}) is a fully antisymmetric and translationally invariant wave function in $8-1=7$ coordinates. Solving the RGM equation in Eq.~(\ref{rgm_eq}) with a given Hamiltonian, one obtains the energy $E$ of $^8$Be and $\chi$. The $2\alpha$ boson wave function $\Phi^{\rm (B)}_{2\alpha}(\vc{r})$ representing the relative motion of the two $\alpha$-particles, mapped from the corresponding fermionic $2\alpha$ wave function $\Psi_{2\alpha}$, is given in Eq.~(\ref{boson_wf}),
\begin{equation}
\Phi^{\rm (B)}_{2\alpha}(\vc{r}) = \int {d\vc{r}'} {{N}^{1/2}(\vc{r}, \vc{r}')} \chi(\vc{r}'), \label{eq:8Be_boson}
\end{equation}
Expressions~(\ref{eq:8Be_RGM}) and (\ref{eq:8Be_boson}) have been obtained with very high numerical accuracy since 50 years with excellent results for all low energy properties of $^8$Be~\cite{hiura72}. The radial part of the $2\alpha$ boson wave function $r\Phi_{2\alpha}(\vc{r})$ in the ground state ($J^{\pi}=0^{+}$) is shown in Fig.~\ref{fig:conv_rel_wf} denoted by $n=30$. We see that there exist two nodes, an effect which stems from the Pauli principle.
 
%%%%%%%%%%%%%%%%%%%%%%%%%%%
%%%%%%%%%%%%%%%%%%%%%%%%%%%
 
\begin{figure}[t]
\begin{center}
\includegraphics[width=0.75\hsize]{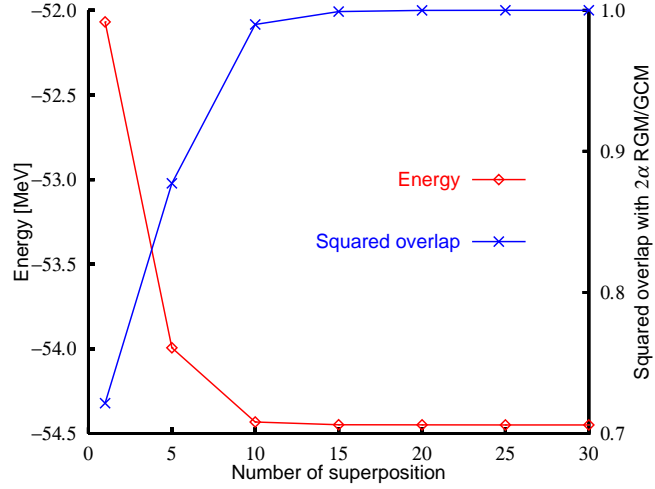}
\caption{(Color online) Binding energy corresponding to $\Psi_{2\alpha}$ (see Eq.~(\ref{eq:8Be_Brink_full_wf})) with the superposition of $n$ Brink wave functions and the squared overlap between the full RGM solution and $\Psi_{2\alpha}$. For $n=1$, a single Brink wave function with optimized $R=3.45$ fm is adopted. Figure is taken from Ref.~\cite{funaki09}.}\label{fig:conv}
\end{center}
\end{figure}

Here we will discuss two approximate forms for $\chi(r)$:~the THSR wave function and the Brink cluster wave function~\cite{brink}. Let us start with the latter. In the Brink wave function, the $\alpha$ particles are placed at certain positions in space. In the case of $^8$Be, placing the $2\alpha$ particles at the positions of $\vc{D}/2$ and $-\vc{D}/2$, respectively,  this leads to
\begin{eqnarray}
\chi^{\rm Brink}(r) &=&  {\widehat P}^{J=0} \exp\Big[-\Big({\vc R}_1-{\vc D}/2 - ({\vc R}_2 + {\vc D}/2)\Big)^2/b^2\Big] \nonumber \\
                      &=& {\widehat P}^{J=0} \exp \Big[ -\frac{1}{b^2}(\vc{r}-\vc{D})^2 \Big], \label{eq:8Be_Brink}
\end{eqnarray}
where ${\widehat P}^{J=0}$ denotes the projection operator onto spin $J=0$. Though this kind of geometrical, crystal-like viewpoint of the cluster structure works well for many cases, for instance, parity-violating $^{12}$C+$\alpha$, $^{16}$O+$\alpha$, and $^{40}$Ca+$\alpha$ structures in $^{16}$O, $^{20}$Ne and $^{44}$Ti, respectively~\cite{ptp_supple_68,44Ti_ohkubo,44Ti_horiuchi}, and also when additionally neutrons are involved~\cite{itagaki04}, it is on the contrary known since several decades that this picture fails for the description of the famous Hoyle state, i.e. the $0_2^+$ state in $^{12}$C (see Sec.~\ref{sec:4}). The ansatz of the two $\alpha$ particles being placed at a distance $\vc{D}$ from one another seems reasonable, since the Quantum Monte Carlo calculation with realistic two-nucleon and three-nucleon potentials in Ref.~\cite{qmc} indeed indicates that the two $\alpha$'s are about $4$ fm apart. Obviously, the parameter ${\vc D}$ can be varied to find the optimal position of the $\alpha$-particles. The result of such a procedure is shown in Fig.~\ref{fig:conv_rel_wf} with the line denoted by $n=1$  taking the optimal value $D=3.45$ fm ($b$ is kept fixed at its free space value, $b=1.36$ fm). Qualitatively such a ``Brink'' wave function follows the full variational solution (line denoted by $n=30$). However, in the outer part, for instance in the exponentially decaying tail quite strong differences appear. The squared overlap with the exact solution is 0.722. Of course,the Brink wave functions also can serve as a basis and it is interesting to study the convergence properties. We, therefore, write for the $^8$Be wave function appearing in Eq.~(\ref{rgm_eq})
\begin{eqnarray}
\Psi_{2\alpha}={\cal A}[\chi(r) \phi_{\alpha_1}\phi_{\alpha_2}]=\sum_i f_i \Psi_{2\alpha}^{\rm Brink} (r,D^{(i)},b), \label{eq:8Be_Brink_full_wf} \\
\Psi_{2\alpha}^{\rm Brink} (r,D^{(i)},b) = {\cal A}\left[ \chi_{D^{(i)}}^{\rm Brink}(r) \phi_{\alpha_1}\phi_{\alpha_2} \right]
\end{eqnarray}
where the $D^{(i)}$ indicate the various positions of the $\alpha$-particles and $f_i$ are the expansion coefficients. The convergence of the squared overlap with the exact solution is studied where we take for the positions $D^{(1)}=1$ fm, $D^{(2)}=2$ fm, $\cdots$, $D^{(n)}=n$ fm. We start with $n=5$. In Fig.~\ref{fig:conv} the convergence rate is shown as a function of $n$ for the squared overlap and for the energy. The point of $n=1$ is with the optimized single Brink wave function $(D^{(1)}=3.45\ {\rm fm})$. We see that the convergence is not extremely fast but for $n=20$ the squared overlap with the full RGM solution amounts to $0.9999$. Also energy is converged to within $10^{-4}$. In Fig.~\ref{fig:conv_rel_wf} we show the convergence of the $2\alpha$ boson wave function $r\Phi_{2\alpha}(r)$. In the insert we see that there is still a slight change in the far tail going from $n=25$ to $n=30$.
 
%%%%%%%%%%%%%%%%%%%%%%%%%%%
%%%%%%%%%%%%%%%%%%%%%%%%%%%
 
Let us now investigate the THSR ansatz for $\chi(r)$. There it is assumed from the beginning that the $\alpha$'s are delocalised and a single Gaussian $e^{-r^2/B^2}$ centered at the origin with, however, a large width $B^2=b^2+2\beta^2$, with $\beta$ a variational parameter, is taken. Very much improved results over the single component Brink wave function are obtained. With $\beta=3.24$ fm the squared overlap becomes $97.24~\%$. However, practically $100~\%$ accuracy, compared with the exact solution, can be achieved starting with a slightly improved ansatz, i.e. with an axially symmetric deformed Gaussian which is then projected on the ground-state spin $J=0$ (projections on $J=2,4$ yield the rotational band of $^8$Be)~\cite{funaki_8be},
\begin{eqnarray}
\hspace{-7mm}\chi^{\rm THSR}(r) & = & {\widehat P}^{J=0} \exp \Big( -\frac{r_\perp^2}{b^2+2\beta_\perp^2}-\frac{r_z^2}{b^2+2\beta_z^2} \Big)
 \label{eq:thsr_8be}\nonumber \\
& \propto &  \frac{\exp( -r^2/B_\perp^2 )}{ir}
 {\rm Erf} \Big(i \frac{(B_z^2-B_{\perp}^2)^{1/2}}{B_{\perp}B_z} r \Big), \label{eq11}
\end{eqnarray}
with $B_i^2=b^2+2\beta_i^2$ and $r_\perp^2=r_x^2+r_y^2$, and ${\rm Erf}(x)$ the error function. The second line of Eq.~(\ref{eq11}) is obtained from a simple calculation.
 
Such an intrinsically deformed ansatz is, of course, physically motivated by the observation of the rotational spectrum of $^8$Be indicating a large value of the corresponding moment of inertia. The minimization of the energy yields $\beta_\perp=\beta_x=\beta_y=1.78$ fm and $\beta_z=7.85$ fm. With these numbers, the squared overlap between the exact $\Psi_{2\alpha}$ and $\Psi_{2\alpha}^{\rm THSR}$ is with $0.9999$ extremely precise. In Fig.~\ref{fig:conv_rel_wf} we also show that the THSR wave function agrees very well even far out in the tail with the ``exact'' solution with $30$ ``Brink'' components.
 
As seen above, the single component, two parameter THSR ansatz, Eq.~(\ref{eq11}), for the relative wave function of two alpha's seems to grasp the physical situation extremely well. The most important part of this wave function is the outer one beyond some 3 fm. There, the two alpha's are in an $S$ wave of essentially Gaussian shape. The corresponding harmonic oscillator frequency is estimated to $\hbar \omega \sim 2$ MeV. Therefore, as long as the two alpha's do not overlap strongly, they swing in a very low frequency harmonic oscillator mode in a wide and delocalized fashion, reminiscent of a weakly bound gas like state. Inside the region $r < 2$-$3$ fm where the two alpha's heavily overlap, because of the strong action of the Pauli principle, the relative wave function has two nodes and small amplitude, as shown in Fig.~\ref{fig:conv_rel_wf}. Contrary to the outer part of the wave function determined dynamically, the behavior of the relative wave function in this strongly overlapping region is determined kinematically, solely reflecting the $r$-dependence of the norm kernel in Eq.~(\ref{eq:RGM_kernel}). This is clearly seen from the fact that both THSR and Brink wave functions have very nearly the same behavior in this region. Thus, we found that the alpha's in $^8$Be move practically as pure bosons in a relative $0S$ state of very low frequency as long as they do not come into one another's way, that is as long as they do not overlap. One should stress that this picture holds after projection on good total momentum and good spin, that is in the laboratory frame. It is equally true, as already mentioned, that in the intrinsic frame $^8$Be can be described as a strongly deformed two alpha structure, see ansatz (\ref{eq11}), reminiscent of a dumbbell.

%%%%%%%%%%%%%%%%%%%%%%%%%%%%%%%%%%%%%%%%%%%
\section{Alpha-gas like states in light nuclei}\label{sec:4}
%%%%%%%%%%%%%%%%%%%%%%%%%%%%%%%%%%%%%%%%%%%
 
%%%%%%%%%%%%%%%%%%%%%%%%%%%%%%%%%%%%%%%%%%%
\subsection{$^{12}$C case}\label{subsec:4-1}
%%%%%%%%%%%%%%%%%%%%%%%%%%%%%%%%%%%%%%%%%%%
 
\begin{figure}[t]
\begin{center}
\includegraphics[width=0.70\hsize]{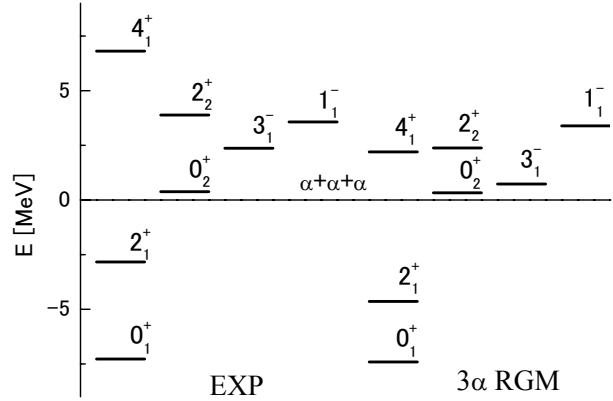}
\caption{Experimental energy spectra of $^{12}$C~\cite{ajzenberg86} together with the calculated ones using the $3\alpha$ RGM~\cite{kamimura}.}
\label{fig:energy_spectra_12c}
\end{center}
\end{figure}
 
\begin{figure}[t]
\begin{center}
\sidecaption
\includegraphics[width=0.50\hsize]{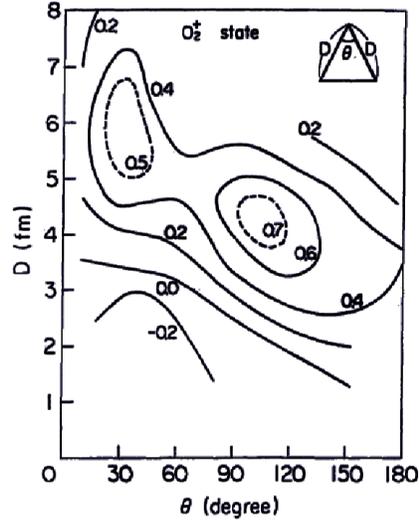}
\caption{Structure of the $0_2^+$ state shown by the overlap between the Brink-type cluster wave function of the isosceles configuration and the exact $0_2^+$ wave function. Figure adopted from Ref.~\cite{uegaki,ptp_supple_68}.}\label{fig:uegaki}
\end{center}
\end{figure}
 
The $\alpha$ cluster nature of $^{12}$C has been studied by many authors using various approaches. Figure~\ref{fig:energy_spectra_12c} shows the energy spectrum of $^{12}$C~\cite{ajzenberg86}. The $0^+_2$ state, located near the $3\alpha$ breakup threshold, is called the Hoyle state~\cite{hoyle,fowler}, which plays an astrophysically crucial role in the synthesis of $^{12}$C in the universe. Its small excitation energy of $7.65$ MeV is very difficult to explain by the shell model, even  using the most modern non-core shell model approach~\cite{nocore,navratil09}. The fully microscopic $3\alpha$ cluster models~\cite{uegaki,kamimura}, however, succeeded in the 1970s in explaining the observed data such as the small excitation energy and the inelastic form factor of the $(e,e')$ reaction etc., together with the structures of the ground-band states ($0^{+}_{1}-2^{+}_{1}-4^{+}_{1}$), $2^{+}_2$, and negative-parity states ($3^{-}_{1}-1^{-}_{1}$). The cluster model studies with the $3\alpha$ GCM (generator coordinate method)~\cite{uegaki} and $3\alpha$ RGM~\cite{kamimura} showed that the Hoyle state has a weakly interacting gas like $3\alpha$-cluster structure  with a very large radius (about $1/3$ of the ground-state density), whereas the ground state has a shell-model-like compact structure.
 
This $3\alpha$ gas-like nature of the Hoyle state is demonstrated in Fig.~\ref{fig:uegaki}, in which the overlap between a Brink-type wave function and the full RGM solution obtained by solving the $3\alpha$ RGM equation in Eq.~(\ref{rgm_eq}) is shown. The overlap is quite poor and in the best case the squared overlap reaches only about $50\%$.  This means that the $0^+_2$ state has a distinct clustering and has no definite spacial or geometrical configuration.  The situation is also pointed out in a recent work~\cite{chernykh07}, in which about $55$ components of the Brink-type wave functions are needed to reproduce accurately the full RGM solution for the Hoyle state. However, this Hoyle-state wave function is shown to be almost completely equivalent to a ''single THSR wave function`` as discussed in next section.

%%%%%%%%%%%%%%%%%%%%%%%%%%%%%%%%%%%%%%%%%%%%%%%%%%%%%%%%%%%%%%%%%%%%%%%%%%%%%
\begin{table}[t]
\begin{center}
\caption{Comparison of the total energies, r.m.s. radii $(R_{\rm r.m.s.})$, and monopole strengths $(M(0_2^+\rightarrow 0_1^+))$ for $^{12}$C given by solving Hill-Wheeler equation based on Eq.~(\ref{eq:2}) and by Ref.~\cite{kamimura}. The effective two-nucleon force Volkov No.~2~\cite{volkov65} was adopted in the two cases for which the $3\alpha$ threshold energy is calculated to be $-82.04$ MeV.}
\label{tab:thsr_12c}
\begin{tabular}{ccccc}
\hline\hline
 &  & \hspace*{5mm}{THSR w.f.}\hspace*{5mm} & \hspace*{5mm}{\raisebox{-1.8ex}[0pt][0pt]{$3\alpha$ RGM \cite{kamimura}}}\hspace*{5mm} & \hspace*{5mm}{\raisebox{-1.8ex}[0pt][0pt]{Exp.}}\hspace*{5mm} \\
 &  & (Hill-Wheeler) &  &  \\
\hline
\raisebox{-1.8ex}[0pt][0pt]{$E$~(MeV)} & $0_1^+$ & $-89.52$ & $-89.4$  & $-92.2$  \\
 & $0_2^+$ & $-81.79$ & $-81.7$  & $-84.6$  \\
\hline
\raisebox{-1.8ex}[0pt][0pt]{$R_{\rm r.m.s.}$~(fm)} & $0_1^+$ &   $\ \ \ 2.40$ &   $\ \ \ 2.40$ &   $\ \ \ 2.44$ \\
 & $0_2^+$ &   $\ \ \ 3.83$ &   $\ \ \ 3.47$ &  \\
\hline
$M(0_2^+\rightarrow 0_1^+)$~(fm$^2$) &  &   $\ \ \ 6.45$ &   $\ \ \ 6.7$ &   $\ \ \ 5.4$  \\
\hline\hline
\end{tabular}
\end{center}
\end{table}
%%%%%%%%%%%%%%%%%%%%%%%%%%%%%%%%%%%%%%%%%%%%%%%%%%%%%%%%%%%%%%%%%%%%%%%%%%%%%
 
%%%%%%%%%%%%%%%%%%%%%%%%%%%%%%%%%%%%%%%%%%%%%%%%%%%%%%%%%%%%%%
\subsubsection{THSR description of the Hoyle state}\label{subsec:4-1-1}
%%%%%%%%%%%%%%%%%%%%%%%%%%%%%%%%%%%%%%%%%%%%%%%%%%%%%%%%%%%%%%
 
\begin{figure}[t]
\begin{center}
%\sidecaption
\includegraphics[scale=0.50,clip]{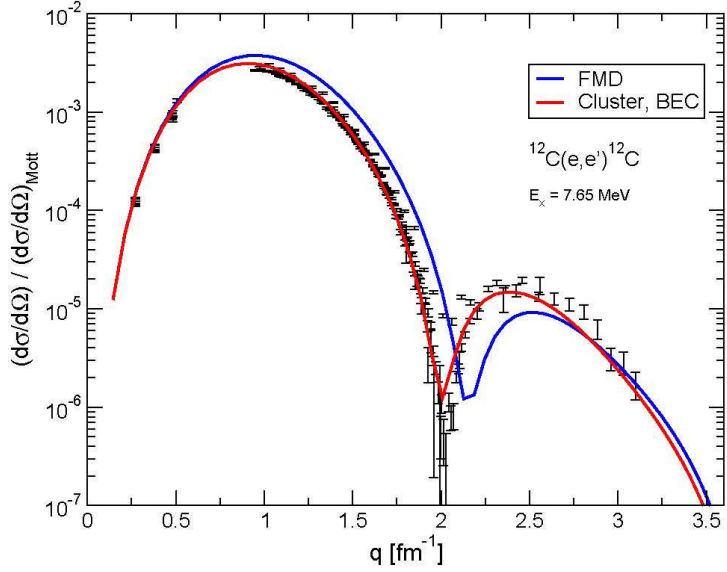}
\caption{(Color online) Comparison of the experimental inelastic form factor of $^{12}$C$(e,e')$ with the RGM (denoted by cluster), THSR (BEC) and FMD calculations. Figure is adopted from Ref.~\cite{chernykh07}.}\label{fig:neff}
\end{center}
\end{figure}

The total wave function for $^{12}$C in the THSR description is obtained by solving the Hill-Wheeler equation based on Eqs.~(\ref{eq:hwwf}), (\ref{eq:thsr}) and (\ref{eq:hw}).
Table~\ref{tab:thsr_12c} shows the results of the energies, r.m.s. radii, and monopole strengths in the THSR description together with those of the full $3\alpha$ RGM calculation and the data. One can see that the THSR description succeeds to reproduce the properties of the two $0^+$ states. Inspecting the r.m.s. radii, the Hoyle state has a volume 3 to 4 times larger than that of the ground state of $^{12}$C. The inelastic form factor of $^{12}$C from the ground state to the Hoyle state in the THSR description is displayed in Fig.~\ref{fig:neff}. We reproduce very accurately the experimental data.
 
In order to study how good a single $3\alpha$ THSR wave function reproduces the full RGM solutions, we use the THSR wave function with axially symmetric deformation,  presented as
\begin{eqnarray}
&&\Psi_{3\alpha}(\beta_{\perp},\beta_z)=\mathcal{A}\left\{ \chi_{3\alpha}^{\rm THSR}(\beta_{\perp},\beta_z) \phi_{\alpha}\phi_{\alpha}\phi_{\alpha} \right\}, \label{eq:tshr_wf_df}\\
&&\chi_{3\alpha}^{\rm THSR}(\beta_{\perp},\beta_z) = \exp \left[ -2 \sum_{i=1}^{2} \mu_{i} \left( \frac{\xi^{2}_{i\perp}}{b^2+2\beta^{2}_{\perp}} + \frac{\xi^{2}_{iz}}{b^2+2\beta^{2}_{z}}\right) \right],\label{eq:tshr_wf_df_xi}
\end{eqnarray}
where $\bm{\xi}_{1,2}$ are the two Jacobi coordinates with $\mu_1=1/2$ and $\mu_2=2/3$, and $\beta_{\perp}$ and $\beta_{z}$ are the deformation parameters with $\beta_{\perp}=\beta_x=\beta_y$. The wave function with good total spin $J=0$ is written as
\begin{eqnarray}
\Psi^{J=0}_{3\alpha}(\beta_{\perp},\beta_z) = \widehat{P}^{J=0} \Psi_{3\alpha}(\beta_{\perp},\beta_z), \label{eq:thsr_deformed_projected}
\end{eqnarray}
where $\widehat{P}^{J}$ is the angular momentum projection operator.
In what concerns the THSR wave function for the description of the Hoyle state, the situation is slightly more complicated than in the $^8$Be case by the fact that the loosely bound $3\alpha$ configuration is now no longer the ground state but the $0_2^+$ state at $7.65$ MeV excitation energy (as a side remark, let us mention that usually Bose-Einstein condensates of cold atoms also are not the ground states of the systems which are given by small crystals) . As discussed in Sec.~\ref{sec:2}, the wave function (\ref{eq:thsr_deformed_projected}) has the dominant configuration of the ground state in the limit of $\beta_{\perp}=\beta_{z}=0$. Thus, in order to discuss the Hoyle state, we have to use the $3\alpha$ wave function $\widetilde{\Psi}^{J=0}_{3\alpha}$ which is orthogonal to the ground state, expressed as
\begin{eqnarray}
\widetilde{\Psi}^{J=0}_{3\alpha}(\beta_{\perp},\beta_z) = \widehat{P}^{J=0} \widehat{P}^{\rm g.s}_{\perp} \Psi_{3\alpha}(\beta_{\perp},\beta_z),\label{eq:thsr_deformed_projected_Hoyle}
\end{eqnarray}
where $\widehat{P}^{\rm g.s}_{\perp}$ keeps the wave function in Eq.~(\ref{eq:thsr_deformed_projected_Hoyle}) to be orthogonal to the ground-state wave function, i.e.~$\widehat{P}^{\rm g.s}_{\perp} = 1 - {|0^{+}_1\rangle}{\langle 0^{+}_1|}$.
 
\begin{figure}[t]
\begin{center}
\includegraphics[width=0.49\hsize]{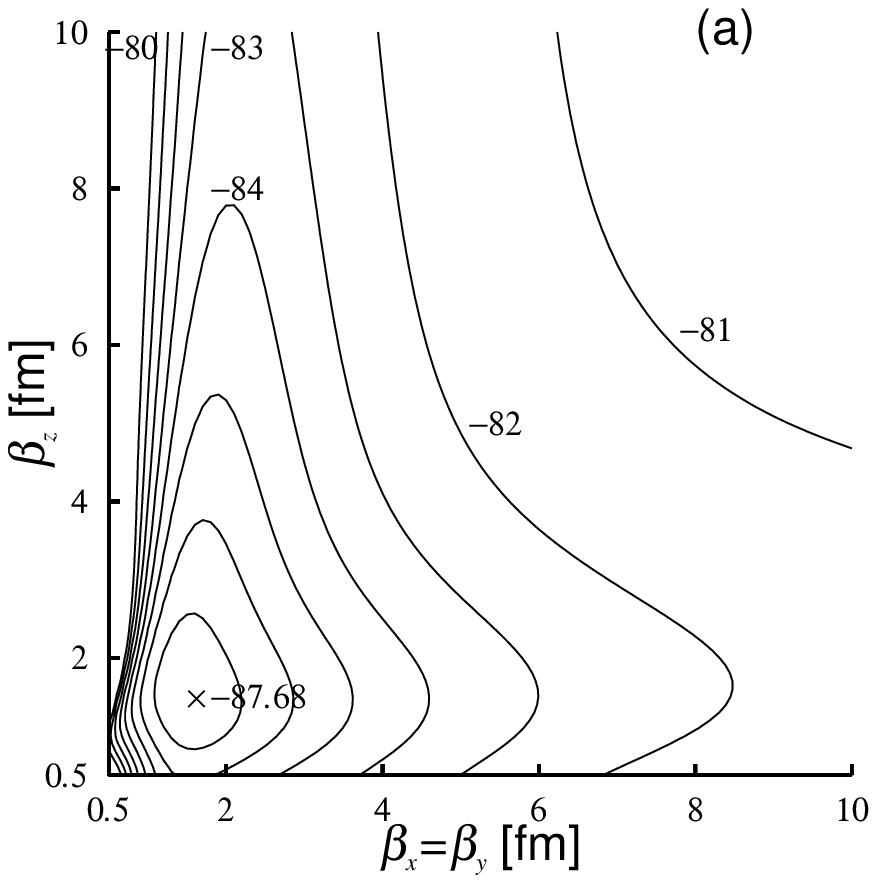}
%\hspace*{1mm}
\includegraphics[width=0.49\hsize]{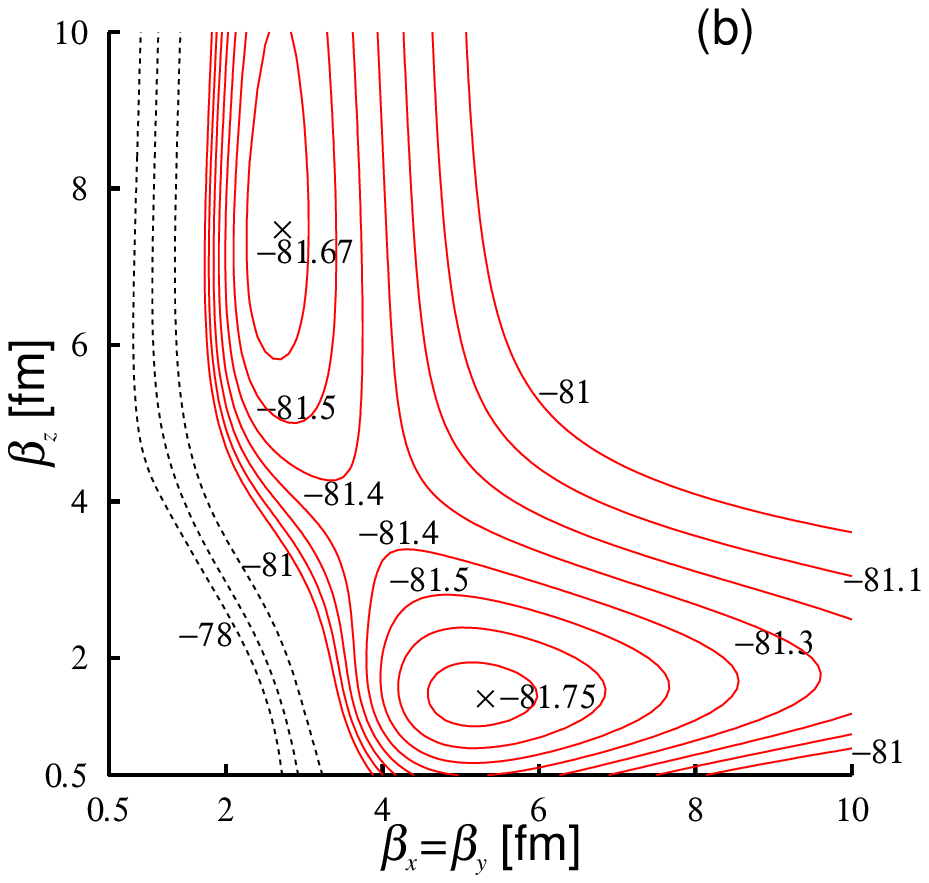}
\caption{(Color online) Contour map of the energy surface in the two parameter space $(\beta_{\perp},\beta_{z})$ for (left)~$\Psi^{J=0}_{3\alpha}(\beta_{\perp},\beta_z)$ in Eq.~(\ref{eq:thsr_deformed_projected}) and for (right)~$\widetilde{\Psi}^{J=0}_{3\alpha}(\beta_{\perp},\beta_z)$ in Eq.~(\ref{eq:thsr_deformed_projected_Hoyle}) orthogonal to the ground state.}\label{fig:counter_map_gs_hoyle}
\end{center}
\end{figure}
 
On the left side of Fig.~\ref{fig:counter_map_gs_hoyle}, we show the contour map of the energy surface corresponding to the state (\ref{eq:thsr_deformed_projected}) in the two parameter space $(\beta_{\perp},\beta_{z})$, defined as
\begin{eqnarray}
E(\beta_{\perp},\beta_{z}) = \frac{\langle \Psi^{J=0}_{3\alpha}(\beta_{\perp},\beta_z) | H | \Psi^{J=0}_{3\alpha}(\beta_{\perp},\beta_z) \rangle}{\langle \Psi^{J=0}_{3\alpha}(\beta_{\perp},\beta_z) | \Psi^{J=0}_{3\alpha}(\beta_{\perp},\beta_z) \rangle},
\end{eqnarray}
where $H$ is the microscopic Hamiltonian of $^{12}$C used in the $3\alpha$ RGM calculation.
One sees a minimum at $\beta_{\perp}=1.5$ fm and $\beta_z=1.5$ fm, which means a spherical shape. The minimum energy of $-87.68$ MeV is about $1.7$ MeV higher than the total energy of $-89.4$ MeV obtained by the full $3\alpha$ RGM calculation (see Table~\ref{tab:thsr_12c}). When the Hill-Wheeler equation in Eq.~(\ref{eq:hw}) is solved in the two-parameter space of $\beta_{\perp}$ and $\beta_{z}$, we can reproduce the total energy of the RGM result.
 
On the right side of Fig.~\ref{fig:counter_map_gs_hoyle}, the contour map of the energy surface corresponding to the state (\ref{eq:thsr_deformed_projected_Hoyle}) orthogonal to the ground state is displayed, where we use the ground-state solution of the Hill-Wheeler equation in the two-parameter space of $\beta_{\perp}$ and $\beta_{z}$. We see an energy minimum at $\beta_{\perp}=5.2$ fm and $\beta_z=1.5$ fm in the prolate region of the map and a second energy minimum at $\beta_{\perp}=2.6$ fm and $\beta_z=7.5$ fm in the oblate region. The minimum energy value is $-81.75$ MeV. This value is almost the same as the total energy of $-81.67$ MeV obtained by the full $3\alpha$ RGM (see Table~\ref{tab:thsr_12c}). The minimum energy of $-81.75$ MeV is close to the second minimum energy of $-81.67$ MeV, and there is a valley with an almost flat bottom connecting these two minima. This means that the energy of the spherical configuration is only slightly higher than that of the deformed configuration, that is, the energy gain due to the deformation is small.
 
\begin{figure}[t]
\begin{center}
\includegraphics[width=0.55\hsize]{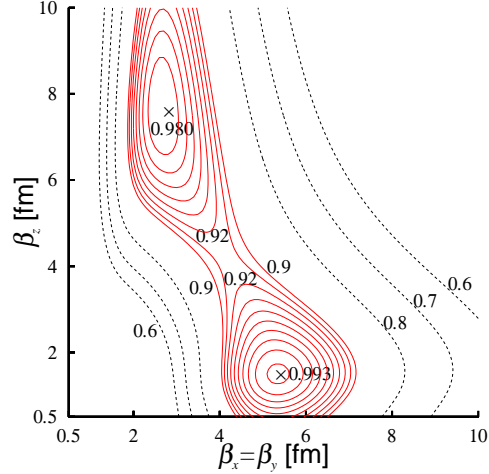}
\caption{(Color online) Contour map of the squared overlap of the normalized THSR wave function $\widetilde{\Psi}^{J=0}_{3\alpha}(\beta_{\perp},\beta_z)$ in Eq.~(\ref{eq:thsr_deformed_projected_Hoyle}), orthogonal to the ground state, with the full RGM solution.}
\label{fig:olpsfc_ort_vk2}
\end{center}
\end{figure}
 
\begin{figure}[t]
\begin{center}
\includegraphics[width=0.55\hsize]{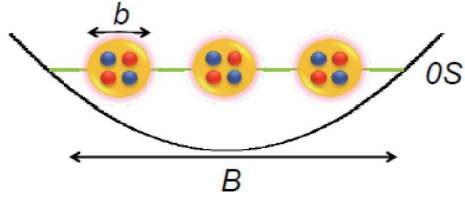}
\caption{(Color online)~Pictorial representation of the THSR wave function for $n=3$ ($^{12}$C). The three $\alpha$-particles are trapped in the $0S$-state of a wide harmonic oscillator $(B)$ and the four nucleons of each $\alpha$ are confined in the $0s$-state of a narrow one $(b)$. All nucleons are antisymmetrized.}
\label{fig:cartoon}
\end{center}
\end{figure}
 
A very remarkable result from the right side of Fig.~\ref{fig:counter_map_gs_hoyle} is that the wave function at the minimum energy point ($\beta_{\perp}=5.3$ fm and $\beta_z=1.5$ fm) has $99.3$~\% squared overlap with the full RGM solution (see Fig.~\ref{fig:olpsfc_ort_vk2}), although the spherical wave function ($\beta_{\perp}=\beta_{z}=4.0$ fm) gives already a squared overlap of $92$~\%. The THSR wave function Eq.~(\ref{eq:tshr_wf_df}) is of Gaussian type with a wide extension, centered at the origin. It is completely different from a Brink type wave function with the three $\alpha$-particles placed at definite values in space. A slight improvement of Eq.~(\ref{eq:thsr_deformed_projected_Hoyle}) can still be achieved in taking the $\beta_i$ parameters as Hill-Wheeler coordinates and superpose a couple of wave functions of the type (\ref{eq:thsr_deformed_projected_Hoyle}) with different width parameters. Practically $100$\% squared overlap with the wave function of the full RGM result is then achieved. It should be pointed out that the superposition of several Gaussians of the type (\ref{eq:thsr_deformed_projected_Hoyle}) does not at all change the physical content of the THSR wave function as a wide extended distribution centered around the origin. Therefore, the Hoyle state can be seen as three almost inert $\alpha$-particles moving in their own mean field potential, to good approximation given by a wide harmonic oscillator, whereas the $\alpha$'s are represented by four nucleons captured in narrow harmonic potentials. The situation is given as a cartoon in Fig.~\ref{fig:cartoon}.

%%%%%%%%%%%%%%%%%%%%%%%%%%%%%%%%%%%%%%%%%%%%%%%%%%%%%%%%%%%%%%
\subsubsection{Influence of antisymmetrization and orthogonalization}\label{subsec:4-1-2}
%%%%%%%%%%%%%%%%%%%%%%%%%%%%%%%%%%%%%%%%%%%%%%%%%%%%%%%%%%%%%%
 
A crucial question is whether for the Hoyle state the THSR wave function (\ref{total_wf_rgm}) with (\ref{eq:thsr_xi}) can be considered to good approximation as a product state of $\alpha$ particles condensed with their c.o.m. motion into the $0S$ orbital. For this, one has to quantify the influence of the antisymmetrizer $\mathcal{A}$ in Eq.~(\ref{total_wf_rgm}).  A direct way to measure the influence of antisymmetrization is to study the following expectation value of the antisymmetrizer $\mathcal{A}$,
\begin{equation}
N(B) = \frac{\langle B|{\cal A}|B\rangle}{\langle B | B \rangle},
\end{equation}
where $|B\rangle$ is the THSR wave function in Eq.~(\ref{eq:6}) without the antisymmetrization, that is, just the product state $\psi_{\alpha_1}\psi_{\alpha_2}\psi_{\alpha_3}$ in Eq.~(\ref{eq:7}).
The normalization of the antisymmetrizer $\mathcal{A}$ is chosen so that $N(B)$ becomes unity in the limit where the intercluster overlap disappears, i.e.~for the with parameter $B \rightarrow \infty$.
 
\begin{figure}[t]
\begin{center}
\includegraphics[scale=0.60]{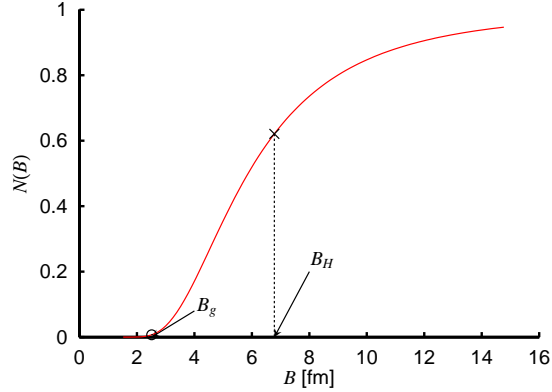}
\caption{(Color online) Expectation value of the antisymmetrization operator for the product state $|B\rangle$. The value at the optimal $B$ values, $B_g$ for the ground state and $B_H$ for the Hoyle state, are denoted by a circle and a cross, respectively.}
\label{fig:thsr_effect_antisymmetrization}
\end{center}
\end{figure}
 
The result of $N(B)$ is shown in Fig.~\ref{fig:thsr_effect_antisymmetrization} as a function of the width parameter $B$. We chose, as optimal values of $B$ for describing the ground and Hoyle states, $B=B_g=2.5$ fm and $B=B_H=6.8$ fm, for which the normalized THSR wave functions give the best approximation of the ground state $0^+_1$ and the Hoyle state $0^+_2$, respectively, which are obtained by solving the Hill-Wheeler equation~(\ref{eq:hw}). The  squared overlaps are $0.93$ and $0.78$, respectively. From Fig.~\ref{fig:thsr_effect_antisymmetrization} we find that $N(B_H)\sim 0.62$ and $N(B_g)\sim 0.007$. These results indicate that the influence of the antisymmetrization is strongly reduced in the Hoyle state compared with the influence in the ground state. An important point in the present consideration is that the THSR wave function at $B=B_H$ is not automatically orthogonal to the ground state. This is contrary to the situation with condensed cold bosonic atoms, for which the density is so low that the overlap of the electron clouds can, on average, be totally neglected. In the present case, the squared overlap of $|{\rm THSR}(B=B_H)\rangle$ with $|{\rm THSR}(B=B_g)\rangle$ (or with the ground state $0^+_1$ obtained by solving the Hill-Wheeler equation) is less than $0.12$. This small value indicates that the orthogonality with the ground state is nearly realized.
 
\begin{figure}[t]
\begin{center}
\includegraphics[scale=0.60]{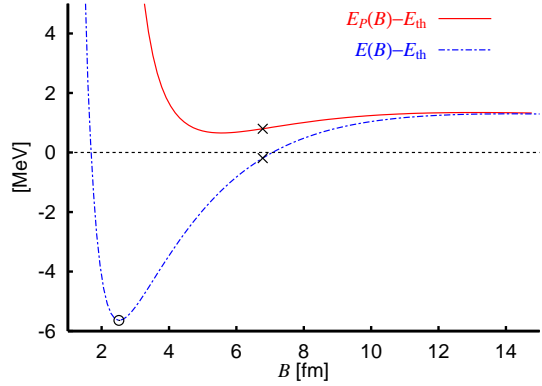}
\caption{(Color online) Energy curve in the orthogonal space to the ground state, denoted by $E_P(B)$, together with $E(B)$. The values at the optimal $B$ values, $B_g$ and $B_H$ for the ground state and Hoyle state, respectively, are marked by a circle and a cross, respectively.}
\label{fig:energy_curve_THSR}
\end{center}
\end{figure}
 
An explicit orthogonalization with $|{\rm THSR}(B)\rangle$ to the ground state $0^+_1$ obtained by solving the Hill-Wheeler equation gives non-negligible effects for a  quantitative description of the Hoyle state with the THSR wave function. As mentioned in Sec.~\ref{subsec:4-1-1}, the normalized THSR wave function orthogonal to the ground state $0^+_1$ ($\sim\hat{P}^{\rm (g.s)}_{\perp} |{\rm THSR}(B)\rangle$) gives a squared overlap of $0.92$ (for $B=6.1$ fm) with the $0^+_2$ state obtained by solving the Hill-Wheeler equation, although the squared overlap using the normalized THSR wave function without the orthogonalization gives already a value of $0.78$. In addition, as shown in Fig.~\ref{fig:energy_curve_THSR}, the energy curves for the THSR wave function,
\begin{equation}
E(B)=\frac{\langle {\rm THSR}(B) | H | {\rm THSR}(B)\rangle}{\langle {\rm THSR}(B) | {\rm THSR}(B)\rangle},
\end{equation}
indicates a minimum corresponding to the ground state at $B \sim B_g$, but the second minimum corresponding to the Hoyle state is not present.
This is due to the fact that the THSR state with $B=B_H$, $ | {\rm THSR}(B=B_H)\rangle$, still includes the ground-state component of about $10$~\%, as mentioned above. In fact, if one calculates the energy taking into account the explicit orthogonalization to the ground state,
\begin{equation}
E_{P}(B) = \frac{\langle {\hat{P}^{\rm (g.s)}_{\perp}}{\rm THSR}(B) | H | {\hat{P}^{\rm (g.s)}_{\perp}}{\rm THSR}(B)\rangle} {\langle {\hat{P}^{\rm (g.s)}_{\perp}}{\rm THSR}(B) | {\hat{P}^{\rm (g.s)}_{\perp}}{\rm THSR}(B)\rangle},
\end{equation}
there appears the minimum corresponding to the Hoyle state at $B \sim B_H$, as shown in Fig.~\ref{fig:energy_curve_THSR}. Thus, the small admixture of the ground-state components to the Hoyle state is never negligible, and explicit elimination by $\hat{P}^{\rm (g.s)}_{\perp}$ plays an essential role to describe the Hoyle state. It is true that the effect of the antisymmetrization is not negligible even for the Hoyle state in the sense that the projection operator $\hat{P}^{\rm (g.s)}_{\perp}$ excludes the compact ground-state components which are strongly subject to the antisymmetrizer. Nevertheless, it is worth emphasizing that as a result of the explicit orthogonalization to the ground state, the Hoyle state can not have a compact structure but has a dilute density, for which, in the end, the effect of antisymmetrization is small.

\begin{figure}[t]
\begin{center}
\includegraphics[width=0.60\hsize]{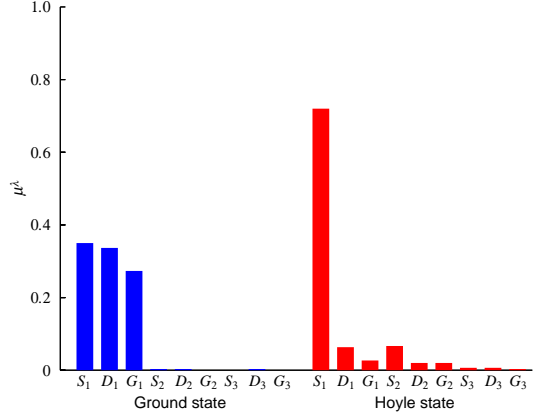}
\caption{(Color online) Occupation of the single-$\alpha$ orbitals of the Hoyle state of $^{12}$C compared with the ground state}
\label{fig:12C_occupation_probability}
\end{center}
\end{figure}
 
\begin{figure}[t]
\begin{center}
\includegraphics[width=0.65\hsize]{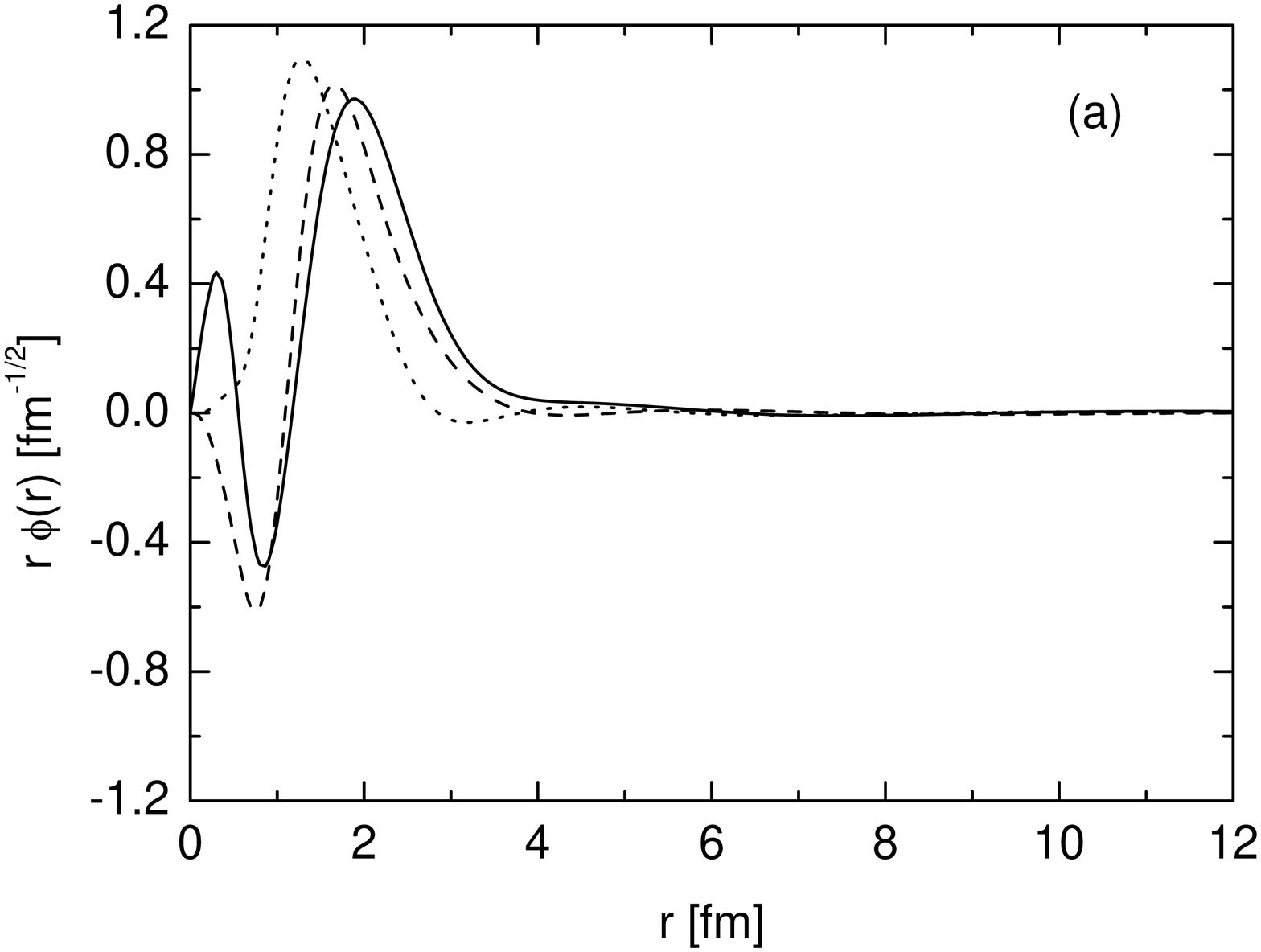}\\%\hspace*{-5mm}
\includegraphics[width=0.65\hsize]{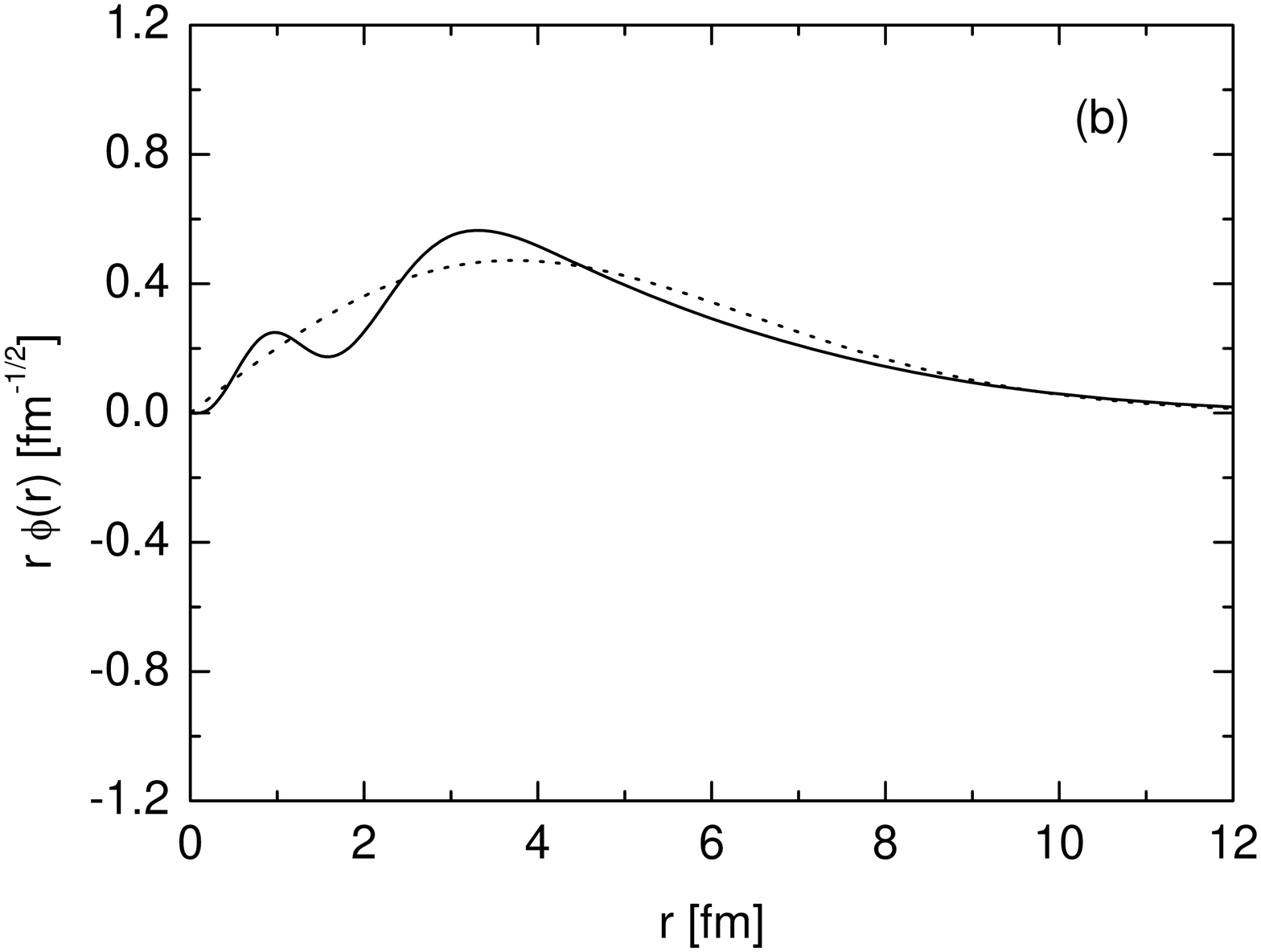}
\caption{Radial parts of the single $\alpha$ orbits, (a)~$S_1$ (solid line), $D_1$ (dashed) and $G_1$ (dotted), in the $0^+_1$ state, and (b)~the $S_1$ (solid) orbit in the $0^+_2$ state compared with an $S$-wave Gaussian function (dotted), $r\varphi_{0s}$, with the size parameter $B=3.6$ fm (see text)~\cite{yamada05}. Note that all the radial parts in figures are multiplied by $r$.}\label{fig:12C_alpha_orbits}
\end{center}
\end{figure}

%%%%%%%%%%%%%%%%%%%%%%%%%%%%%%%%%%%%%%%%%%%%%%%%%%%%%%%%%%%%%%%%%%%%%%%%%%%%%%%%%
\subsubsection{Alpha-particle occupation probabilities, momentum distribution, and the de Broglie wave length in the Hoyle state}\label{subsec:4-1-3}
%%%%%%%%%%%%%%%%%%%%%%%%%%%%%%%%%%%%%%%%%%%%%%%%%%%%%%%%%%%%%%%%%%%%%%%%%%%%%%%%%
 
Direct quantities indicating how well the Hoyle state is described by a product state of three $\alpha$'s are the $\alpha$-particle occupation probabilities and single particle orbits, which are obtained by diagonalizing the internal single $\alpha$-particle density matrix $\rho^{(1)}_{\rm int}(\bm{q},\bm{q}')$ defined in Eq.~(\ref{eq:one_particle_density_matrix}). The occupation of the single-$\alpha$ orbits of the Hoyle state is shown in Fig.~\ref{fig:12C_occupation_probability}. One finds that the $\alpha$ particles occupy the $S_1$ orbit to over $70$~\%, and those for other orbits are very small.  This means that each of the three $\alpha$ particles in the $0^+_2$ state is in the $S_1$ orbit with occupation probability as large as about $70$~\%. The radial behavior of the $S_1$ orbit is illustrated with the solid line in Fig.~\ref{fig:12C_alpha_orbits}(b). We see no nodal behavior but small oscillations in the inner region ($r<4$ fm) and a long tail up to $r\sim$10 fm. For reference, the radial behavior of the $S$-wave Gaussian function, $\varphi_{0s}(r)=N_{0s}(B)\exp(-r^2/(2B^2))$, is drawn with the dashed line in Fig.~\ref{fig:12C_alpha_orbits}(b), where the size parameter $B$ is chosen to be $3.6$ fm, and $N_{0s}(B)$ denotes the normalization factor. The radial behavior of the $S_1$ orbit is similar to that of the $S$-wave Gaussian function, in particular, in the outer region ($r>4$ fm), whereas a slight oscillation of the former around the latter can be seen in the inner region ($r<4$ fm). Thus, the Hoyle state can be described as the product state of $(0S)_{\alpha}^{3}$ being realized with a probability of over $70$~\%.
 
In the case of the ground state of $^{12}$C, the $\alpha$-particle occupations are equally shared between $S_1$, $D_1$ and $G_1$ orbits (see Fig.~\ref{fig:12C_occupation_probability}), thus invalidating a condensate picture for the ground state. These occupancies can be explained quite well from the following fact:~The ground state has as main configuration  the SU(3) shell model wave function $(\lambda,\mu)=(04)$. Figure~\ref{fig:12C_alpha_orbits}(a) demonstrates the radial parts for the $S_1$-, $D_1$- and $G_1$-orbits, the number of nodes of which are two, one and zero, respectively. Reflecting the SU(3) character, the radial behavior of the three orbits is similar to those of the harmonic oscillator wave functions ($u_{NL}$) with $Q=4$, $u_{02}$, $u_{21}$ and $u_{40}$, respectively, where $N$ ($L$) denotes the number of nodes (orbital angular momentum). We see that the radial parts of the single $\alpha$-particle orbits oscillate strongly in the inside region ($r<4$ fm). This is due to the important Pauli blocking effect for the ground state with its compact shell-model-like structure.
 
Another important quantity to demonstrate the $3\alpha$ condensate nature of the Hoyle state is the momentum distribution of a single-$\alpha$ particle. It is defined as a double Fourier transformation of the internal single-$\alpha$ density matrix $\rho^{(1)}_{\rm int}(\bm{q},\bm{q}')$ defined in Eq.~(\ref{eq:one_particle_density_matrix}),
\begin{eqnarray}
\rho(k)=\int d\vc{q}' d\vc{q} \frac{e^{i\vc{k}\cdot\vc{q}'}}{(2\pi)^{3/2}} \rho^{(1)}_{\rm int}(\vc{q},\vc{q}')\frac{e^{-i\vc{k}\cdot\vc{q}}}{(2\pi)^{3/2}},\hspace*{7mm}\int d\vc{k}\rho(k)=1,\label{eq:rho_k}
\end{eqnarray}
Let us remind that $\rho(k)$ would have a $\delta$-function like peak around $k=0$ for an ideal dilute condensed state in homogeneous infinite matter.
 
\begin{figure}[t]
\begin{center}
\includegraphics[width=0.70\hsize]{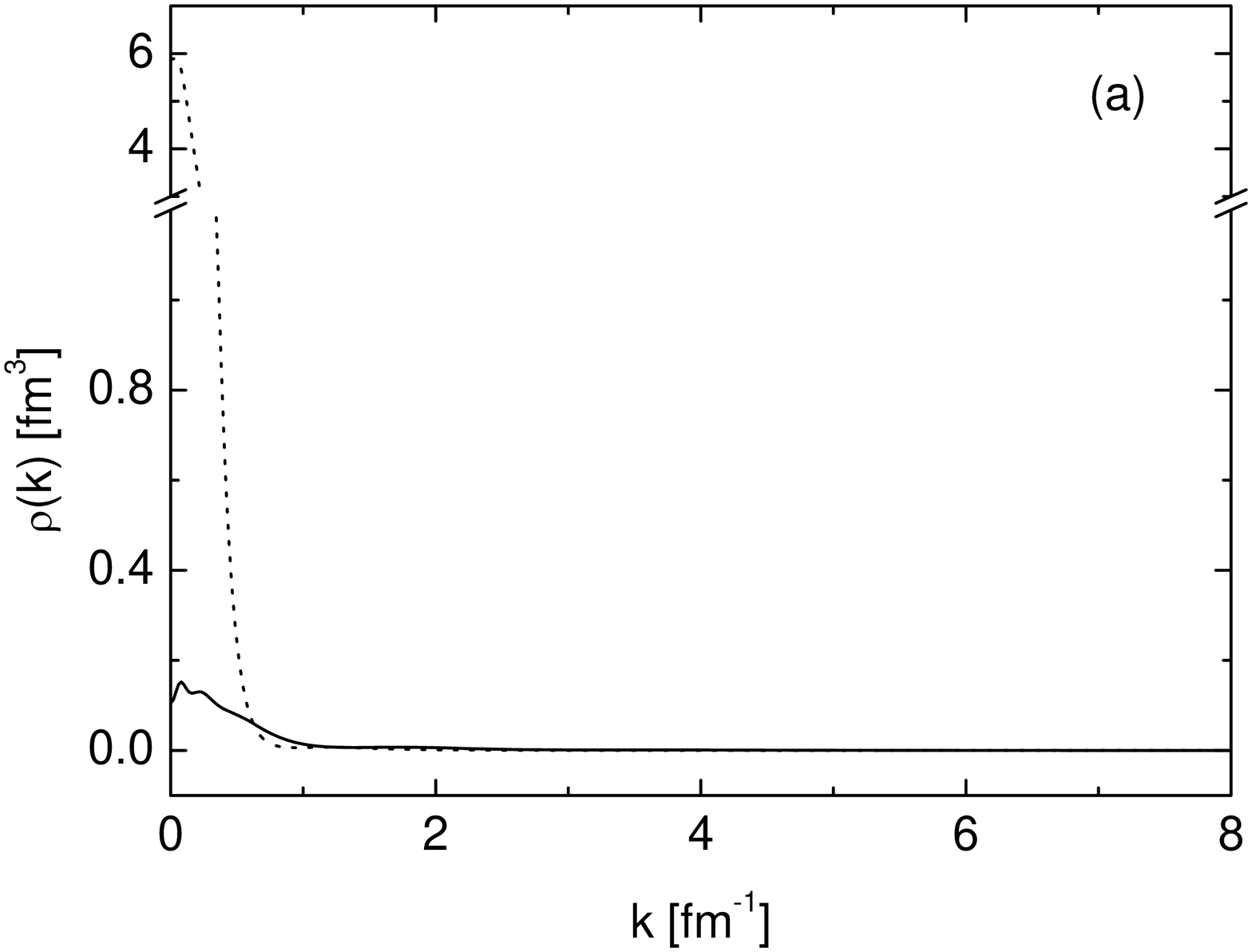}\\
\includegraphics[width=0.70\hsize]{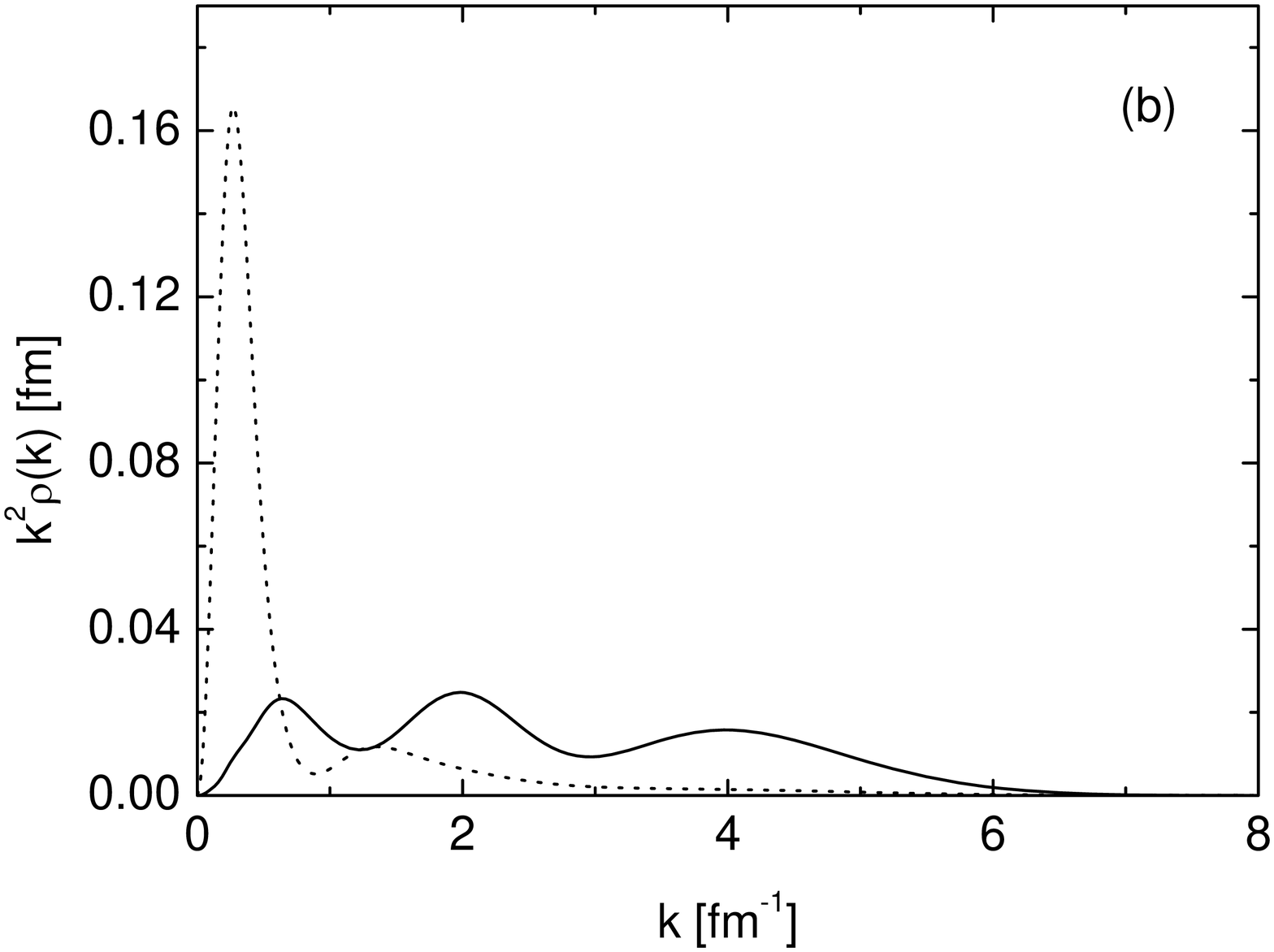}
\caption{Momentum distribution of the $\alpha$ particle in $^{12}$C, (a)~$\rho(k)$ and (b)~$k^2\rho(k)$, for the $0^+_1$ (solid line) and $0^+_2$ (dotted) states~\cite{yamada05}. }
\label{fig:momentum_distribution_12C}
\end{center}
\end{figure}
 
The momentum distributions of the $\alpha$ particle, $\rho(k)$ and $k^2\times\rho(k)$, are shown for the $0^+_1$ and $0^+_2$ states in Fig.~\ref{fig:momentum_distribution_12C}. Reflecting the dilute structure of the Hoyle, we see a strong concentration of the momentum distribution in the $k<1$ fm$^{-1}$ region, and the behavior of $\rho(k)$ is of the $\delta$-function type, similar to the momentum distribution of the dilute neutral atomic condensate states at very low temperature trapped by an external magnetic field~\cite{dalfovo99}. On the other hand, the ground state has higher momentum components up to $k\sim6$ fm$^{-1}$ as seen from the behavior of $k^2\times\rho(k)$ reflecting the compact structure. The above results for the radial behavior of the $S_1$ orbit, occupation probability and momentum distribution for the $0^+_2$ state again lead us to conclude that this state is of the $3\alpha$ condensate character with as much as about $70 \%$ occupation probability.
 
The de Broglie wave length of the $\alpha$'s moving in the Hoyle state is an interesting quantity. It can be estimated from the resonance energy of $^8$Be being roughly $100$ keV. Otherwise, one can estimate the kinetic energy of the $\alpha$-particles from a bosonic mean field picture using the Gross-Pitaevskii equation~\cite{yamada04} (see Sec.~\ref{subsec:4-3}). The mean field potential of $\alpha$-particles in the Hoyle state (see Fig.~\ref{fig:gross_pitaevskii}) indicates the position of the single $\alpha$ particle energy (180~keV). The kinetic energy of the single $\alpha$ particle is calculated to be 380~keV. From this, the de Broglie wave length $\lambda = 2\pi(\frac{2M_{\alpha}}{\hbar^2}E_{\alpha})^{-1/2}$ is, therefore, estimated to be of a lower limit of approximately $20$ fm. A more reliable estimate of the de Broglie wave length is to use the expectation value of $k^2$ for the wave number $k$ of the $\alpha$ particle in the Hoyle state, evaluated from the momentum distribution of the alpha particle, $\rho(k)$, in Fig.~\ref{fig:momentum_distribution_12C}, obtained by a $3\alpha$ OCM calculation~\cite{yamada05}. The result is $\lambda=2\pi/\sqrt{\langle k^2 \rangle}\sim 20$ fm, consistent with the previous value. These estimates all indicate that the de Broglie wave length is much longer than the inter $\alpha$-particle distance, contrary to what is claimed in Ref.~\cite{zinner07}, using qualitative arguments.

%%%%%%%%%%%%%%%%%%%%%%%%%%%%%%%%%%%%%%%%%%%%%%%%%%%%%%%%%%%%%%%%%%%
\subsubsection{Family of the Hoyle state: $2^{+}_{2}$ and $0^{+}_{3}$}\label{subsec:4-1-4}
%%%%%%%%%%%%%%%%%%%%%%%%%%%%%%%%%%%%%%%%%%%%%%%%%%%%%%%%%%%%%%%%%%%
 
In the previous section, we found that the Hoyle state has a dilute $3\alpha$-condensate-like structure with a main configuration of $(0S)^{3}_{\alpha}$. Then, an excited state of the Hoyle state, for example, a $2^+$ state with $(0S)^{2}_{\alpha}(0D)_{\alpha}$, may exist somewhat higher up in energy than the Hoyle state.  Itoh et al.~observed the $2^{+}_{2}$ state at $2.6\pm0.3$ MeV above the $3\alpha$ threshold with a width of $1.0\pm0.3$~MeV by measuring $\alpha$ particles decaying from excited $^{12}$C states with inelastic $\alpha$ scattering~\cite{itoh04}. This state was quite recently confirmed by experiment with a high-energy-resolution magnetic spectrometer~\cite{freer09}.
 
\begin{figure}[t]
\begin{center}
%\sidecaption
\includegraphics[width=0.7\hsize,angle=270]{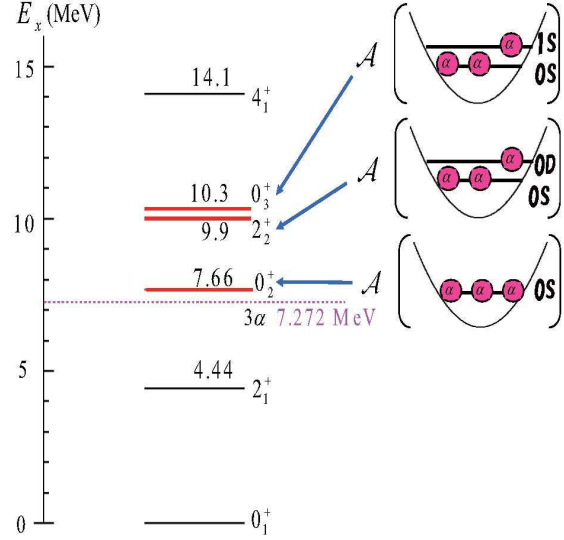}
\caption{(Color online) Theoretical interpretation of the $0^+_2$, $2^+_2$ and $0^+_3$ states.}
\label{fig:interpretation_12C}
\end{center}
\end{figure}
 
A deformed calculation using the THSR wave function in Eqs.~(\ref{eq:tshr_wf_df}) and (\ref{eq:tshr_wf_df_xi}) was performed for the $2^+$ state of $^{12}$C. Projecting on good angular momentum with a treatment of resonances yields the position of the $2^{+}_{2}$-state in $^{12}$C ($2.1$~MeV above $3\alpha$ threshold) which is in good agreement with the experimental value~\cite{itoh04,funaki05}. Also the calculated width ($0.64$~MeV) gives a quite reasonable estimate of the data. Detailed investigation of the wave function of the $2_2^+$-state shows that it can essentially be described in lifting out of the condensate state with the three $\alpha$'s in the $0S$-orbit, one $\alpha$-particle in the next $0D$-orbit. It is tempting to imagine that the $0_3^+$-state which, experimentally, is almost degenerate with the $2_2^+$-state, is obtained by lifting one $\alpha$-particle into the $1S$-orbit. Preliminary theoretical studies~\cite{kurokawa05} indicate that this scenario might indeed apply. However, the width of the $0_3^+$ state is very broad ($\sim$ 3 MeV), rendering a theoretical treatment rather delicate. Further investigations are necessary to validate or reject this picture which is shown graphically in Fig.~\ref{fig:interpretation_12C}.  Anyway, it would be quite satisfying, if the triplet of states, ($0_2^+,2_2^+, 0_3^+$) could all be explained from the $\alpha$-particle perspective, since those three states are {\it precisely} the ones which cannot be explained within a (no core) shell model approach~\cite{nocore,navratil09}.

%%%%%%%%%%%%%%%%%%%%%%%%%%%%%%%%%%%%%%%%%%%%%%%%%%%%%%%%%%%%%%%%%%%
\subsubsection{Precursors of a $3\alpha$ condensate state: $3^{-}_{1}$ and $1^{-}_{1}$}\label{subsec:4-1-5}
%%%%%%%%%%%%%%%%%%%%%%%%%%%%%%%%%%%%%%%%%%%%%%%%%%%%%%%%%%%%%%%%%%%
 
The $3^-$ state at $E_{3\alpha}=2.37$~MeV measured from the $3\alpha$ threshold in Fig.~\ref{fig:energy_spectra_12c} is interesting from the point of view of the dilute $\alpha$ condensation. If the state is a condensate with all of the $3\alpha$ particles in the $P$ orbit, there is the possibility of a superfuid with a vortex line, similar to the rotating dilute atomic condensate at very low temperature~\cite{dalfovo99}. Thus, it is an intriguing problem to study the structure within the $3\alpha$ OCM. The OCM~\cite{yamada05} reproduces well the energy of the $3^{-}_{1}$ state as well as the $1^{-}_{1}$ state ($E_{3\alpha}=3.57$~MeV) with respect to the $3\alpha$ threshold.
 
The calculated nuclear radius of the $3^-$ state is 2.95 fm, the value of which is larger than that for the ground state ($0^+_1$), while it is smaller than that for the $0^+_2$ state. This suggests that the structure of the $3^-$ state is intermediate between the shell-model-like compact structure ($0^+_1$) and the dilute $3\alpha$ structure ($0^+_2$). The occupation probabilities of the single-$\alpha$ orbits of the state are $44.7 \%$ for $P_1$-orbit and $27.9 \%$ for $F_1$-orbit. Although the concentration of the single $P_1$ orbit amounts to about $50 \%$, the radial behavior of the single-$\alpha$ orbit has two nodes in the inner region. However, the amplitude of the inner oscillations is significantly smaller than that for the ground state in Fig.~\ref{fig:12C_alpha_orbits}(a)~\cite{yamada05}. The small oscillations indicate a weak Pauli-blocking effect, and thus, we can see the precursor of a $3\alpha$ condensate state~\cite{yamada05}, although the $3^-$ state is not an ideal rotating dilute $3\alpha$ condensate.
 
As for the $1^{-}_1$ state, the calculated nuclear radius, 3.32 fm, is larger than that of the ground state and the $3^-_1$ state but is still smaller than that of the $0^+_2$ one. The occupation probabilities of the $\alpha$ particles in the $1^-_1$ state are $35~\%$ for $P_1$ orbit and $16~\%$ for $F_1$ orbit. Thus, there is no concentration of the occupation probability to a single orbit like in the $0^+_2$ state. Since the $\alpha$ particles in the $1^-_1$ state are distributed over several orbits, the state is not of the dilute $\alpha$-condensate type. On the other hand, the radial behavior of the $P_1$ orbit has two nodes in the inner region, the behavior of which is rather similar to the $2P$ harmonic oscillator wave function. However, the $F_1$ orbit has a $F$-wave Gaussian-type behavior. Also the oscillatory behavior of the $F_1$ orbit for $ 0 < r < 2 $ fm is similar to the one of the $S_1$ orbit in the $0^+_2$ state in Fig.~\ref{fig:12C_alpha_orbits}. These interesting behaviors of the $F_1$ orbit indicate some signal of dilute $\alpha$ condensation, reflecting the relatively large nuclear radius (3.32 fm) of the $1^-_1$ state.

%%%%%%%%%%%%%%%%%%%%%%%%%%%%%%%%%%%%%%%%%%%
\subsection{$^{16}$O case}\label{subsec:4-2}
%%%%%%%%%%%%%%%%%%%%%%%%%%%%%%%%%%%%%%%%%%%
 
In the previous section, we showed that the Hoyle state, which has about one third of saturation density, can be described, to good approximation, as a product state of three $\alpha$-particles, condensed, with their c.o.m. motion, into the lowest $\alpha$ mean field $0S$-orbit~\cite{matsumura04,yamada05}. These novel aspects indicate that the Hoyle state has a $3\alpha$-condensate-like structure. The establishment of the novel aspect of the Hoyle state naturally leads us to speculate about the possibility of $4\alpha$-particle condensation in $^{16}$O.
 
The experimental $0^+$ spectrum of $^{16}$O up to about the $4\alpha$ disintegration threshold is shown in Fig.~\ref{fig:energy_spectra_4a_ocm_thsr}. In the past, the $0^{+}_{1}$ (g.s), $0^{+}_{2}$ and $0^{+}_{3}$ states up to about $13$ MeV excitation energy has very well been reproduced with a semi-microscopic cluster model, i.e.~the $\alpha + ^{12}$C OCM (Orthogonality Condition Model)~\cite{Suz76}. In particular, this model calculation, as well as that of an $\alpha+^{12}$C GCM (Generator-Coordinate-Method) one~\cite{baye2}, demonstrates that the $0_2^+$ state at $6.05$ MeV and the $0_3^+$ state at $12.05$ MeV have $\alpha + ^{12}$C structures~\cite{Hor68} where the $\alpha$-particle orbits around the $^{12}$C$(0_1^+)$-core in an $S$-wave and around the $^{12}$C$(2_1^+)$-core in a $D$-wave, respectively. Consistent results were later obtained by the $4\alpha$ OCM calculation within the harmonic oscillator basis~\cite{Kat92}. However, the model space adopted in Refs.~\cite{Suz76,Kat92,baye2} is not sufficient to account simultaneously for the $\alpha+ ^{12}$C and the $4\alpha$ gas-like configurations. On the other hand, the $4\alpha$-particle condensate state was first investigated in Ref.~\cite{thsr} and its existence was predicted around the $4\alpha$ threshold with the THSR wave function. While the THSR wave function can well describe the dilute $\alpha$ cluster states as well as shell model like ground states, other structures such as $\alpha + ^{12}$C clustering can not be treated and are only incorporated in an average way. Thus, it is important to explore the $4\alpha$ condensate without any a priori assumption with respect to the structure of the $4\alpha$ system. For this purpose, a full four-body $4\alpha$ OCM calculation with Gaussian basis functions was performed~\cite{funaki08}. This model space is large enough to cover the $4\alpha$ gas, the  $\alpha +^{12}$C cluster, as well as the shell-model configurations.  In this section, we first present the results of the $4\alpha$ OCM calculation, and then discuss a recent analysis with the THSR wave function for the $4\alpha$ system.
 
%%%%%%%%%%%%%%%%%%%%%%%%%%%%%%%%%%%%%%%%%%%
\subsubsection{$4\alpha$ OCM analysis}\label{subsec:4-2-1}
%%%%%%%%%%%%%%%%%%%%%%%%%%%%%%%%%%%%%%%%%%%
 
\begin{figure}[t]
\begin{center}
\sidecaption
\includegraphics[width=60mm]{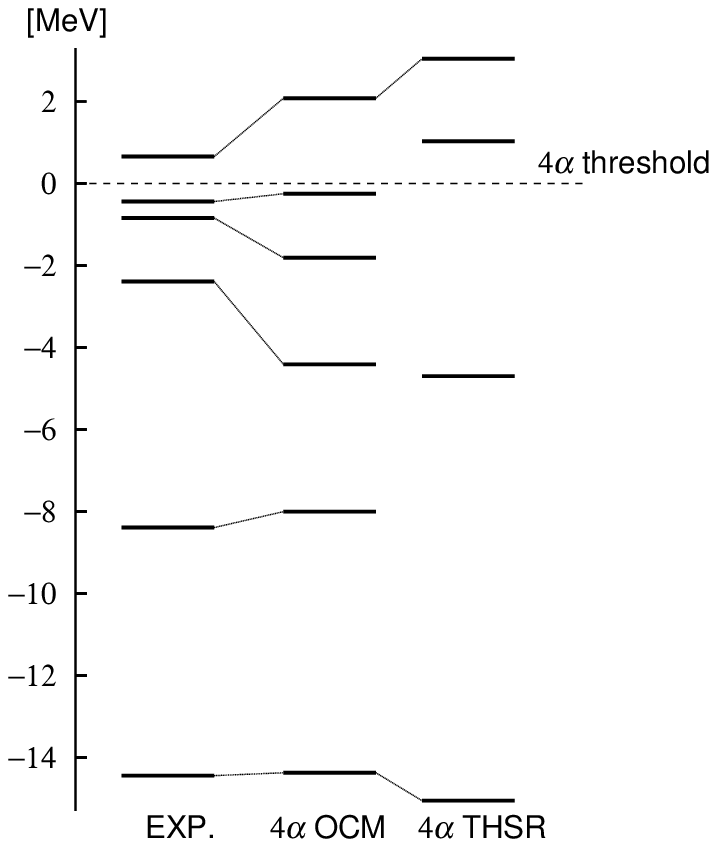}
\caption{Comparison of energy spectra among experiment, the $4\alpha$ OCM calculation~\cite{funaki08}, and the THSR treatment~\cite{funaki10}.  Dotted line denotes the $4\alpha$ threshold. Experimental data are taken from Ref.~\cite{ajzenberg86} and from Ref.~\cite{wakasa} for the $0_4^+$ state.}
\label{fig:energy_spectra_4a_ocm_thsr}
\end{center}
\end{figure}
 
The $4\alpha$ OCM Hamiltonian was presented in Eq.~(\ref{hamiltonian_ocm}). The effective $\alpha$-$\alpha$ interaction $V_{2\alpha}^{\rm eff}$ is constructed by the folding procedure from an effective two-nucleon force~\cite{mhn} including the Coulomb interaction.  One should note that the folded $\alpha$-$\alpha$ potential reproduces the $\alpha$-$\alpha$ scattering phase shifts and energies of the $^8$Be ground state and of the Hoyle state. The three-body force $V_{3\alpha}^{\rm eff}$ was phenomenologically introduced so as to fit the ground state energy of $^{12}$C. In addition, the phenomenological four-body force $V_{4\alpha}$ was adjusted to the ground state energy of $^{16}$O. The origin of the three-body and four-body forces is considered to be deducible from the state dependence of the effective nucleon-nucleon interaction and the additional Pauli repulsion between more than two $\alpha$-particles. However, they are short-range, and hence only act in compact configurations. The expectation values of those forces is less than 10~\% of the one of the corresponding two-body term.
 
The $J^\pi=0^+$ energy spectrum obtained by the $4\alpha$ OCM is shown in Fig.~\ref{fig:energy_spectra_4a_ocm_thsr}. We can reproduce the full spectrum of $0^+$ states up to about the $4\alpha$ disintegration threshold, and tentatively make a one-to-one correspondence of those states with the six lowest $0^+$ states of the experimental spectrum. In view of the complexity of the situation, the agreement is considered to be very satisfactory.
 
\begin{table}[t]
\begin{center}
\caption{Energies ($E-E^{\rm th}_{4\alpha}$), r.m.s.~radii ($R$), and monopole transition matrix elements to the ground state [$M({\rm E}0)$]. $R_{\rm exp.}$ and $M({\rm E}0)_{\rm exp.}$ are the corresponding experimental data. The finite size of $\alpha$ particle is taken into account in $R$ and $M({\rm E}0)$ (see Ref.~\cite{yamada05} for details).}
\label{tab:4a_ocm}
\begin{tabular}{cccccccccc}
\hline\hline
           & \multicolumn{4}{c}{$4\alpha$ OCM} & &  \multicolumn{4}{c}{Experiment} \\
           & $E-E^{\rm th}_{4\alpha}$ & {$R$} & \hspace{1mm}{$M({\rm E}0)$}\hspace{1mm} & {$\Gamma$} & \hspace*{5mm} & $E-E^{\rm th}_{4\alpha}$ &\hspace{1mm}$R_{\rm exp.}$\hspace{1mm} & \hspace{1mm}$M({\rm E}0)_{\rm exp.}$\hspace{1mm} &  {$\Gamma_{\rm exp.}$} \\
\hline
\hspace{3mm}$0_1^+$\hspace{3mm} & $-14.37$ & $2.7$ &        &              &  & $-14.44$ & $2.71\pm0.02$ &                     &           \\
$0_2^+$                                 & $-8.00$   & $3.0$ & $3.9$ &              &  & $-8.39$ &                    &   $3.55\pm 0.21$  & \\
$0_3^+$                                 & $-4.41$   & $3.1$ & $2.4$ &              &  & $-2.39$ &                     & $4.03\pm 0.09$  & \\
$0_4^+$                                 & $-1.81$   & $4.0$ & $2.4$ & $\sim0.15$ &  & $-0.84$ &                    &  no data            & $0.6$    \\
$0_5^+$                                 & $-0.248$  & $3.1$ & $2.6$ & $\sim0.05$ &  & $-0.43$ &                    &  $3.3\pm0.7$      & $0.185$  \\
$0_6^+$                                 & $2.08$     & $5.6$ & $1.0$ & $\sim0.05$ &  & $0.66$  &                    &  no data            & $0.166$ \\
\hline\hline
\end{tabular}
\end{center}
\end{table}
 
We show in Table~\ref{tab:4a_ocm} the calculated r.m.s. radii and monopole transition matrix elements to the ground state, together with the corresponding experimental values. The r.m.s. radius of the ground state is reproduced well, and those for the other five $0^+$ states are by about $10~\%$ or more larger than the ground state. The $M({\rm E}0)$ values for the $0_2^+$ and $0_5^+$ states are consistent with the corresponding experimental values. The $M$(E0) value for the $0_3^+$ state is accurate only  within a factor of two.

\begin{figure}[t]
\begin{center}
\includegraphics[width=33mm]{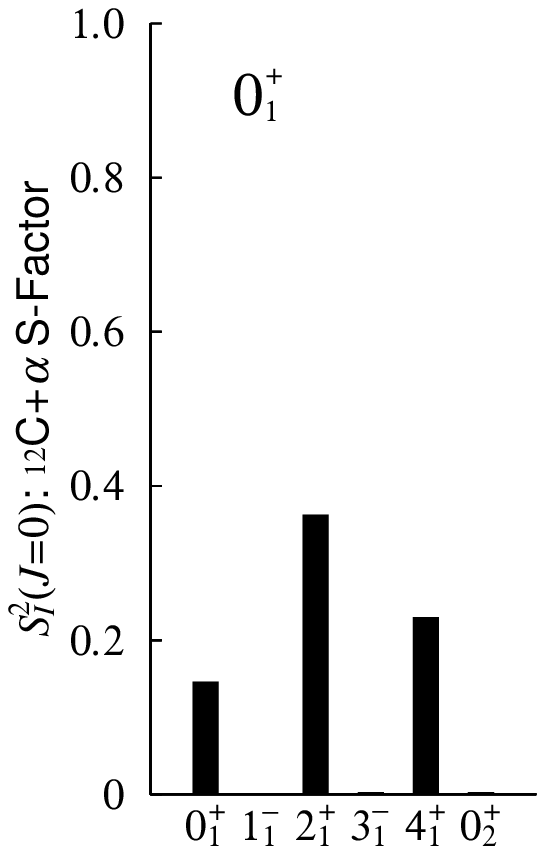}
\hspace*{5mm}
\includegraphics[width=33mm]{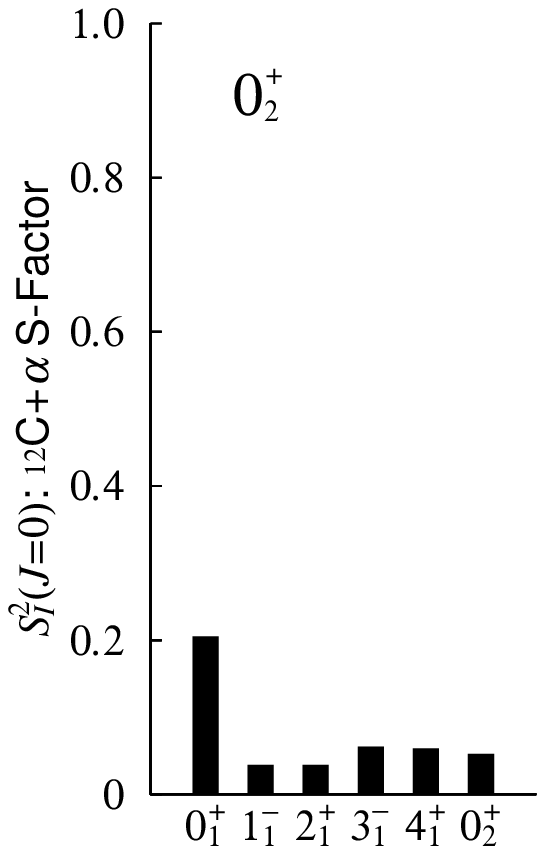}
\hspace*{5mm}
\includegraphics[width=33mm]{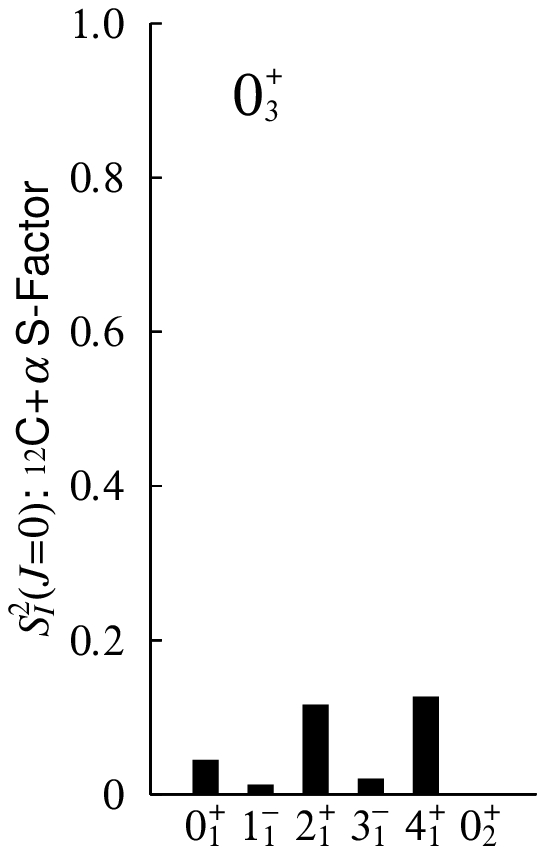}
\\
\includegraphics[width=33mm]{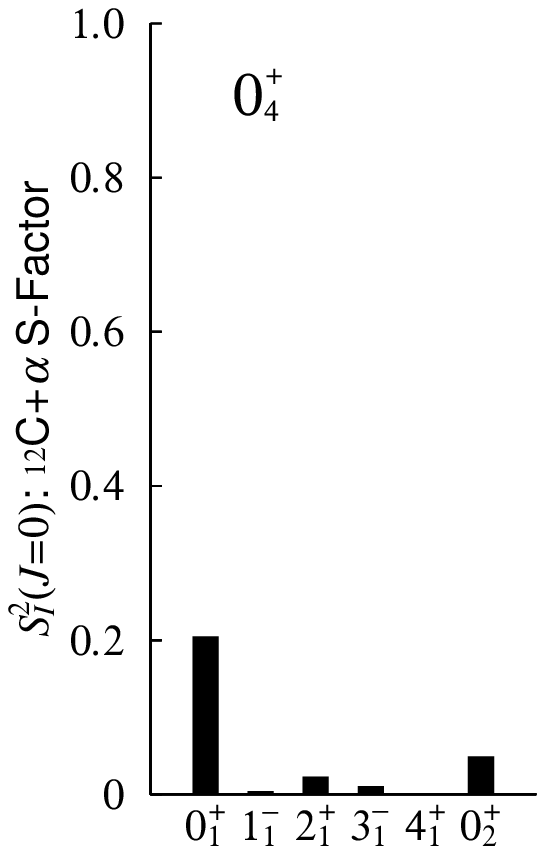}
\hspace*{5mm}
\includegraphics[width=33mm]{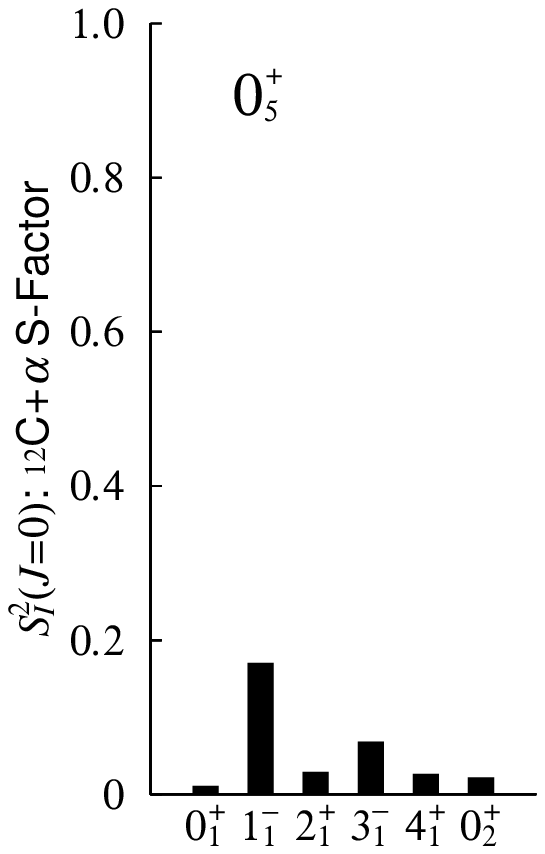}
\hspace*{5mm}
\includegraphics[width=33mm]{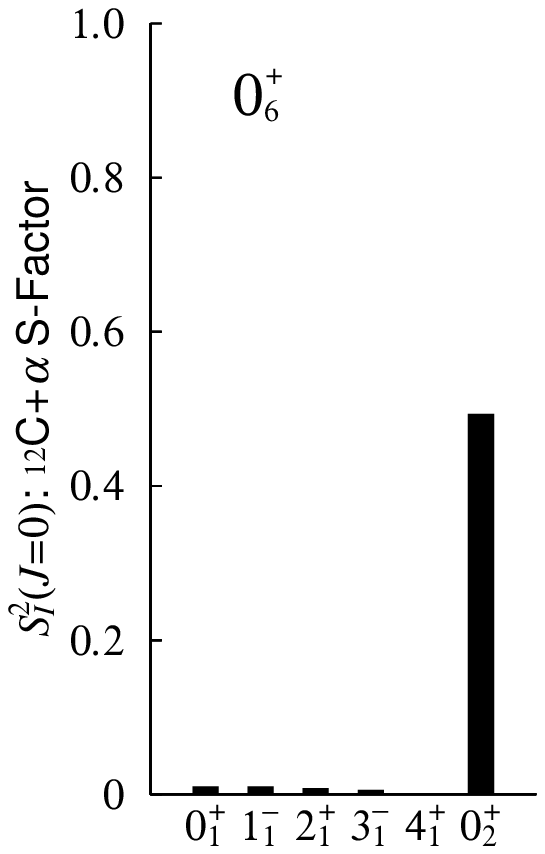}
\caption{Spectroscopic factors of the $\alpha + ^{12}$C$(L^{\pi}_n)$ channels ($L^{\pi}_n=0^{+}_{1},1^{-}_{1},2^{+}_{1},3^{-}_{1},4^{+}_{1},0^{+}_{2}$) in the six $0^{+}$ sates of $^{16}$O.}
\label{fig:s2_factors_0+_states_16O}
\end{center}
\end{figure}

In order to analyze the obtained wave functions, it is useful to study the overlap amplitude $\mathcal{Y}(r)$ and spectroscopic factor $S^2$, which are defined as follows:
\begin{eqnarray}
&&\mathcal{Y}(r) = \left\langle \left[ \frac{\delta(r^\prime-r)}{r^{\prime 2}} Y_{L}(\hat{\vc{r}^\prime})\Phi_{L}(^{12}{\rm C}) \right]_{0} |  \Psi(0_6^+) \right\rangle, \label{eq:rwa}\\
&&S^{2}=\int_{0}^{\infty} {dr} \left[ r\mathcal{Y}(r) \right]^2,\label{eq:s2_factor}
\end{eqnarray}
where $\Phi_{L}(^{12}{\rm C})$ is the wave function of $^{12}$C, given by the $3\alpha$ OCM calculation~\cite{yamada05}, and $r$ is the relative distance between the center-of-mass of $^{12}$C and the $\alpha$ particle. From this quantity we can see how large is the component in a certain $\alpha$+$^{12}$C channel which is contained in the wave functions obtained by the $4\alpha$ OCM. The results of $S^2$ factors are shown in Fig.~\ref{fig:s2_factors_0+_states_16O}. Since the ground state has a closed shell structure with the dominant component of SU(3)$(\lambda,\mu)=(0,0)$, the values of the $S^2$ factors for $0^+_1$ in Fig.~\ref{fig:s2_factors_0+_states_16O} can be explained by the SU(3) nature of the state. As mentioned above, the structures of the $0_2^+$ and $0_3^+$ states are well established as having $\alpha + ^{12}$C$(0_1^+)$ and $\alpha + ^{12}$C$(2_1^+)$ cluster structures, respectively. These structures of the $0_2^+$ and $0_3^+$ states are confirmed by the $4\alpha$ OCM calculation. In fact, one sees that the $S^2$ factors for the $\alpha + ^{12}$C$(0_1^+)$ and $\alpha + ^{12}$C$(2_1^+)$ channels are dominant in the $0_2^+$ and $0_3^+$ states, respectively.
 
On the contrary, the structures of the observed $0_4^+$, $0_5^+$ and $0_6^+$ states in Fig.~\ref{fig:energy_spectra_4a_ocm_thsr} have, in the past, not clearly been understood, since they have never been discussed with the previous cluster model calculations~\cite{Suz76,baye2,Kat92}. Although Ref.~\cite{thsr} predicts the $4\alpha$ condensate state around the $4\alpha$ threshold, it is not clear to which of those states it corresponds to. As shown in Fig.~\ref{fig:energy_spectra_4a_ocm_thsr}, the $4\alpha$ OCM calculation succeeded, for the first time, to reproduce the $0_4^+$, $0_5^+$ and $0_6^+$ states, together with the $0_1^+$, $0_2^+$ and $0_3^+$ states. From the analyses of the overlap amplitudes and the $S^2$ factors (see Fig.~\ref{fig:s2_factors_0+_states_16O}), the $4\alpha$ OCM showed that the $0_4^+$ and $0_5^+$ states mainly have $\alpha + ^{12}$C$(0_1^+)$ structure with higher nodal behavior and an $\alpha + ^{12}$C$(1^-)$ structure, respectively. The monopole strength of the $0^+_5$ state is reproduced nicely within the experimental error.
 
\begin{figure}[t]
\begin{center}
\includegraphics[scale=0.6]{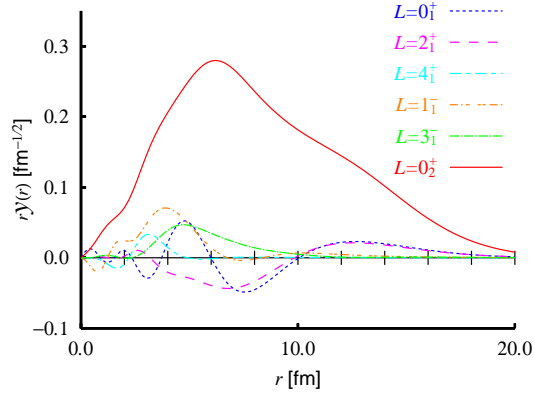}
\end{center}
\caption{(Colors online) Overlap amplitudes multiplied by $r$, defined by Eq.~(\ref{eq:rwa}), for the $0_6^+$ state in $^{16}$O.}
\label{fig:rwa_6th_0+}
\end{figure}
 
In Table~\ref{tab:4a_ocm}, the largest r.m.s. radius is about 5 fm for the $0_6^+$ state. Comparing with the relatively smaller r.m.s. radii of the $0_4^+$ and $0_5^+$ states, this large size suggests that the $0_6^+$ state may be composed of a weakly interacting gas of $\alpha$ particles of the condensate type. In addition, the $0_6^+$ state has a large overlap amplitude with the $\alpha + ^{12}$C$(0^+_2)$ channel with a $S^2$ factor of about $0.5$  (see Fig.~\ref{fig:s2_factors_0+_states_16O}), whereas the amplitudes in the other channels are much suppressed  (see Fig.~\ref{fig:rwa_6th_0+}). The amplitude in the Hoyle-state channel has no oscillations and a long tail stretches out to $\sim 20$ fm.
 
\begin{figure}[t]
\begin{center}
\includegraphics[scale=0.7]{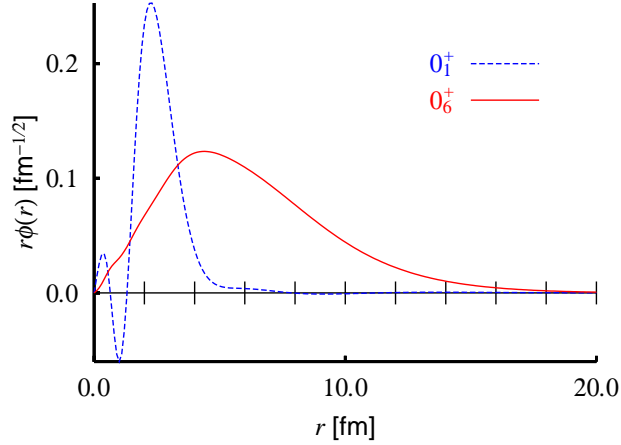}
\end{center}
\caption{(Color online) Radial parts of single-$\alpha$ orbits with $L=0$ belonging to the largest occupation number, for the ground and $0_6^+$ states.}
\label{fig:single_alpha_orbit_6th_0+_16o}
\end{figure}
 
\begin{figure}[t]
\begin{center}
\includegraphics[width=70mm]{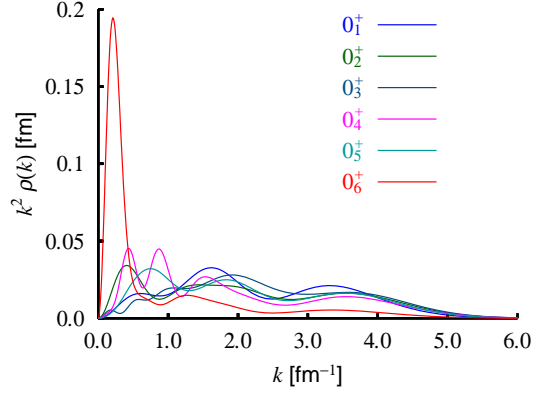}
\caption{(Color online) Momentum distribution of $\alpha$ particles multiplied by $k^2$, $k^2\rho(k)$, in the six $0^+$ states of $^{16}$O.
}\label{fig:momentum_dis_16o}
\end{center}
\end{figure}

While a large size is generally necessary for forming an $\alpha$ condensate, the best way for its identification is to investigate the single-$\alpha$ orbit and its occupation probability, which can be obtained by diagonalizing the one-body ($\alpha$) density matrix as defined in Eq.~(\ref{eq:one_particle_density_matrix})~\cite{matsumura04,yamada05,suzuki_02,suzuki_08,yamada08_obdm,yamada09_obdm}. As a result of the calculation of the $L=0$ case, a large occupation probability of $61~\%$ of the lowest $0S$-orbit is found for the $0_6^+$ state, whereas the other five $0^+$ states all have appreciably smaller values, at most $25~\%$ ($0^+_2$). The corresponding single-$\alpha$ $S$ orbit is shown in Fig.~\ref{fig:single_alpha_orbit_6th_0+_16o}. It has a strong spatially extended behavior without any node $(0S)$. This behavior is very similar to that of the overlap amplitude of $0_6^+$ for the $\alpha + ^{12}$C$(0^+_2)$ channel shown in Fig.~\ref{fig:rwa_6th_0+}. These results indicate that $\alpha$ particles are condensed into a very dilute $0S$ single-$\alpha$ orbit, see also Refs.~\cite{yamada05,ropke2}. In addition, Figure~\ref{fig:momentum_dis_16o} shows the momentum distribution  $k^2\rho(k)$ of $\alpha$ particles in the six $0^+$ states defined in Eq.~(\ref{eq:rho_k}). One sees that the momentum distribution of the $0^+_6$ state concentrates strongly in the narrow region $k<1$~fm$^{-1}$, and the behavior is quite similar to that of the Hoyle state in Fig.~\ref{fig:momentum_distribution_12C}. Thus, the $0^+_6$ state clearly has, according to our calculation, a  $4\alpha$ condensate character.
 
Comparing the single-$\alpha$ orbit of the $0^{+}_{6}$ state in Fig.~\ref{fig:single_alpha_orbit_6th_0+_16o} with that of the Hoyle state shown in Fig.~\ref{fig:12C_alpha_orbits}, one can see an almost identical shape. This is also an important indication that  the $0^+_6$ state has $\alpha$-particle condensate nature. Of course, the extension is slightly different because of the smallness of the system. The nodeless character of the wave function is very pronounced and only some oscillations with small amplitude are present in $^{12}$C, reflecting the weak influence of the Pauli principle between the $\alpha$'s, as discussed in Sec.~\ref{subsec:4-1}. On the contrary, due to the much reduced ground-state radii, the ``$\alpha$-like'' clusters strongly overlap in $^{12}$C and $^{16}$O, producing strong amplitude oscillations which take care of antisymmetrization between clusters~\cite{yamada05,funaki08}. In fact, on sees in Fig.~\ref{fig:single_alpha_orbit_6th_0+_16o} that the single-$\alpha$ orbit for the ground state has maximum amplitude at around $3$ fm and oscillations in the interior with two nodal $(2S)$ behavior, due to the Pauli principle and reflecting the shell-model configuration (also see Fig.~\ref{fig:12C_alpha_orbits} for the ground state of $^{12}$C).
 
The $\alpha$ decay width constitutes a very important information to identify the $0_6^+$ state from the experimental point of view. The width $\Gamma_L$ can be estimated, based on the $R$-matrix theory~\cite{r-matrix},
\begin{eqnarray}
&&\Gamma_L =2P_L(a) \cdot \gamma^2_L(a), \nonumber \\
&&P_L(a)=\frac{ka}{F_L^2(ka)+G_L^2(ka)}, \nonumber \\
&&\gamma^2_L(a)=\theta^2_L(a)\gamma^2_{\rm W}(a), \nonumber \\
&&\gamma^2_{\rm W}(a)=\frac{3\hbar^2}{2\mu a^2}, \hspace*{5mm}\theta^2_L(a)=\frac{a^3}{3}\mathcal{Y}_L^2(a),\label{eq:gamma}
\end{eqnarray}
where $k$, $a$ and $\mu$ are the wave number of the relative motion, the channel radius, and the reduced mass, respectively, and $F_L$, $G_L$, and $P_L(a)$ are the regular and irregular Coulomb wave functions and the corresponding penetration factor, respectively. The reduced width of $\theta^2_L(a)$ is related with the overlap amplitude $\mathcal{Y}(r)$ defined in Eq.~(\ref{eq:rwa}). In Table~\ref{tab:widths_16o}, we show the partial $\alpha$ decay widths $\Gamma_L$ of the $0_6^+$ state decaying into the $\alpha+ ^{12}{\rm C}(0_1^+)$, $\alpha+ ^{12}{\rm C}(2_1^+)$ and $\alpha+ ^{12}{\rm C}(0_2^+)$ channels, and also the total $\alpha$ decay width which is obtained as a sum of the partial widths, and reduced widths $\theta_L^2(a)$ defined in Eq.~(\ref{eq:rwa}). Experimental values are all taken as given by the decay energies. Thus the excitation energy of the calculated $0_6^+$ state is assumed to be the experimental value, i.e.~15.1 MeV.
 
\begin{table}[t]
\begin{center}
\caption{Partial $\alpha$ widths in $^{16}$O$^\ast$ decaying into possible channels and the total width. The reduced widths defined in Eq.~(\ref{eq:rwa}) are also shown. $a$ is the channel radius.}\label{tab:widths_16o}
\begin{tabular}{ccccc}
\hline\hline
 & $^{12}{\rm C}(0_1^+)+\alpha$ & $^{12}{\rm C}(2_1^+)+\alpha$ & $^{12}{\rm C}(0_2^+)+\alpha$ & \raisebox{-1.8ex}[0pt][0pt]{Total} \\
 & ($a=8.0$ fm) & ($a=7.4$ fm) & ($a=8.0$ fm) &    \\
\hline
$\Gamma_L$ (keV) & 26 & 8 & $2\times 10^{-7}$ & $34$  \\
$\theta^2_L(a)$ & 0.006 & 0.004 & 0.15 &   \\
\hline\hline
\end{tabular}
\end{center}
\end{table}
 
The obtained very small total $\alpha$ decay width of 34 keV, in reasonable agreement with the corresponding experimental value of 160 keV, indicates that this state is unusually long lived. The reason of this fact can be explained in terms of the present analysis as follows: Since this state has a very exotic structure composed of gas-like four alpha particles, the overlap between this state and $\alpha + ^{12}{\rm C}(0_1^+)$ or $\alpha + ^{12}{\rm C}(2_1^+)$ wave functions with a certain channel radius becomes very small, as this is, indeed, indicated by small $\theta_L^2(a)$ values, 0.006 and 0.004, respectively, and therefore by small $\gamma_L^2(a)$ values. These largely suppress the decay widths expressed by Eq.~(\ref{eq:gamma}) in spite of the large values of the penetration factors caused by large decay energies 7.9 MeV and 3.5 MeV into these two channels, $\alpha + ^{12}{\rm C}(0_1^+)$ and $\alpha + ^{12}{\rm C}(2_1^+)$, respectively. On the other hand, the decay into the $\alpha + ^{12}{\rm C}(0_2^+)$ channel is also suppressed due to the very small penetration caused by the very small decay energy 0.28 MeV, even though the corresponding reduced width takes a relatively large value $\theta_L^2(a)=0.15$. This is natural since the $0_2^+$ state of $^{12}$C has a gas-like three-alpha-particle structure. It is very likely that the above mechanism holds generally for the alpha condensate states in heavier $n\alpha$ systems, and therefore the alpha condensate states can also be expected to exist in heavier systems as a relatively long lived resonance.
 
As for the decay widths of the $0^+_4$ and $0^+_5$ states, as evaluated by Eq.~(\ref{eq:gamma}), they are shown in Table~\ref{tab:4a_ocm}. The calculated width of the $0_4^+$ state is $\sim 150$ keV, which is quite a bit larger than that found for the $0_5^+$ state $\sim 50$ keV. Both are qualitatively consistent with the corresponding experimental data, $600$ keV and $185$ keV, respectively. We should note that our calculation consistently reproduces the ratio of the widths of the $0_4^+$, $0_5^+$, and $0_6^+$ states, i.e. about $3:1:1$, respectively, though the magnitudes of the widths are underestimated by about a factor of four with respect to the experimental values (see Table~\ref{tab:4a_ocm}). The reason why the width of the $0_4^+$ state is larger than that of the $0_5^+$ state, though the $0_4^+$ state has lower excitation energy, is due to the fact that the former has a much larger component of the $\alpha+ ^{12}$C$(0_1^+)$ decay channel, reflecting the characteristic structure of the $0_4^+$ state. The $4\alpha$ condensate state, thus, should not be assigned to the $0_4^+$ or $0_5^+$ state~\cite{funaki08,funaki10} but very likely to the $0_6^+$ state.

%%%%%%%%%%%%%%%%%%%%%%%%%%%%%%%%%%%%%%%%%%%
\subsubsection{THSR wave function analysis}\label{subsec:4-2-2}
%%%%%%%%%%%%%%%%%%%%%%%%%%%%%%%%%%%%%%%%%%%
 
As mentioned already, the first investigation of the $4\alpha$-particle condensate state was performed with the THSR wave function~\cite{thsr}. It was conjectured that  $4\alpha$-particle condensation should occur around the $4\alpha$ threshold. As mentioned in Sec.~\ref{subsec:2-2}, the THSR wave function allows only for two limiting configurations, that is a pure Slater determinant for $B=b$ and a pure $\alpha$-particle gas for $B\gg b$. Asymptotic configurations like $\alpha + ^{12}{\rm C}(0_1^+)$ are absent. Thus, the THSR wave function can well describe the dilute $\alpha$ cluster states as well as shell model like ground states, whereas other structures such as $\alpha + ^{12}$C($0^{+}_{1}$) clustering may be  only incorporated in an average way.  In this section, we see whether or not the counterpart of the $0^+_6$ state obtained by the $4\alpha$ OCM calculation can also be found with the THSR wave function, and then we study how well the $4\alpha$ condensate state is described with the THSR wave function.
 
\begin{table}[t]
\begin{center}
\caption{Energies $E-E^{\rm th}_{4\alpha}$, r.m.s. radii $R_{\rm rms}$, monopole matrix elements $M(E0)$, and $\alpha$ decay widths $\Gamma$, obtained within the $4\alpha$  THSR framework, where $E^{\rm th}_{4\alpha}=4E_{\alpha}$ denotes the $4\alpha$ threshold energy, with $E_\alpha$ the binding energy of the $\alpha$ particle~\cite{funaki10}.}
\label{tab:4a_thsr}
\begin{tabular}{ccccc}
\hline\hline
 & \hspace*{1mm}$E-E^{\rm th}_{4\alpha}$ [MeV]\hspace*{1mm} & \hspace*{1mm}$R_{\rm rms}$ [fm]\hspace*{1mm} & \hspace*{1mm}$M(E0)$ [fm$^2$]\hspace*{1mm} & \hspace*{1mm}$\Gamma$ [MeV]\hspace*{1mm} \\
\hline
$(0_1^+)_{\rm THSR}$ & $-15.05$ & 2.5 &  &        \\
$(0_2^+)_{\rm THSR}$ & $-4.7$ & 3.1 & 9.8 &       \\
$(0_3^+)_{\rm THSR}$ & $1.03$ & 4.2 & 2.5 &  1.6 \\
$(0_4^+)_{\rm THSR}$ & $3.04$ & 6.1 & 1.2 &  0.14 \\
\hline\hline
\end{tabular}
\end{center}
\end{table}

The microscopic wave function of $^{16}$O with the THSR ansatz is described in Eqs.~(\ref{eq:hwwf}) and (\ref{eq:thsr}), and the eigenenergies and eigenstates are obtained by solving the Hill-Wheeler equation Eq.~(\ref{eq:hw}).  The Hamiltonian is given in Eq.~(\ref{eq:H_for_THSR}), where the effective nucleon-nucleon interaction called F1~\cite{tohsaki_F1} was adopted. The resulting $0^+$ spectrum is shown in Fig.~\ref{fig:energy_spectra_4a_ocm_thsr}. Hereafter, we assign the four $0^+$ states as $(0_1^+)_{\rm THSR}$--$(0_4^+)_{\rm THSR}$. In Table~\ref{tab:4a_thsr}, the energies, r.m.s. radii, monopole transition matrix elements $M(E0)$ to the ground state, and $\alpha$-decay widths of the $(0_1^+)_{\rm THSR}$--$(0_4^+)_{\rm THSR}$ states are displayed. The $(0_3^+)_{\rm THSR}$ state has a large r.m.s. radius of $4.2$ fm and the $(0_4^+)_{\rm THSR}$ state has an even larger one of $6.1$ fm. They are comparable to the values for the $(0_4^+)_{\rm OCM}$ and $(0_6^+)_{\rm OCM}$ states in the $4\alpha$ OCM calculation, respectively, where the six $0^+$ states obtained by the $4\alpha$ OCM calculation in Table~\ref{tab:4a_ocm} are labeled as $(0_1^+)_{\rm OCM}$--$(0_6^+)_{\rm OCM}$. On the other hand, the $M(E0)$ values of the $(0_3^+)_{\rm THSR}$ and $(0_4^+)_{\rm THSR}$ states well agree with those of the $(0_4^+)_{\rm OCM}$ and $(0_6^+)_{\rm OCM}$ states, respectively. This suggests that the $(0_4^+)_{\rm THSR}$ state corresponds to the $(0_6^+)_{\rm OCM}$ state, and hence to the $15.1$ MeV state.

\begin{figure}[t]
\begin{center}
\includegraphics[width=70mm]{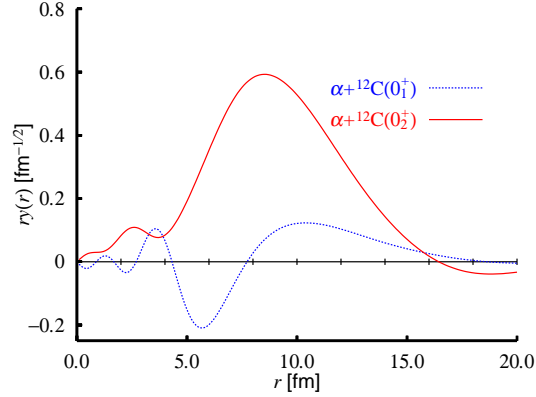}
\caption{(Color online) Overlap amplitude multiplied by $r$ for the $(0_4^+)_{\rm THSR}$ state in the $\alpha + {^{12}{\rm C}(0_1^+,0_2^+)}$ channels, defined in Eq.~(\ref{eq:rwa}).}
\label{fig:rwa_thsr_4th_0+}
\end{center}
\end{figure}

\begin{figure}[t]
\begin{center}
\includegraphics[width=70mm]{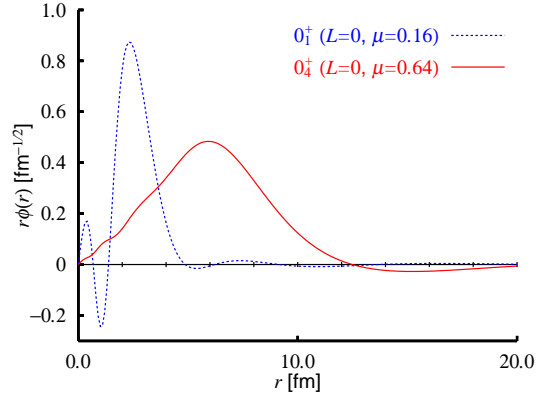}
\caption{(Color online) Radial parts of single-$\alpha$-particle orbits with $L=0$ and $n_L=1$ for the $(0_1^+)_{\rm THSR}$ (dotted curve) and $(0_4^+)_{\rm THSR}$ (solid) states.}
\label{fig:single_orbit_thsr_4th_0+}
\end{center}
\end{figure}

More quantitative evidences for the $(0_4^+)_{\rm THSR}$ state being the counterpart of the $(0_6^+)_{\rm OCM}$ state are presented from the analyses of the overlap amplitudes of the $\alpha + {^{12}{\rm C}(0_1^+,0_2^+)}$ channels and the one-body density matrix for the $\alpha$ particle etc.~with the wave function of the $(0_4^+)_{\rm THSR}$ state. In Fig.~\ref{fig:rwa_thsr_4th_0+}, we show the overlap amplitudes of the $\alpha + {^{12}{\rm C}(0_1^+,0_2^+)}$ channels for the $(0_4^+)_{\rm THSR}$ state. The radial behaviors of the $\alpha + {^{12}{\rm C}(0_1^+,0_2^+)}$ channels are very similar to those of the $(0_6^+)_{\rm OCM}$ case in Fig.~\ref{fig:rwa_6th_0+}. In addition, we found that the single-$\alpha$ $S$ orbit occupancy in the $(0_4^+)_{\rm THSR}$ state is as large as $64~\%$, which is comparable to that in the $(0_6^+)_{\rm OCM}$ state, and the radial behavior of the single-$\alpha$ $S$ orbit in the former state is illustrated in Fig.~\ref{fig:single_orbit_thsr_4th_0+}, and is similar to that in the latter in Fig.~\ref{fig:single_alpha_orbit_6th_0+_16o}.
 
In conclusion we found that the $(0^{+}_{4})_{\rm THSR}$ state corresponds to the $(0^{+}_{6})_{\rm OCM}$ state and is most appropriately considered to be the $4\alpha$ condensate state. This further gives us a strong support that the $4\alpha$ condensate state exists around the $4\alpha$ breakup threshold and is very likely to correspond to the observed $0^{+}_{6}$ state at $15.1$ MeV.
 
%%%%%%%%%%%%%%%%%%%%%%%%%%%%%%%%%%%%%%%%%%%%%%%%%%%%
\subsection{Heavier $4n$ nuclei: Gross-Pitaevskii Equation}\label{subsec:4-3}
%%%%%%%%%%%%%%%%%%%%%%%%%%%%%%%%%%%%%%%%%%%%%%%%%%%%
 
In principle, one could go on, increasing the number of $\alpha$-particles, as for $^{20}$Ne, $^{24}$Mg, etc.~ and study their structure with use of the THSR wave function or within the OCM framework. However, one easily imagines that the complexity of the calculations quickly becomes prohibitive. In order to get a rough idea what happens for more $\alpha$-particles, drastic approximations have to be performed. One such approximation is to consider the $\alpha$-particles as ideal inert bosons and to treat them in mean field approximation. This then leads to the Gross-Pitaevskii Equation (GPE)~\cite{gross} which is widely employed in the physics of cold atoms~\cite{dalfovo99}. One interesting question that can be asked in this connection is: How many $\alpha$'s can maximally exist in a self-bound $\alpha$-gas state?  Seeking an answer, it is interesting to investigate it schematically using an effective $\alpha$-$\alpha$ interaction within an $\alpha$-gas mean-field calculation of the Gross-Pitaevskii type~\cite{yamada04}.
 
In the mean field approach, the total wave function of the condensate $n\alpha$-boson system is represented as
\begin{eqnarray}
\Phi(n\alpha)=\prod_{i=1}^{n}\varphi(\vc{r}_i),
\end{eqnarray}
where $\varphi$ is the normalized single-$\alpha$ wave function of the {\it i}-th $\alpha$ boson. Then, the equation of motion for the $\alpha$ boson, called the Gross-Pitaevskii equation, is of the non-linear Schr\"odinger-equation type,
\begin{eqnarray}
&&-\frac{\hbar^2}{2m}\left(1-\frac{1}{n}\right)\bm{\nabla}^2\varphi(\vc{r}) + U(\vc{r})\varphi(\vc{r}) = \varepsilon\varphi(\vc{r}),\label{GPE}\\
&&U(\vc{r})=(n-1)\int d\vc{r}'\left|\varphi(\vc{r}')\right|^2 {V_{\alpha\alpha}(\vc{r}',\vc{r})},\label{GPE_alpha_pot}
\end{eqnarray}
where $m$ stands for the mass of the $\alpha$ particle and $U$ is the mean-field potential of $\alpha$-particles. The center-of-mass kinetic energy correction, $1-\frac{1}{n}$, is taken into account together with the finite number corrections, $n-1$. In the present study, only the $S$-wave state is solved self-consistently with the iterative method. The effective $2\alpha$  interaction $\upsilon_2$ is taken of Gaussian-type including a repulsive density-dependent term, to account for the Pauli repulsion at short distances, which is of similar form as the Gogny interaction (known as an effective $NN$ potential)~\cite{gogny} used in nuclear mean-field calculations
\begin{eqnarray}
&&{V_{\alpha\alpha}(\vc{r},\vc{r}')}={V_{0}}\exp\left[-0.7^2(\vc{r}-\vc{r}')^2\right] - 130\exp\left[-0.475^2(\vc{r}-\vc{r}')^2\right]\nonumber\\
&&\hspace*{1.5cm}+(4\pi)^2g\delta(\vc{r}-\vc{r}')\rho\left(\frac{\vc{r}+\vc{r}'}{2}\right) + {V_{\rm Coul}}(\vc{r}-\vc{r}'),
\label{DD_pot}
\end{eqnarray}
where the units of $\upsilon_2$ and $r$ are MeV and fm, respectively, and $\rho$ denotes the density of the $n\alpha$ system. The folded Coulomb potential ${V_{\rm Coul}}$ is presented as
\begin{eqnarray}
{V_{\rm Coul}}(\vc{r}-\vc{r}')=\frac{4e^2}{|\vc{r}-\vc{r}'|}{\rm erf}(a|\vc{r}-\vc{r}'|),
\label{Coulomb_pot}
\end{eqnarray}
 
The Gaussian-potential part in Eq.~(\ref{DD_pot}) is based on the Ali-Bodmer potential~\cite{ali66}, which is known to reproduce well the elastic $\alpha$-$\alpha$ scattering phase shift up to about 60 MeV for ${V_0=500}$ MeV. The two parameters of the force, ${V_0}$ and $g$, were adjusted to reproduce the energy (measured from the $3\alpha$ threshold) and r.m.s. radius of the Hoyle state obtained from the THSR analysis.
 
\begin{figure}[t]
\begin{center}
\includegraphics[width=0.60\hsize]{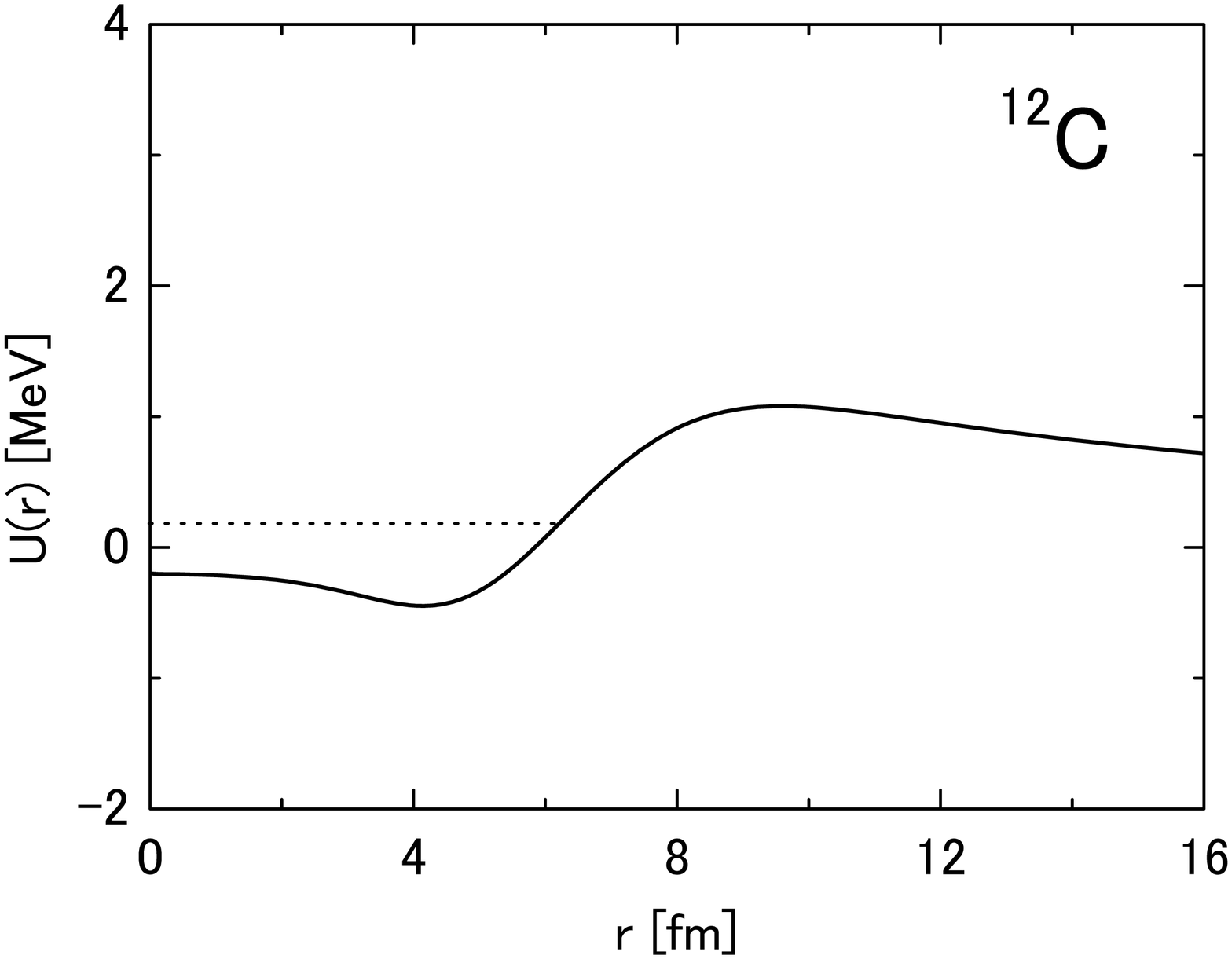}\\
\includegraphics[width=0.60\hsize]{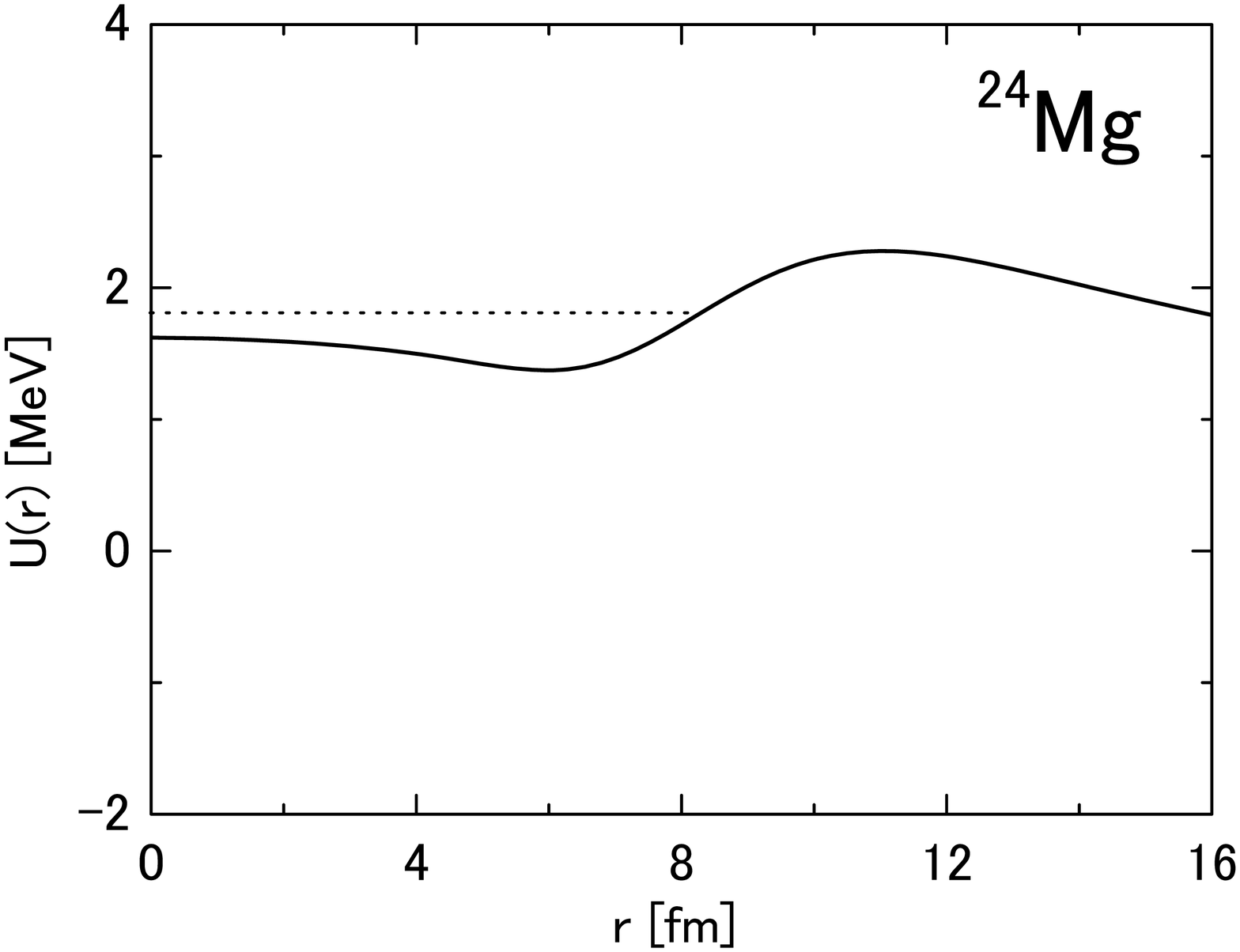}
\caption{Alpha-particle mean-field potential for three $\alpha$'s in $^{12}$C and six $\alpha$'s in $^{24}$Mg.  The dotted line denotes the energy of the single-$\alpha$ $0S$ orbit (from Ref.~\cite{yamada04}).}
\label{fig:gross_pitaevskii}
\end{center}
\end{figure}
 
The corresponding $\alpha$ mean-field potential for three $\alpha$'s of $^{12}$C is shown in Fig.~\ref{fig:gross_pitaevskii}. One sees the $0S$-state lying slightly above threshold but below the Coulomb barrier.  As more $\alpha$-particles are added, the Coulomb repulsion drives the loosely bound system of $\alpha$-particles farther and farther apart. For example, in the case of six $\alpha$'s in $^{24}$Mg, the Coulomb barrier is lower and its position is moved outwards. Thus, eventually the Coulomb barrier should fade away in some limiting nucleus.  According to our estimate~\cite{yamada04}, a maximum of eight to ten $\alpha$-particles can be held together in a condensate.  However, there may be ways to lend additional stability to such systems.  We know that in the case of $^8$Be, adding one or two neutrons produces extra binding without seriously disturbing the pronounced $\alpha$-cluster structure.  Therefore, one has reason to speculate that adding a few neutrons to a many-$\alpha$ state may stabilize the condensate. But again, state-of-the-art microscopic investigations are necessary before anything definite can be said about how extra neutrons will influence an $\alpha$-particle condensate.
 
\begin{figure}[t]
\begin{center}
\includegraphics[width=0.45\hsize]{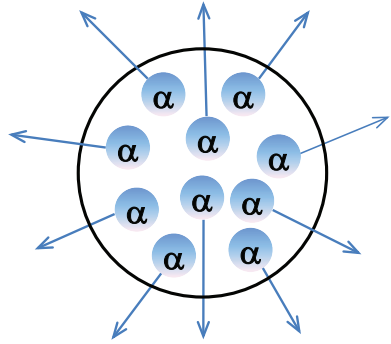}
\caption{(Color online) Cartoon of a Coulomb explosion of $10$ $\alpha$-particles from $^{40}$Ca.}
\label{fig:cartoon_10_alphas}
\end{center}
\end{figure}
 
Concerning excitation of condensate states with many $\alpha$ particles, heavy ion reactions and Coulomb excitation may be appropriate tools. As an ideal case let us imagine that $^{40}$Ca has been excited by Coulomb excitation to a state of about $60$ MeV. Coulomb excitation favors $0^{+}$-states and $60$ MeV is the threshold for disintegration into $10$ $\alpha$-particles. Since the Coulomb barrier is  absent for ten $\alpha$'s, this state may perform a Coulomb explosion of a $10$ $\alpha$ particle coherent state. A cartoon of such a scenario is sketched in Fig.~\ref{fig:cartoon_10_alphas}. With heavy ion reactions, experiments with coincident measurements are being analyzed to detect multi $\alpha$ events by Borderie et al.~\cite{Bordeier}. W.~v.~Oertzen et al.~seem to have detected an enhancement of multi $\alpha$ decay out of $\alpha$-condensates in compound states of heavier $N=Z$ nuclei, see Ref.~\cite{Oertzen}.
 
%%%%%%%%%%%%%%%%%%%%%%%%%%%%%%%%%%%%%%%%%%%%%%%%%%%%
\subsection{Hoyle-analogue states in non-$4n$ nuclei: $^{11}$B and $^{13}$C}\label{subsec:4-4}
%%%%%%%%%%%%%%%%%%%%%%%%%%%%%%%%%%%%%%%%%%%%%%%%%%%%
 
In the previous sections, we discussed the $\alpha$-gas-like states in $4n$ nuclei. On the other hand, one can also expect cluster-gas states composed of alpha and triton clusters (including valence neutrons etc.) around their cluster disintegrated thresholds in $A \not= 4n$ nuclei, in which all clusters are in their respective $0S$ orbits, similar to the Hoyle state with $(0S_{\alpha})^3$. The states, thus, can be called {\it Hoyle-analogue} states in non-self-conjugated nuclei. It is an intriguing subject to investigate whether or not the Hoyle-analogue states exist in $A \not= 4n$ nuclei, as for example, $^{11}$B, composed of $2\alpha$ and $1t$ clusters as well as $^{13}$C, composed of $3\alpha$ and $1n$.
 
The structure of $3/2^-$ and $1/2^+$ ($1/2^-$ and $1/2^+$) states in $^{11}$B ($^{13}$C) up to around the $2\alpha$+$t$ ($3\alpha$+$n$) threshold were investigated by the $2\alpha$+$t$ OCM~\cite{yamada10} $3\alpha$+$n$ OCM~\cite{yamada08_IJMPE}) for $^{11}$B ($^{13}$C) combined with the Gaussian expansion method. The model space for the $2\alpha$+$t$ ($3\alpha$+$n$) OCM is large enough to cover the $2\alpha+t$ ($3\alpha+n$) gas, the $^{7}$Li+$\alpha$ and $^{8}$Be+$t$ ($^{9}$Be+$\alpha$ and $^{12}$C+$n$) clusters, as well as the shell-model configurations. As well known, the $\alpha-t$ and $\alpha-n$ potentials have a strong parity dependence~\cite{ptp_supple_68}. In the odd waves they are strongly attractive to produce the bound states (for $\alpha-t$) and resonant states ($\alpha-t$ and $\alpha-n$), while the even ones are weakly attractive and have no ability to produce any resonant states up to $E_{x}\sim 15$ MeV~\cite{ajzenberg_A=7}. Thus, the gas-like states in $^{11}$B and $^{13}$C might appear in their even-parity states.
 
\begin{figure}[t]
\begin{center}
\includegraphics[width=0.70\hsize]{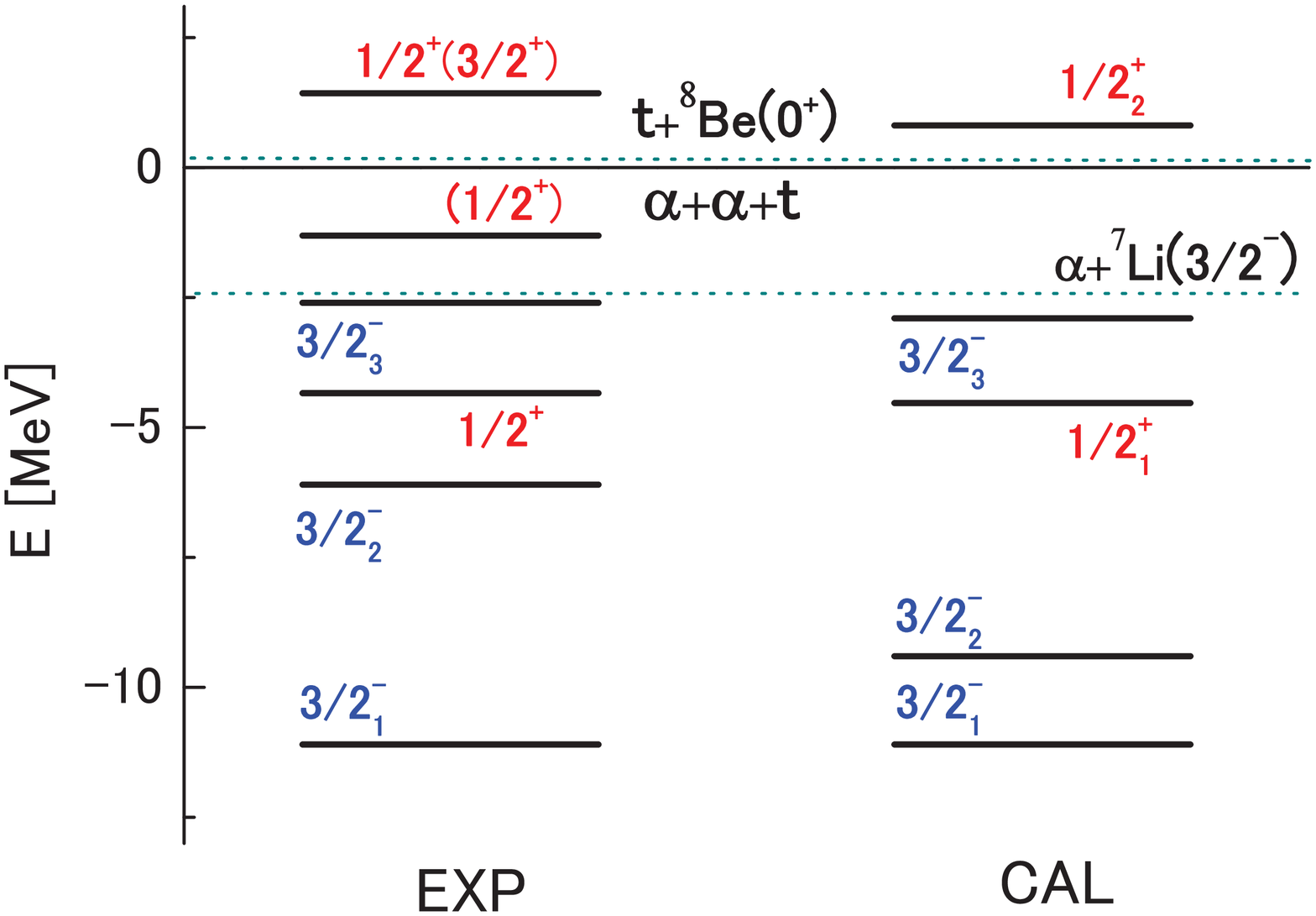}
\caption{(Color online) Calculated energy levels of $3/2^{-}$ and $1/2^{+}$ states in $^{11}$B with respect to the $2\alpha+t$ threshold, together with the experimental data.}
\label{fig:11B_energy_levels}
\end{center}
\end{figure}

The energy levels of $3/2^{-}$ and $1/2^+$ states in $^{11}$B are shown in Fig.~\ref{fig:11B_energy_levels}. The $3/2^{-}_1$ state is the ground state with a shell-model-like structure. The calculated nuclear radius is $R=2.22$~fm ($R^{\rm exp}=2.43\pm0.11$~fm). The dominant configuration of this state is SU(3)$[f](\lambda,\mu)_L=[443](1,3)_1$ with $Q=7$ harmonic oscillator quanta (95~\%), having the main angular momentum channel of $(L,S)_J=(1,\frac{1}{2})_{\frac{3}{2}}$.  On the other hand, also the $3/2^{-}_{2}$ state has a shell-model-like structure.
 
The $3/2^{-}_{3}$ state appears at $E_x=8.2$~MeV ($E=-2.9$~MeV referring to the $2\alpha+t$ threshold). The radius of $3/2^{-}_3$ is 3.00~fm. This value is by about $30$~\% larger than that of the ground state of $^{11}$B, and the $\alpha - \alpha$ r.m.s. distance (distance between $^{8}$Be($2\alpha$) and $t$) is 4.47~fm (3.49~fm). Thus, $3/2^{-}_{3}$ has a $2\alpha+t$ cluster structure. A characteristic feature of $3/2^{-}_{3}$ is that the isoscalar monopole transition rate $B$(E0:IS) is as large as $96\pm 16$~fm$^4$, comparable to that of the Hoyle state ($120\pm 9$~fm$^4$)$^{8}$. The present model ($92$~fm$^4$) reproduces well the data. It is interesting to study whether $3/2^{-}_{3}$ possesses an $\alpha$ gas nature like the Hoyle state. To this purpose, we study the single-cluster orbits and their occupation probabilities in the $3/2^{-}_{3}$ state by solving the eigenvalue equation of the single-cluster density matrices.
 
\begin{figure}[t]
\begin{center}
\includegraphics[width=0.48\hsize]{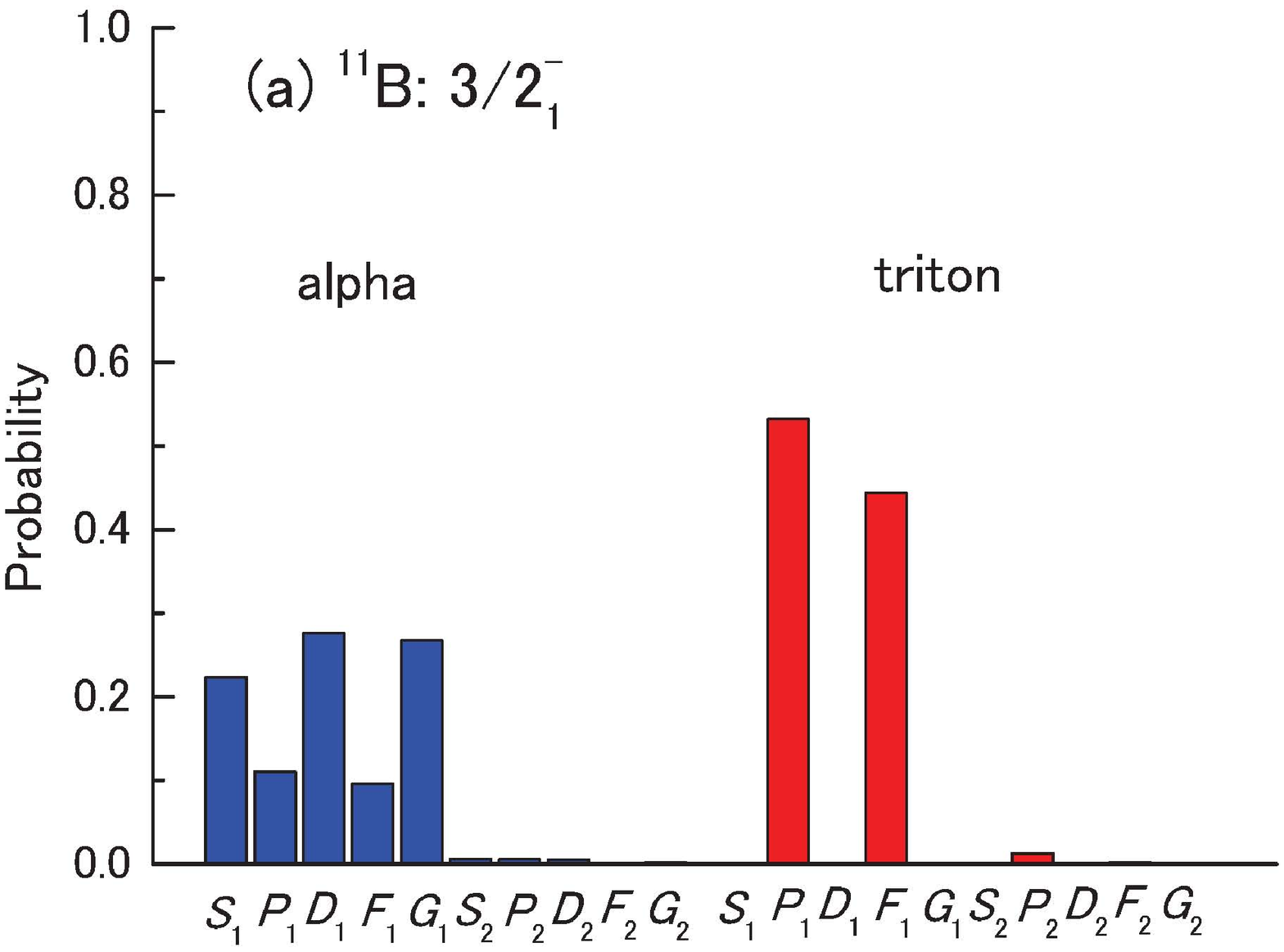}
\hspace*{2mm}
\includegraphics[width=0.48\hsize]{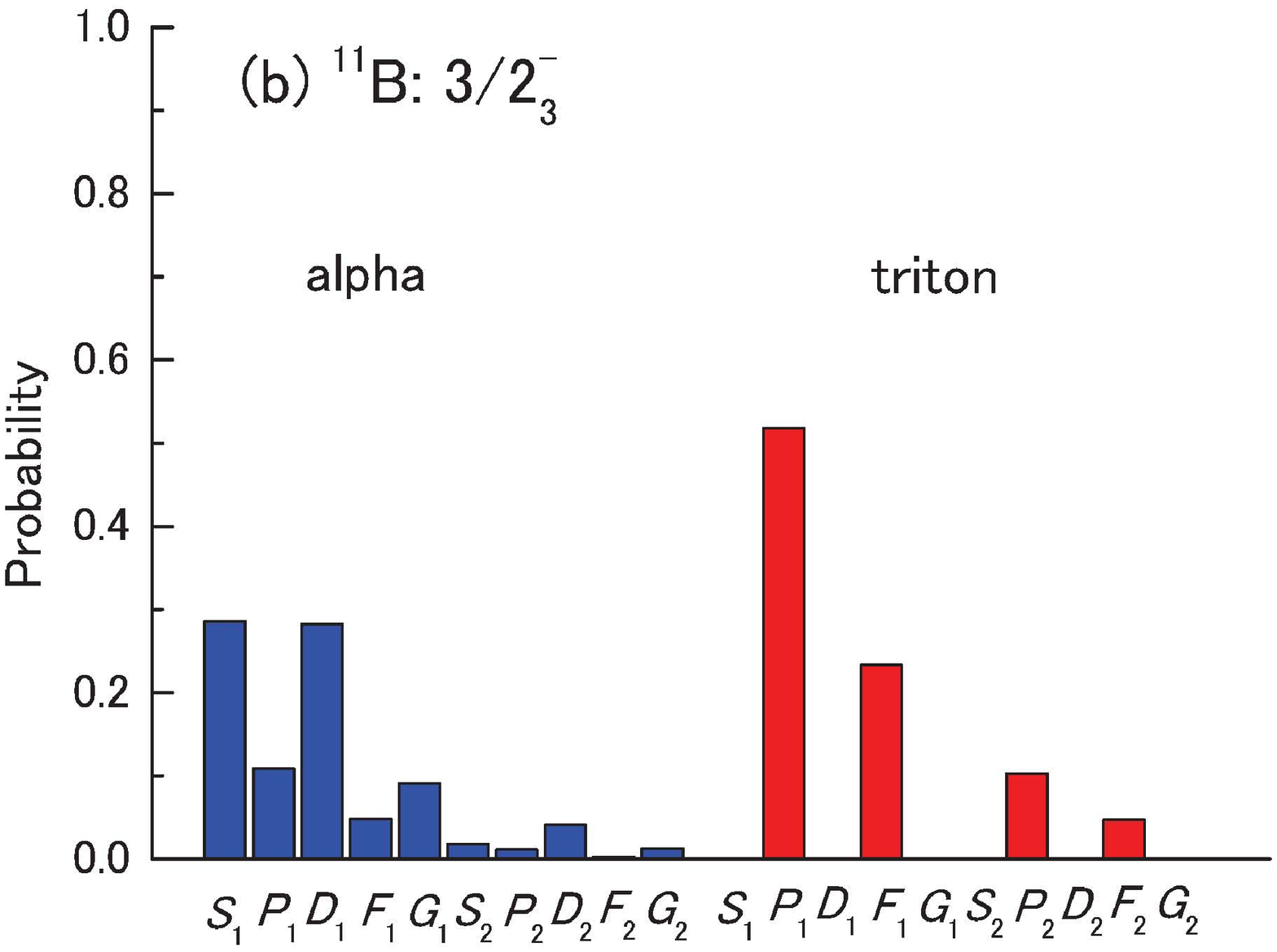}\\
\vspace*{2mm}
\includegraphics[width=0.48\hsize]{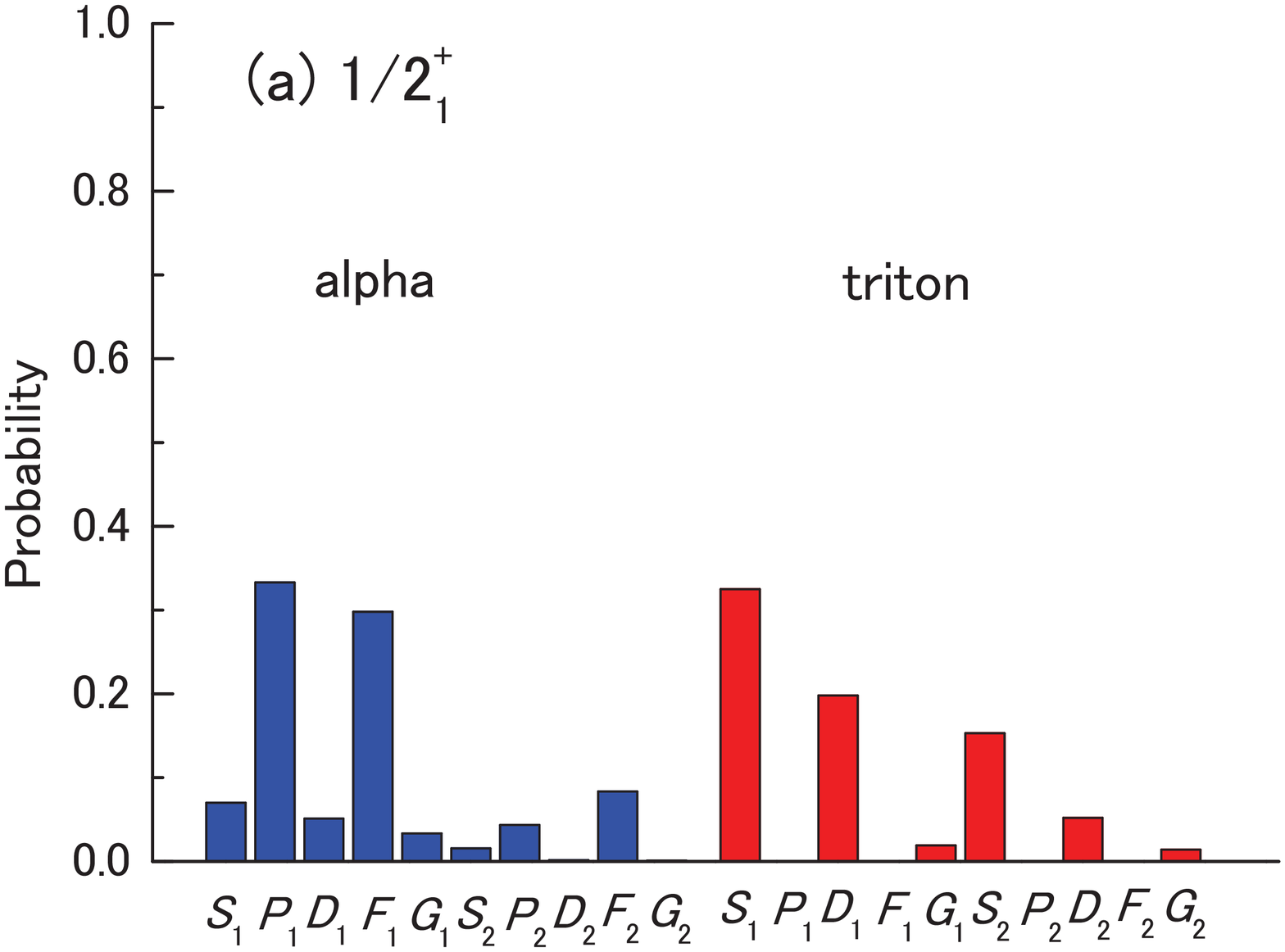}
\hspace*{2mm}
\includegraphics[width=0.48\hsize]{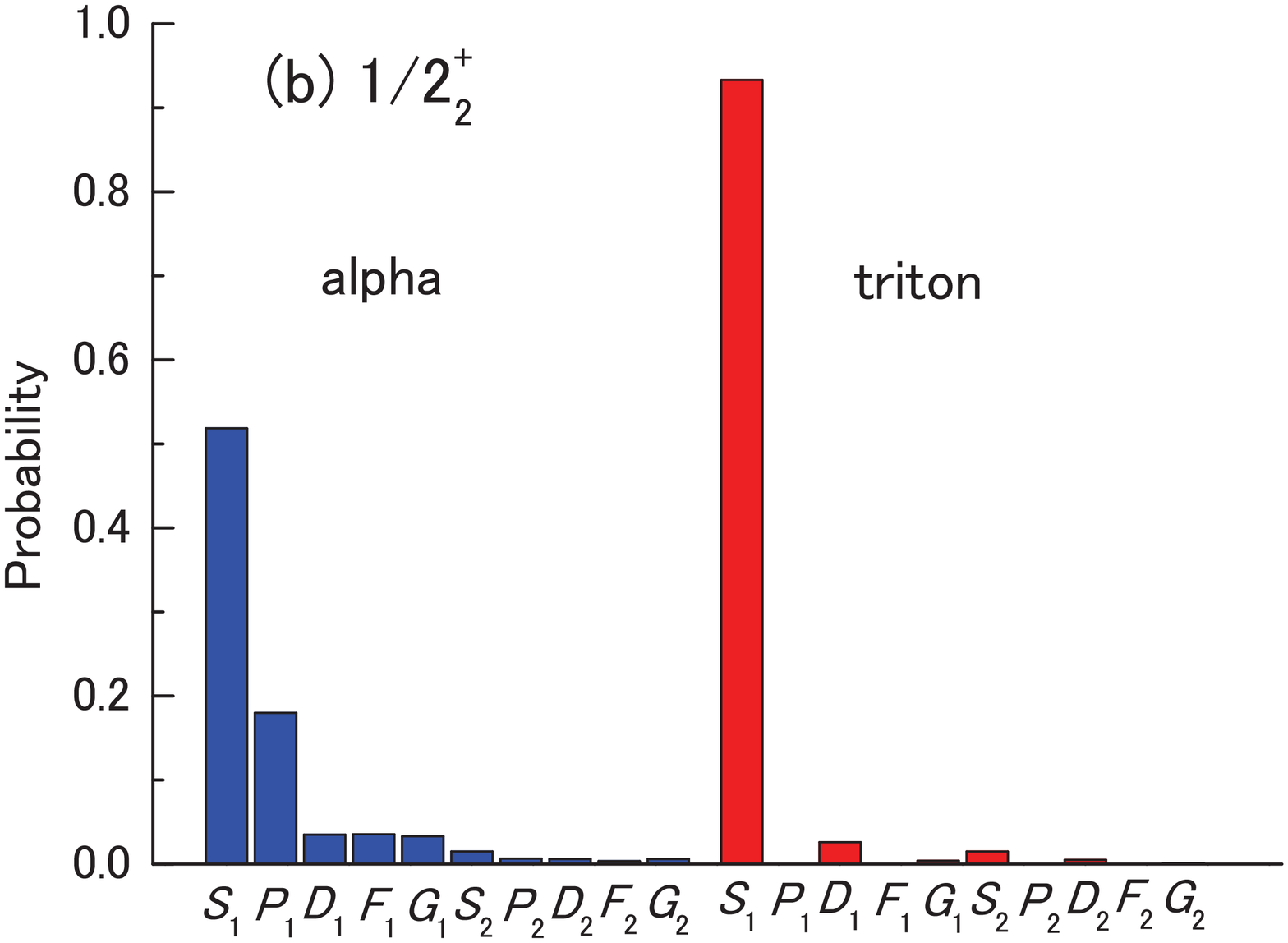}
\caption{(Color online) Occupation probabilities of the $\alpha$ ($t$) orbits for the (a)~$3/2^{-}_{1}$ and (b)~$3/2^{-}_{3}$ states in $^{11}$B.}
\label{fig:occupation_probabilities_11B}
\end{center}
\end{figure}
 
Figure~\ref{fig:occupation_probabilities_11B} shows the occupation probabilities of the {\it n}-th $L$-wave single-$\alpha$-particle (single-$t$-particle) orbit in the $3/2^-_1$ and $3/2^-_3$ states. In the $3/2^-_1$ state, the occupation probabilities of $\alpha$ particles spread out in several orbits, and those of $t$ orbits concentrate mainly on two orbits. These results originate from the SU(3) nature of the $3/2^-_1$ state as mentioned above. On the other hand, in the $3/2^-_3$ state, there also is no concentration of $\alpha$ occupation probability on a single orbit. This result is in contrast with those of the Hoyle state (see Fig.~\ref{fig:12C_occupation_probability}). Consequently the $3/2_{3}^{-}$ state can not be identified as the analogue of the Hoyle state. The reason why the $3/2_{3}^{-}$ state is not of Hoyle-type is  as follows:~The $3/2^{-}_{3}$ state is bound by $2.9$ MeV with respect to the $2\alpha+t$ threshold, while the Hoyle state is located by $0.38$~MeV above the $3\alpha$ threshold. This extra binding energy of $3/2_{3}^{-}$ with respect to the $2\alpha+t$ threshold suppresses strongly the development of the gas-like $2\alpha+t$ structure.
 
As for the $1/2^+$ states, the $1/2^+_1$ state appears as a bound state at $E_{x}^{\rm exp}=6.79$ MeV around the $^7$Li+$\alpha$ threshold. This low excitation energy indicates that $\alpha$-type correlations should play an important role in this state. In fact, we found that the $1/2^+_1$ state with the radius of 3.14~fm has a $^7$Li(g.s)+$\alpha$ structure with $P$-wave relative motion, although the $^7$Li($\alpha+t$) part is rather distorted in comparison with the ground state of $^7$Li. Since the $3/2^{-}_{3}$ state has the largest $S^{2}$ factor for the $^{7}$Li(g.s)+$\alpha$ channel with $S$-wave relative motion, the $1/2_{1}^{+}$ and $3/2^{-}_{3}$ states of $^{11}$B can be interpreted as  parity-doublet partners.
 
In addition to the $1/2_{1}^{+}$ state, the $1/2^{+}_{2}$ state appears as a resonant state at $E_x=11.95$ MeV ($\Gamma=190$~keV) around the $2\alpha+t$ threshold using the complex-scaling method~\cite{aguilar71,kuruppa88,kuruppa90,aoyama06}. The large radius ($R_{N}=5.98$~fm) indicates that the state has a dilute cluster structure. The analysis of the single-cluster properties showed that this state has as main configuration $(0S_{\alpha})^2(0S_t)$ orbital occupation with about $65$\% probability (see Fig.~\ref{fig:occupation_probabilities_11B}). Thus, the $1/2^{+}_{2}$ state can be called the Hoyle-analogue.  Recently, the $1/2^{+}~(3/2^{+})$ state at $E_{x}=12.56$ MeV with $\Gamma=210\pm20$ keV (located at $1.4$ MeV above the $2\alpha+t$ threshold) was observed in the $\alpha$+$^7$Li decay channel~\cite{soic04,curtis05,charity08}. The energy and width of the $12.56$-MeV state are in good correspondence to the present study. The Hoyle-analogue state in $^{11}$B, thus, could be assigned as the $12.56$-MeV state. It should be reminded that the $1/2^{+}_{2}$ state is located by $0.75$ MeV above the $\alpha+\alpha+t$ threshold, while $1/2^{+}_{1}$ is bound by $4.2$ MeV with respect to the three cluster threshold. The latter binding energy leads to a suppression of the development of  the gas-like $\alpha+\alpha+t$ structure in $1/2^{+}_{1}$, whereas it is generated with a large nuclear radius  in the $1/2^{+}_{2}$ state because of its appearance above the three-body threshold.
 
%%%%%%%%%%%%%%%%%%%%%%%%%%%%%%%%%%%%%%%%%%%
\begin{figure}[t]
\begin{center}
\includegraphics[width=0.75\hsize]{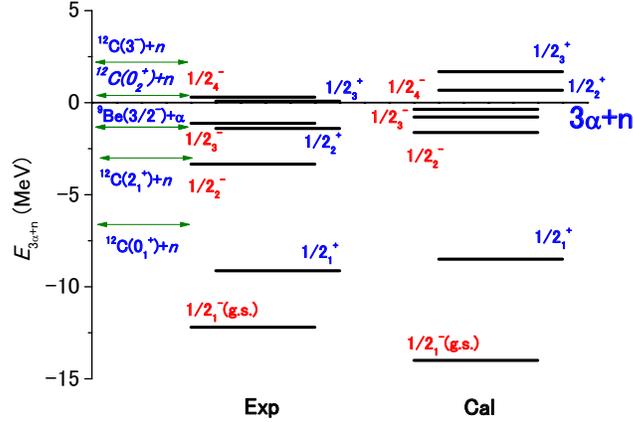}
\caption{(Color online) Calculated energy levels of $1/2^{-}$ and $1/2^{+}$ states in $^{13}$C with respect to the $3\alpha+t$ threshold, together with the experimental data.}
\label{fig:13C_energy_levels}
\end{center}
\end{figure}
%%%%%%%%%%%%%%%%%%%%%%%%%%%%%%%%%%%%%%%%%%%
 
The calculated energy spectrum of $1/2^-$ states in $^{13}$C is shown in Fig.~\ref{fig:13C_energy_levels}. The four $1/2^-$ energy levels are in good correspondence with  the experimental data. The ground state ($1/2^-_1$) is described as having a shell-model configuration. The calculated nuclear radius (2.39~fm) agrees with the data (2.44~fm). The three isoscalar monopole transition strengths, $M(1/2^{-}_{1}-1/2^{-}_{2})=4.2$, $M(1/2^{-}_{1}-1/2^{-}_{3})=5.6$, and $M(1/2^{-}_{1}-1/2^{-}_{4})=8.2$, are also consistent with experiment~\cite{kawabata08}, $6.1\pm0.5$, $4.2\pm0.4$, and $4.9\pm0.4$, respectively, in units of fm$^2$. The nuclear radii for the three excited $1/2^-$ states are $3.36$, $2.96$, and $3.19$~fm, respectively.
 
From the analysis of the radial behavior of the overlap amplitudes referring to the $^{12}$C+$n$ and $^9$Be+$\alpha$ channels, the $1/2^-_2$ and $1/2^-_3$ states are characterized as having  large components of $^{12}$C(g.s,$2^+$)+$n$ and $^{12}$C($3^-$)+$n$ with $^9$Be(g.s)+$\alpha$, respectively. On the other hand,  the $1/2_4^-$ state contains a somewhat large component of the $^{12}$C(Hoyle)+$n$ channel together with $^{12}$C($2^+$)+$n$ and $^9$Be(g.s)+$\alpha$. However, this state does not have as large an $\alpha$ condensate component as the Hoyle state in $^{12}$C. This is due to strong attraction of the odd-wave $\alpha-n$ potentials which induces the coupling of $^{12}$C($2^+$)+$n$ and $^9$Be(g.s)+$\alpha$ structures with the $^{12}$C(Hoyle)+$n$ configuration, and disturbs significantly the structure of the $3\alpha$ condensate in $^{13}$C. In the mirror nucleus $^{13}$N, the $3\alpha$+$p$ OCM analysis gives qualitatively similar results to those of the present $^{13}$C case~\cite{yamada11}.
 
As for the $1/2^+$ states of $^{13}$C, the energy spectrum of the first three $1/2^+$ states correspond well with the data (see Fig.~\ref{fig:13C_energy_levels}). We found that the $1/2_1^+$ state has a main configuration [$^{12}{\rm C(g.s)}\otimes{2s_{1/2}}$]. Reflecting the fact that the neutron binding energy of the $1/2_1^+$ state with respect to the $^{12}$C(g.s.)+$n$ threshold is as small as $1.9$ MeV, this state has a neutron-halo-like structure. In fact the calculated nuclear radius of this state ($2.68$ fm) is larger than that of the ground state ($2.39$ fm), and this enhancement of the radius comes from the neutron-halo-like structure.
 
On the other hand, the $1/2^+_2$ state has a dominant configuration of the extra neutron coupled with the Hoyle state, with  non-negligible mixing of $^9$Be(g.s,$1/2_1^-$)+$\alpha$ channels. The nuclear radius is about 4.0 fm, which is smaller than that of the Hoyle state in the $3\alpha$ OCM calculation~\cite{yamada05}. We found that the size of the $3\alpha$ part in this state is reduced by about 15~\% in comparison with that of the Hoyle state. The occupation probability of $\alpha$ particle in $0S$ orbit in this state is less than 30~\%, which is much smaller than that for the Hoyle state.
 
The $1/2_{3}^{+}$ state around the $3\alpha+n$ threshold has the nuclear radius of 5.40~fm with a dilute $\alpha$ condensate feature, in which $3\alpha$ particles occupy an identical $0S$ orbit with $55~\%$ probability. This state has a rather large overlap with the $^9$Be($1/2_1^+$)+$\alpha$ channel as well as with the $^{12}$C(Hoyle)+$n$ one. It is noted that the $^9$Be($1/2_1^+$) state is known to have a neutron-halo-like structure (or $2\alpha+n$ gas-like structure). Thus, these results suggest that the $1/2_{3}^{+}$ state is a candidate for the Hoyle-analogue state.

With this we terminate our consideration of cluster and condensate aspects in finite nuclei. An important connection with the finite systems is given by clustering, for instance $\alpha$ particle clustering and condensation in infinite matter. In the next section we turn to these issues.
 
%%%%%%%%%%%%%%%%%%%%%%%%%%%%%%%%%%%%%%%%%%%%%%%
\section{Clusters in nuclear matter and $\alpha$-particle condensation}\label{sec:5}
%%%%%%%%%%%%%%%%%%%%%%%%%%%%%%%%%%%%%%%%%%%%%%%
 
%%%%%%%%%%%%%%%%%%%%%%%%%%%%%%%%%%%%%%%%%%%%%%%
\subsection{Nuclear clusters in the medium}
%%%%%%%%%%%%%%%%%%%%%%%%%%%%%%%%%%%%%%%%%%%%%%%
 
Of course, it is also interesting and important to study how $\alpha$-clusters behave and actually condensate in infinite symmetric and asymmetric nuclear matters. This not only in regard to better understand what finally happens in a finite nucleus but in collapsing and compact stars one may speculate about the existence of a macroscopic $\alpha$-particle condensate. So let us first consider the modification an $\alpha$ particle undergoes when it is embedded in a nuclear medium.
 
Medium modifications of single-particle states as well as of few-nucleon states become of importance with increasing density of nuclear matter. The self-energy of an $A$-particle cluster can in principle be deduced from contributions describing the single-particle self-energies as well as medium modifications of the interaction and the vertices. A guiding principle in incorporating medium effects is the construction of consistent (``conserving'') approximations, which treat medium corrections in the self-energy and in the interaction vertex at the same level of accuracy. This can be achieved in a systematic way using the Green functions formalism~\cite{KKER}.  At the mean-field level, we have only the Hartree-Fock self-energy $\Gamma^{\rm HF}(1) = \sum_2 {\bar V(12,12)} f(2)$ together with the Pauli blocking factors, which modify the two-nucleon interaction from ${V(12,1'2')}$ to ${V(12,1'2')[1 - f(1) - f(2)]}$, with $f(1)=[1+\exp({\varepsilon}^{\rm HF}(1)-\mu)/T]^{-1}$ and ${\bar V(12,12)=V(12,12)-V(12,21)}$.  In the case of the two-nucleon system ($A=2$), the effective wave equation which includes those corrections is presented in the following form~\cite{RMS,RMS_1},
\begin{eqnarray}
&&\left[\varepsilon^{\rm HF}(1)+\varepsilon^{\rm HF}(2)-E_{2,P}\right] {\psi_{2,P}(12)} \nonumber \\
    &&\hspace*{20mm} + \frac{1}{2}\sum_{1'2'}[1-f(1)-f(2)]\,\,{{\bar V}(12,1'2')}  {\psi_{2,P}(1'2')} = 0.
\label{two_part_bind}
\end{eqnarray}
This effective wave equation describes bound states as well as scattering states. The onset of pair condensation is achieved when the binding energy $E_{2,P=0}$ coincides with $2\mu$, where $P$ denotes the total momentum of the two-nucleon system. It is noted that the Gor'kov equation in BCS theory of superfluidity is a special case of Eq.~(\ref{two_part_bind}).
 
Similar equations have been derived from the Green function approach for the case $A = 3$ and $A = 4$, describing triton/helion ($^3$He) nuclei as well as $\alpha$-particles in nuclear matter. The effective wave equation contains in mean field approximation the Hartree-Fock self-energy shift of the single-particle energies as well as the Pauli blocking of the interaction. We give the effective wave equation for $A=4$,
\begin{eqnarray}
\label{four_part_bind}
&& \left[\varepsilon^{\rm HF}(1)+\varepsilon^{\rm HF}(2)+ \varepsilon^{\rm HF}(3)+\varepsilon^{\rm HF}(4)
  -E_{4,P}\right]
\psi_{4,P}(1234) \nonumber\\ && +
\frac{1}{2}\sum_{i<j}\sum_{1'2'3'4'}[1-f(i)-f(j)]{{\bar V}(ij,i'j')}\prod_{k\neq i,j}\delta_{k,k'}
 \psi_{4,P}(1'2'3'4')=0.
\label{EWE}
\end{eqnarray}
A similar equation is obtained for $A=3$, which is an equation for a fermionic cluster.
 
\begin{figure}[t]
\begin{center}
\includegraphics[width=0.6\hsize]{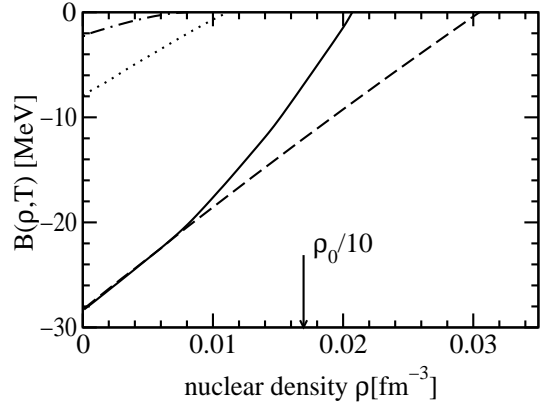}
\caption{Shift of binding energy of the light clusters ($d$ - dash dotted, $t/h$ - dotted, and $\alpha$ - dashed: perturbation theory, full line:~non-perturbative Faddeev-Yakubovski equation) in symmetric nuclear matter as a function of density for given temperature $T = 10$ MeV~\cite{beyer00}.}
\label{shifts}
\end{center}
\end{figure}

The effective wave equation has been solved using separable potentials for $A=2$ by integration. For $A=3,4$ we can use a {\it Faddeev approach}~\cite{beyer00}.  The shifts of binding energy can also be calculated approximately via perturbation theory.  In Fig.~\ref{shifts} we show the shift of the binding energy of the light clusters ($d, t/h$ and $\alpha$) in symmetric nuclear matter as a function of density for temperature $T$ = 10 MeV~\cite{beyer00}.
 
It is found that the cluster binding energy decreases with increasing density.  Finally, at the {\it Mott density} $\rho_{A,P}^{\rm Mott}(T)$ the bound state is dissolved. The clusters are not present at higher densities,  merging into the nucleonic medium.  For a given cluster type characterized by $A,n$, we can also introduce the Mott momentum $P^{\rm Mott}_{A}(\rho,T)$ in terms of the ambient temperature $T$ and nucleon density $\rho$, such that the bound states exist only for $P \ge P^{\rm Mott}_{A}(\rho,T)$.  We do not present an example here, but it is intuitively clear that a cluster with high c.o.m. momentum with respect to the medium is less affected by the Pauli principle than a cluster at rest, because the overlap of the bound state wave function in momentum space and the Fermi distribution function becomes smaller.

%%%%%%%%%%%%%%%%%%%%%%%%%%%%%%%%%%%%%%%%%%%%%%%
\subsection{Four-particle condensates and quartetting in nuclear matter}
\label{subsec:four_particle_condensates}
%%%%%%%%%%%%%%%%%%%%%%%%%%%%%%%%%%%%%%%%%%%%%%%
 
In general, it is necessary to take into account {\it all bosonic clusters} to gain a complete picture of the onset of superfluidity. As is well known, the deuteron is weakly bound as compared to other nuclei.  Higher $A$-clusters can arise that are more stable.  In this section, we will consider the formation of $\alpha$-particles, which are of special importance because of their large binding energy per nucleon ($\sim 7$ MeV).  We will not include tritons or helions, which are fermions and not so tightly bound.  Moreover, we will not consider nuclei in the iron region, which have even larger binding energy per nucleon than the $\alpha$-particle and thus constitute, in principle, the dominant component at low temperatures and densities. However, the latter are complex structures of many particles and are strongly affected by the medium as the density increases for given temperature, so that they are assumed not to be of relevance in the density region considered here.
 
The in-medium wave equation for the four-nucleon problem has been solved using the Faddeev-Yakubovski technique, with the inclusion of Pauli blocking, see also below.  The binding energy of an $\alpha$-like cluster with zero c.o.m.\ momentum vanishes at around $\rho_0/10$, where $\rho_0 \simeq 0.16$ nucleons/fm$^3$ denotes the saturation density of isospin-symmetric nuclear matter, see Fig.~\ref{shifts}.  Thus, the four-body bound states make no significant contribution to the composition of the system above this density.  Given the medium-modified bound-state energy $E_{4,P}$, the bound-state contribution to the EOS is
\begin{equation}
\rho_4(\beta,\mu) = \sum_P\left[e^{\beta(E_{4,P} - 2 \mu_p-2 \mu_n)} -1\right]^{-1}\,.
\end{equation}
We will not include the contribution of the excited states nor that of scattering states.  Because of the large specific binding energy of the $\alpha$ particle, low-density nuclear matter is predominantly composed of $\alpha$ particles. This observation underlies the concept of $\alpha$ matter and its relevance to diverse nuclear phenomena~\cite{akaishi69,brink73,tohsaki89,tohsaki96,takemoto04}.
 
As exemplified by Eq.~(\ref{EWE}), the effect of the medium on the properties of an $\alpha$ particle in mean-field approximation (i.e., for an uncorrelated medium) is produced by the Hartree-Fock self-energy shift and Pauli blocking. The shift of the $\alpha$-like bound state has been calculated using perturbation theory~\cite{RMS,RMS_1} as well as by solution of the Faddeev-Yakubovski equation \cite{beyer00}. It is found that the bound states of clusters $d$, $t$, and $h$ with $A<4$ are already dissolved at a Mott density $\rho_\alpha^{\rm Mott} \approx \rho_0/10$, see Fig.~\ref{shifts}. Since Bose condensation only is of relevance for $d$ and $\alpha$, and the fraction of $d$, $t$ and $h$ becomes low compared with that of $\alpha$ with increasing density, we can neglect the contribution of them to an equation of state. Consequently, if we further neglect the contribution of the four-particle scattering phase shifts in the different channels, we can now construct an equation of state $\rho(T, \mu) =\rho^{\rm free}(T, \mu) + \rho^{{\rm bound}, d}(T, \mu) +\rho^{{\rm bound}, \alpha}(T, \mu)$ such that $\alpha$-particles determine the behavior of symmetric nuclear matter at densities below $\rho_\alpha^{\rm Mott}$ and temperatures below the binding energy per nucleon of the $\alpha$-particle. The formation of deuteron clusters alone gives an incorrect description because the deuteron binding energy is small, and, thus, the abundance of $d$-clusters is small compared with that of $\alpha$-clusters. In the low density region of the phase diagram, $\alpha$-matter emerges as an adequate model for describing the nuclear-matter equation of state.
 
With increasing density, the medium modifications -- especially Pauli blocking -- will lead to a deviation of the critical temperature $T_c(\rho)$ from that of an ideal Bose gas of $\alpha$-particles (the analogous situation holds for deuteron clusters, i.e., in the isospin-singlet channel)~\cite{beyer00}.
 
Symmetric nuclear matter is characterized by the equality of the proton and neutron chemical potentials, i.e., $\mu_p=\mu_n=\mu$. Then an extended Thouless condition based on the relation for the four-body T-matrix (in principle equivalent to Eq.~(\ref{EWE}) at eigenvalue 4$\mu$)
\begin{eqnarray}
{\rm T}_4(1234,1''2''3''4'', 4 \mu)& =& \frac{1}{2}\sum_{1'2'3'4'} \Biggl\{
  \frac{{{\bar V}(12,1'2')}[1-f(1)-f(2)] }{ 4
    \mu - E_1-E_2-E_3-E_4 }\delta(3,3')\delta(4,4')\nonumber\\
&& \qquad \qquad + {\rm cycl.} \Biggr\}
{\rm T}_4(1'2'3'4',1''2''3''4'', 4 \mu)
\label{eq:4body_T_matrix}
\end{eqnarray}
serves to determine the onset of Bose condensation of $\alpha$-like clusters, noting that the existence of a solution of this relation signals a divergence of the four-particle correlation function. An approximate solution has been obtained by a variational approach, in which the wave function is taken as Gaussian incorporating the correct solution for the two-particle problem~\cite{roepke98}.
 
\begin{figure}[t]
\begin{center}
\includegraphics[width=60mm]{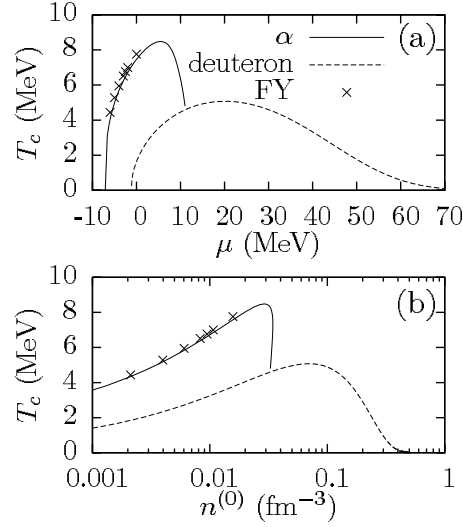}
\caption{\label{fig2}
Critical temperature of alpha and deuteron condensations as functions of (a)~chemical potential and (b)~density of free nucleons~\cite{slr09}. Crosses ($\times$) correspond to the solution of Eq.~(\ref{EWE}) with the Malfliet-Tjon interaction (MT I-III) using the Faddeev-Yakubovski method.}
\end{center}
\end{figure}
 
On the other hand, Eq.~(\ref{eq:4body_T_matrix}), respectively Eq.~(\ref{EWE}) at eigenvalue $4\mu$, has also been solved numerically exactly by the Faddeev-Yakubovsky method employing the Malfliet-Tjon force~\cite{MT}.  The results for the critical temperature of $\alpha$-condensation is presented in Fig.~\ref{fig2} as a function of the chemical potential $\mu$ (see also Ref.~\cite{roepke98}). The exact solution could only be obtained for negative $\mu$, i.e.~when there exists a bound cluster. It is, therefore, important to try yet another approximate solution of the in-medium four-body equation. Since the $\alpha$-particle is strongly bound, we make a momentum projected mean field ansatz for the quartet wave function~\cite{km05,schuck08,sg99}
\begin{equation}
\Psi_{1234}= (2\pi)^3 \delta^{(3)}({\bf k}_1 +{\bf k}_2 + {\bf k}_3 + {\bf k}_4) \prod_{i=1}^4\varphi({\bf k}_i)\chi^{ST},
\label{eq3}
\end{equation}
where $\chi^{ST}$ is the spin-isospin function which we suppose to be the one of a scalar ($S=T=0$). We will not further mention it from now on. We work in momentum space and $\varphi({\bf k})$ is the as-yet unknown single particle $0S$ wave function. In position space, this leads to the usual formula~\cite{Ring_Schuck} $\Psi_{1234} \rightarrow \int d^3R \prod_{i=1}^4 \tilde\varphi({\bf r}_i - {\bf R})$ where $\tilde\varphi({\bf r}_i)$ is the Fourier transform of $\varphi({\bf k}_i)$. If we take for $\varphi({\bf k}_i)$ a Gaussian shape, this gives: $\Psi_{1234} \rightarrow \exp[-c\sum_{1\leq i<k \leq 4} ({\bf r}_i - {\bf r}_k)^2]$ which is the translationally invariant ansatz often used to describe $\alpha$-clusters in nuclei. For instance, it is also employed in the $\alpha$-particle condensate wave function of Tohsaki, Horiuchi, Schuck, R\"opke (THSR) in Ref.~\cite{thsr}.
 
Inserting the ansatz (\ref{eq3}) into (\ref{EWE}) and integrating over superfluous variables, or minimizing the energy, we arrive at a Hartree-Fock type of equation for the single particle $0S$ wave function $\varphi(k)=\varphi(|{\bf k}|)$ which can be solved. However, for a general two body force ${V_{{\bf k}_1 {\bf k}_2, {\bf k}'_1 {\bf k}'_2}}$, the equation to be solved is still rather complicated. We, therefore, proceed to the last simplification and replace the two body force by a unique separable one, that is
\begin{equation}
{V_{{\bf k}_1 {\bf k}_2, {\bf k}'_1 {\bf k}'_2}} = \lambda e^{-k^2/k_0^2}e^{-k'^2/k_0^2} (2\pi)^3\delta^{(3)}({\bf K}-{\bf K}'),
\label{eq5}
\end{equation}
where ${\bf k}=({\bf k}_1-{\bf k}_2)/2$, ${\bf k}'=({\bf k}_1'-{\bf k}_2')/2$, ${\bf K}={\bf k}_1+{\bf k}_2$, and ${\bf K}'={\bf k}_1'+{\bf k}_2'$. This means that we take a spin-isospin averaged two body interaction and disregard that in principle the force may be somewhat different in the $S,T = 0, 1$ or $1, 0$ channels. It is important to remark that for a mean field solution the interaction only can be an effective one, very different from a bare nucleon-nucleon force. This is contrary to the usual gap equation for pairs, to be considered below, where, at least in the nuclear context, a bare force can be used as a reasonable first approximation.
 
We are now ready to study the solution of Eq.~(\ref{EWE}) for the critical temperature $T_c^{\alpha}$, defined by the point where the eigenvalue equals $4\mu$. For later comparison, the deuteron (pair) wave function at the critical temperature is also deduced from Eqs.~(\ref{EWE}) and (\ref{eq5}) to be
\begin{equation}
\phi(k)= -\frac{1-2f(\varepsilon)}{k^2/m-2\mu}\lambda e^{-k^2/k_0^2}
\int \frac{d^3k'}{(2\pi)^3} e^{-k^2/k_0^2} \phi(k'),
\label{eq8}
\end{equation}
where $\phi(k)$ is the relative wave function of two particles given by $\Psi_{12} \to {\phi(|\frac{{\bf k}_1-{\bf k}_2}{2}|)}$ ${\delta^{(3)}({\bf k}_1+{\bf k}_2)}$, and $\varepsilon=k^2/(2m)$. We also neglected the momentum dependence of the Hartree-Fock mean field shift in Eq.~(\ref{eq8}). It, therefore, can be incorporated into the chemical potential $\mu$. With Eq.~(\ref{eq8}), the critical temperature of pair condensation is obtained from the following equation:
\begin{equation}
1=-\lambda \int \frac{d^3k}{(2\pi)^3}
\frac{1-2f(\varepsilon)}{k^2/m-2\mu} e^{-2k^2/k_0^2}.
\label{eq10}
\end{equation}
 
In order to determine the critical temperature for $\alpha$-particle condensation, we have to adjust the temperature so that the eigenvalue of (\ref{EWE}) and (\ref{eq:4body_T_matrix}) equals $4\mu$. The result is shown in Fig. \ref{fig2}(a). In order to get an idea how this converts into a density dependence, we use for the moment the free gas relation between the density $n^{(0)}$ of uncorrelated nucleons and the chemical potential
\begin{equation}
n^{(0)}=4\int \frac{d^3k}{(2\pi)^3} f(\varepsilon).
\label{eq-density}
\end{equation}
We are well aware of the fact that this is a relatively gross simplification, for instance at the lowest densities, and we intend to generalize our theory in the future so that correlations are included into the density. This may be done along the work of Nozi\'eres and Schmitt-Rink~\cite{ns85}. The two open constants $\lambda$ and $k_0$ in Eq. (\ref{eq5}) are determined so that binding energy ($-28.3$ MeV) and radius ($1.71$ fm) of the free ($f_i=0$) $\alpha$-particle come out right. The adjusted parameter values are: $\lambda=-992$ MeV fm$^{3}$, and $k_0=1.43$ fm$^{-1}$. The results of the calculation are shown in Fig.~\ref{fig2}.
 
\begin{figure}[t]
\begin{center}
\includegraphics[width=60mm]{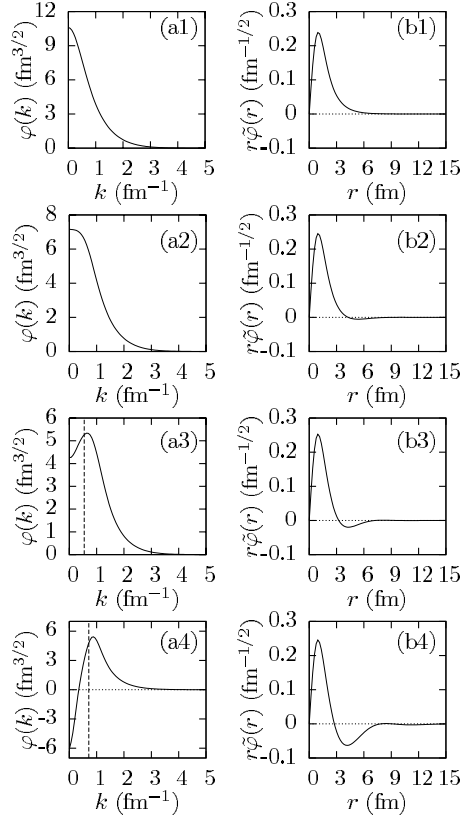}
\end{center}
\caption{\label{fig3}
Single particle wave functions (a1$\sim$a4)~in momentum space $\varphi(k)$ and (b1$\sim$b4)~in position space $r\tilde \varphi(r)$ at chemical potential ($\mu$), critical temperature ($T_c$), and density ($n$), which are obtained by solving Eq.~(\ref{EWE}) with the mean field ansatz (\ref{eq3})~\cite{slr09}:~
for (a1) [(b1)] $\mu=-7.08$ MeV, $T_c=0$ MeV, $n=0$ fm$^{-3}$,
for (a2) [(b2)] $\mu=-2.22$ MeV, $T_c=6.61$ MeV, $n=9.41 \times 10^{-3}$ fm$^{-3}$,
for (a3) [(b3)] $\mu=6.17$ MeV, $T_c=8.45$ MeV, $n=3.07 \times 10^{-2}$ fm$^{-3}$,
and
for (a4) [(b4)] $\mu=10.6$ MeV, $T_c=5.54$ MeV, $n=3.34 \times 10^{-2}$ fm$^{-3}$.
Figs.~(a1) and (b1) correspond to the wave functions for free $\alpha$-particle. The vertical lines in Figs.~(a3) and (a4) are at the Fermi wave length $k_F=\sqrt{2m\mu}$.}
\end{figure}
 
In Fig.~\ref{fig2}, the maximum of critical temperature $T^{\alpha}_{c, {\rm max}}$ is at $\mu=5.5$ MeV, and the $\alpha$-condensation can exist up to  $\mu_{\rm max}=11$ MeV.  It is very remarkable that the results obtained with (\ref{eq3}) for $T_c^{\alpha}$ very well agree with the exact solution of (\ref{EWE}) and (\ref{eq:4body_T_matrix}) using the Malfliet-Tjon interaction (MT I-III)~\cite{MT} with the Faddeev-Yakubovski method also shown by crosses in Fig.~\ref{fig2} (the numerical solution only could be obtained for negative values of $\mu$). This indicates that $T_c^{\alpha}$ is essentially determined by the Pauli blocking factors.
 
In Fig.~\ref{fig2} we also show the critical temperature for deuteron condensation derived from Eq.~(\ref{eq10}). In this case, the bare force is adjusted with $\lambda= -1305$ MeV fm$^3$ and $k_0 = 1.46$ fm$^{-1}$ to get experimental energy ($-2.2$ MeV) and radius ($1.95$ fm) of the deuteron. It is seen that at higher densities deuteron condensation wins over the one of $\alpha$-particle. The latter breaks down rather abruptly at a critical positive value of the chemical potential. Roughly speaking, this corresponds to the point where the $\alpha$-particles start to overlap. This behavior stems from the fact that Fermi-Dirac distributions in the four body case, see Eq.~(\ref{EWE}), can never become step-like, as in the two body case, even not at zero temperature, since the pairs in an $\alpha$-particle are always in motion. As a consequence, $\alpha$-condensation generally only exists as a BEC phase and the weak coupling regime is absent, see also discussion in Sec.~\ref{sebsec:gap_equation}.
 
Figure~\ref{fig3} shows the normalized self-consistent solution of the wave function in momentum space derived from Eq.~(\ref{EWE}) with the mean field ansatz (\ref{eq3}) and the wave function in position space defined by its Fourier transform $\tilde{ \varphi}(r)$. Figures~\ref{fig3}(a1) and (b1) represent the wave functions of the free $\alpha$-particle. The wave function resembles a Gaussian and this shape is approximately maintained as long as $\mu$ is negative, see Fig.~\ref{fig3}(a2). On the contrary, the wave function of Fig.~\ref{fig3}(a3), where the chemical potential is positive, has a dip around $k=0$ which is due to the Pauli blocking effect. For the even larger positive chemical potential of Fig.~\ref{fig3}(a4) the wave function develops a node. The maximum of the wave function shifts to higher momenta and follows the increase of the Fermi momentum $k_F$, as indicated on Fig.~\ref{fig3}.
 
On the other hand, the wave functions in position space in Figs.~\ref{fig3}(b2), (b3) and (b4) develop an oscillatory behavior, as the chemical potential increases. This is reminiscent to what happens in BCS theory for the pair wave function in position space~\cite{m06}.
 
An important consequence of this study is that at the lowest temperatures, Bose-Einstein condensation occurs for $\alpha$ particles rather than for deuterons.  As the density increases within the low-temperature regime, the chemical potential $\mu$ first reaches $-7$ MeV, where the $\alpha$'s Bose-condense.  By contrast, Bose condensation of deuterons would not occur until $\mu$ rises to $-1.1$ MeV.
 
The {\it ``quartetting''} transition temperature sharply drops as the rising density approaches the critical Mott value at which the four-body bound states disappear.  At that point, pair formation in the isospin-singlet deuteron-like channel comes into play, and a deuteron condensate will exist below the critical temperature for BCS pairing up to densities above the nuclear-matter saturation density $\rho_0$, as described in the previous Section. Of course, also isovector {\it n-n} and {\it p-p} pairing develops. The critical density at which the $\alpha$ condensate disappears is estimated to be $\rho_0/3$. Therefore, $\alpha$-particle condensation primarily only exists in the Bose-Einstein-Condensed (BEC) phase and there does not seem to exist a phase where the quartets acquire a large extension as Cooper pairs do in the weak coupling regime.  However, the variational approaches of Ref.~\cite{roepke98} and of Eq.~(\ref{eq3}) on which this conclusion is based represent only a first attempt at the description of the transition from quartetting to pairing.  The detailed nature of this fascinating transition remains to be clarified. Many different questions arise in relation to the possible physical occurrence and experimental manifestations of quartetting: Can we observe the hypothetical ``$\alpha$ condensate'' in nature?  What about thermodynamic stability?  What happens with quartetting in asymmetric nuclear matter?  Are more complex quantum condensates possible?  What is their relevance for finite nuclei?  As discussed, the special type of microscopic quantum correlations associated with quartetting may be important in nuclei, its role in these finite inhomogeneous systems being similar to that of pairing.
 
On the other hand, if at all, $\alpha$-condensation in compact star occurs at strongly asymmetric matter. It is, therefore, important to generalize the above study for symmetric nuclear matter to the asymmetric case. This can be done straight forwardly again using our momentum projected mean field ansatz (\ref{eq3}) generalized to the asymmetric case. This implies to introduce two chemical potentials, one for neutrons and for protons. We also have to distinguish two single particle wave functions in our product ansatz which now reads
\begin{eqnarray}
\psi_{1234}
&\to&
\varphi_p(\vec k_1)\varphi_p(\vec k_2)\varphi_n(\vec k_3)\varphi_n(\vec k_4)
\chi_0
\nonumber \\
&\times&
(2\pi)^3\delta(\vec k_1+\vec k_2+\vec k_3+\vec k_4)
\label{eq-phfwf}
\end{eqnarray}
where $\varphi_{\tau}(\vec k_i)=\varphi_{\tau}(|\vec k_i|)$ is the $s$-wave single particle wave functions for protons ($\tau=p$) and neutrons ($\tau=n$), respectively.  $\chi_0$ is the spin-isospin singlet wave function. This now leads to two coupled equations of the Hartree-Fock type for $\varphi_n$ and $\varphi_p$. For the force we use the same as in the symmetric case.
 
\begin{figure}[t]
\begin{center}
\includegraphics[width=60mm]{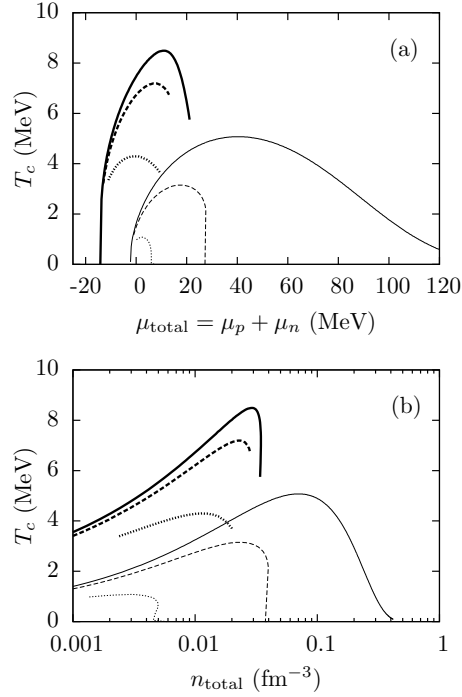}
\caption{\label{fig-ntotalvstc}
Critical temperature as a function of the total chemical potential $\mu_{\rm total}=\mu_p+\mu_n$ (top) and the total free density $n_{\rm total}$ (bottom)~\cite{sogo10}. Thick (thin) lines are for $\alpha$-particle (deuteron). Solid, dashed, and dotted lines are respectively for $\delta=0.0$, $\delta=0.5$, and $\delta=0.9$, where the density ratio $\delta$ is as in Eq.~(\ref{eq-densityratio}).}
\end{center}
\end{figure}
 
Fig.~\ref{fig-ntotalvstc}(a) shows the critical temperature of $\alpha$ condensation as a function of the total chemical potential $\mu_{\rm total}=\mu_p+\mu_n$. We see that $T_c$ decreases as the asymmetry, given by the parameter
\begin{eqnarray}
\delta=\frac{n_n-n_p}{n_n+n_p},
\label{eq-densityratio}
\end{eqnarray}
increases. This is in analogy with the deuteron case (also shown) which already had been treated in Refs.~\cite{afr93,lns01}. On the other hand, in Fig.~\ref{fig-ntotalvstc}(b), it is also interesting to show $T_c$ as a function of the free density which is
\begin{eqnarray}
n^{(0)}_{\rm total}&=&n^{(0)}_p+n^{(0)}_n
\label{eq-totaldensity}\\
n^{(0)}_p&=&2\int \frac{d^3k}{(2\pi)^3}f_p(k) \\
n^{(0)}_n&=&2\int \frac{d^3k}{(2\pi)^3}f_n(k),
\end{eqnarray}
where the factor two in front of the integral comes from the spin degeneracy, and $f_{p,n}(k)=[1+\exp({\hbar^{2}}k^2/2m-\mu_{p,n})]^{-1}$. It should be emphasized, however, that in the above relation between density and chemical potential, the free gas relation is used and correlations in the density have been neglected. In this sense the dependence of $T_c$ on density only is indicative, more valid at the higher density side. The very low density part where the correlations play a more important role shall be treated in a future publication. It should, however, be stressed that the dependence of $T_c$ on the chemical potential as in Fig.~\ref{fig-ntotalvstc}(a), stays unaltered.
 
The fact that for more asymmetric matter the transition temperature decreases, is natural, since as the Fermi levels become more and more unequal, the proton-neutron correlations will be suppressed. For small $\delta$'s, i.e., close to the symmetric case, $\alpha$ condensation (quartetting) breaks down at smaller density (smaller chemical potential) than deuteron condensation (pairing). This effect has already been discussed in our previous work for symmetric nuclear matter~\cite{roepke98,slr09}. For large $\delta$'s, i.e. strong asymmetries,  the behavior is opposite, i.e., deuteron condensation breaks down at smaller densities than $\alpha$ condensation, because the small binding energy of the deuteron can not compensate the difference of the chemical potentials.
 
More precisely, for small $\delta$'s, the deuteron with zero center of mass momentum is only weakly influenced by the density or the total chemical potential as can seen in Fig.~\ref{fig-ntotalvstc}. However, as $\delta$ increases, the different chemical potentials for protons and neutrons very much hinders the formation of proton-neutron Cooper pairs in the isoscalar channel for rather obvious reasons. The point to make here is that because of the much stronger binding per particle of the $\alpha$-particle, the latter is much less influenced by the increasing difference of the chemical potentials. For the strong asymmetry $\delta=0.9$ in Fig.~\ref{fig-ntotalvstc} then finally $\alpha$-particle condensation can exist up to $n_{\rm total}=0.02$~fm$^{-3}$ ($\mu_{\rm total}=9.3$~MeV), while the deuteron condensation exists only up to $n_{\rm total}=0.005$~fm$^{-3}$ ($\mu_{\rm total}=6.0$~MeV).
 
\begin{figure*}[t]
\begin{center}
\includegraphics[width=120mm]{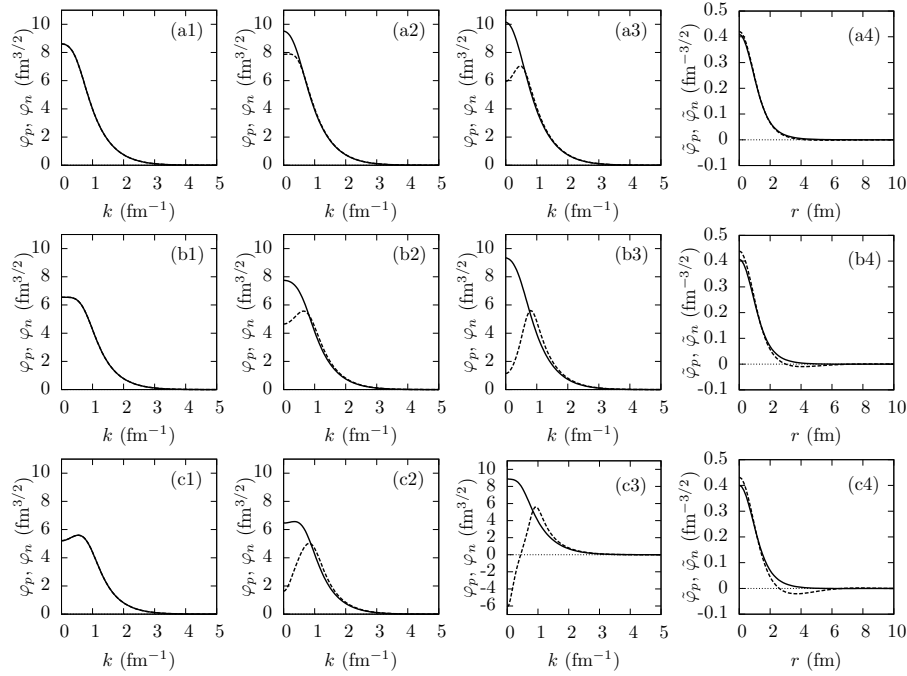}
\caption{\label{fig-spwf}
Momentum-space single particle wave functions for proton $\varphi_p$ (solid line), and for neutron $\varphi_n$ (dashed line) for the critical temperature as a function of $k$ for $\delta=0.0$, $0.5$ $0.9$, and the real-space wave functions for proton $\tilde \varphi_p$ (solid line), for neutron $\tilde \varphi_n$ (dashed line) as a function of $r$ for $\delta=0.9$ derived from the Fourier transform of $\varphi_{p,n}(k)$ with $\tilde \varphi_{p,n}(r)=\int d^3 ke^{i\vec k \cdot \vec r} \varphi_{p,n}(k)/(2\pi)^3$~\cite{sogo10}. The top, middle and bottom figures are for $\mu_{\rm total}=\mu_p+\mu_n\sim-11$~MeV, $\sim 0.0$~MeV, and $\sim 9.0$~MeV, respectively. The wave functions are normalized by $\int d^3 k \varphi_{p,n}^2(k)/(2\pi)^3=1$. The details of data for respective figures are following:
(a1)
$\delta=0.0$,
$\mu_{\rm total}=-11.1$~MeV,
$\mu_p=-5.53$~MeV,
$\mu_n=-5.53$~MeV,
$T_c=4.52$~MeV.
(a2)
$\delta=0.5$,
$\mu_{\rm total}=-11.5$~MeV,
$\mu_p=-8.18$~MeV,
$\mu_n=-3.35$~MeV,
$T_c=4.07$~MeV.
(a3), (a4)
$\delta=0.9$,
$\mu_{\rm total}=-11.0$~MeV,
$\mu_p=-10.8$~MeV,
$\mu_n=-0.163$~MeV,
$T_c=3.35$~MeV.
(b1)
$\delta=0.0$,
$\mu_{\rm total}=0.028$~MeV,
$\mu_p=-0.014$~MeV,
$\mu_n=-0.014$~MeV,
$T_c=7.46$~MeV.
(b2)
$\delta=0.5$,
$\mu_{\rm total}=0.11$~MeV,
$\mu_p=-4.65$~MeV,
$\mu_n=4.76$~MeV,
$T_c=6.74$~MeV.
(b3), (b4)
$\delta=0.9$,
$\mu_{\rm total}=-0.02$~MeV,
$\mu_p=-8.18$~MeV,
$\mu_n=8.16$~MeV,
$T_c=4.29$~MeV.
(c1)
$\delta=0.0$,
$\mu_{\rm total}=8.80$~MeV,
$\mu_p=4.40$~MeV,
$\mu_n=4.40$~MeV,
$T_c=8.44$~MeV.
(c2)
$\delta=0.5$,
$\mu_{\rm total}=8.93$~MeV,
$\mu_p=-1.12$~MeV,
$\mu_n=10.0$~MeV,
$T_c=7.16$~MeV.
(c3), (c4)
$\delta=0.9$,
$\mu_{\rm total}=8.94$~MeV,
$\mu_p=-4.21$~MeV,
$\mu_n=13.2$~MeV,
$T_c=3.72$~MeV.}
\end{center}
\end{figure*}
 
Overall, the behavior of $T_c$ is more or less as expected. We should, however, remark that the critical temperature for $\alpha$-particle condensation stays quite high, even for the strongest asymmetry considered here, namely $\delta$ = 0.9. This may be of importance for the possibility of $\alpha$-particle condensation in neutron stars and supernovae explosions~\cite{ls91,sto98}.
 
We also show the single particle wave functions of protons and neutrons, entering the quartet wave function~(\ref{eq-phfwf}), for various ratios of Fermi surface imbalance and chemical potentials in Fig.~\ref{fig-spwf}. In most cases of Fig.~\ref{fig-spwf}, the momentum-space wave functions with negative chemical potentials are monotonically decreasing whereas the ones with positive chemical potentials have a dip at $k=0$. However, the momentum-space wave functions also develop a dip at $k=0$ even at a negative chemical potential as the asymmetry takes on stronger values. This can be seen in Fig.~\ref{fig-spwf}(a3) and (c2). Furthermore, the neutron wave function in $k$-space with large positive chemical potential develops a node. This behavior is similar to the wave functions in Ref.~\cite{slr09}. As shown in Fig.~\ref{fig-spwf}, the dissymmetry of proton and neutron wave functions increases as $\delta$ increases. As a consequence, the critical temperature decreases, and the $\alpha$ condensation breaks down at a more dilute density, see Fig.~\ref{fig-ntotalvstc}. We also present in Fig.~\ref{fig-spwf}(a4), (b4) and (c4) the proton and neutron wave functions in real space. In spite of the sometimes strong dissymmetry in momentum space, the proton and neutron wave functions are relatively more similar to one another in $r$-space. The neutron wave function develops a node as the total chemical potential $\mu_{\rm total}=\mu_{p}+\mu_{n}$ increases, but the negative values of the wave function remain rather moderate.
 
In conclusion the $\alpha$-particle (quartet) condensation was investigated in homogeneous symmetric nuclear matter as well as in asymmetric nuclear matter.  We found that the critical density at which the $\alpha$-particle condensate appears is estimated to be around ${\rho_0}/3$ in the symmetric nuclear matter, and the $\alpha$-particle condensation can occur only at low density. This result is consistent with the fact that the Hoyle state ($0^+_2$) of $^{12}$C also has a very low density $\rho \sim \rho_0/3$. On the other hand, in the asymmetric nuclear matter, the critical temperature $T_c$ for the $\alpha$-particle condensation was found to decrease with increasing asymmetry. However, $T_c$ stays relatively high for very strong asymmetries, a fact of importance in the astrophysical context. The asymmetry affects deuteron pairing more strongly than $\alpha$-particle condensation. Therefore, at high asymmetries, if at all, $\alpha$-particle condensate seems to dominate over pairing at all possible densities.
 
%%%%%%%%%%%%%%%%%%%%%%%%%%%%%%%%%%%%%%%%%%
\subsection{Reduction of the $\alpha$-condensate with increasing density}
%%%%%%%%%%%%%%%%%%%%%%%%%%%%%%%%%%%%%%%%%%
 
The properties of $\alpha$ matter can be used to frame the discussion of the structure of $n\,\alpha$ nuclei.  As described in the preceding section, computational studies of these nuclei based on THSR cluster states have demonstrated that an $\alpha$ condensate is established at low nucleon density.  More specifically, states lying  near the threshold for decomposition into $\alpha$ particles, notably the ground state of $^8$Be, $^{12}$C in the $0_2^+$ Hoyle state, and corresponding states in $^{16}$O and other $n \alpha$ nuclei are {\it dilute}, being of low mean density and unusually extended for their mass numbers.  We have shown quantitatively within a variational approach that $\alpha$-like clusters are well formed, with the pair correlation function of $\alpha$-like clusters predicting relatively large mean distances.  For example, in determining the sizes of the $^{12}$C nucleus in its $0_1^+$ (ground) state and in its $0_2^+$ excited state, we obtained the r.m.s. radii of 2.44 fm and 3.83 fm, respectively.  The corresponding mean nucleon densities estimated from $36/4 \pi r^3_{\rm rms}$ are close to the nuclear-matter saturation density $\rho_0= 0.16$ nucleon/fm$^3$ in the former state and 0.03 nucleon/fm$^3$ in the latter. The expected low densities of putative alpha-condensate states are confirmed by experimental measurements of form factors \cite{funaki06_epja}.
 
All of our considerations indicate that quartetting is possible in the low-density regime of nucleonic matter, and that $\alpha$ condensates can survive until densities of about 0.03 nucleons/fm$^3$ are reached.  Here, we are in the region where the concept of $\alpha$ matter can reasonably be applied~\cite{JC80,SMS06}. It is then clearly of interest to use this model to gain further insights into the formation of the condensate, and especially the reduction or suppression of the condensate due to repulsive interactions~\cite{ropke2}.  We will show explicitly that in the model of $\alpha$ matter, as in our studies of finite nuclei, condensate formation is diminished with increasing density.  Already within an $\alpha$-matter model based on a simple $\alpha - \alpha$ interaction, we can demonstrate that the condensate fraction -- the fraction of particles in the condensate -- is significantly reduced from unity at a density of 0.03 nucleon/fm$^3$ and essentially disappears approaching nuclear matter-saturation density.
 
The quantum condensate formed by a homogeneous interacting boson system at zero temperature has been investigated in the classic 1956 paper of Penrose and Onsager \cite{PO} who characterize the phenomenon in terms of off-diagonal long-range order of the density matrix.  Here we recall some of their results that are most relevant to our problem.  Asymptotically, i.e., for $|{\vec r} - {\vec r}^\prime | \sim \infty$, the nondiagonal density matrix in coordinate representation can be decomposed as
\begin{equation}
\rho ({\vec r}, {\vec r}') \sim \psi_0^* ({\vec r}) \psi_0 ({\vec r}') + \gamma ({\vec r}- {\vec r}')\,.
\end{equation}
In the limit, the second contribution on the right vanishes, and the first approaches the condensate fraction, formally defined by
\begin{equation}
\rho_0 = \frac{\langle \Psi |a_0^\dagger a_0^{} | \Psi \rangle }{ \langle\Psi | \Psi \rangle } \,.
\end{equation}
Penrose and Onsager showed that in the case of a hard-core repulsion, the condensate fraction is determined by a filling factor describing the ratio of the volume occupied by the hard spheres. They applied the theory to liquid $^4$He, and found that for a hard-sphere model of the atom-atom interaction yielding a filling factor of about 28\%, the condensate fraction at zero temperature is reduced from unity (its value for the noninteracting system) to around 8\%.  (Remarkably, but to some extent fortuitously, this estimate is in rather good agreement with current experimental and theoretical values for the condensate fraction in liquid $^4$He.)
 
To make a similar estimate of the condensate fraction for $\alpha$ matter, we follow Ref.~\cite{sto98} and assume an ``excluded volume'' for $\alpha$ particles of  20 fm$^3$.  At a nucleonic density of $\rho_0/3$, this corresponds to a filling factor of about 28\%, the same as for liquid $^4$He.  Thus, a substantial reduction of the condensate fraction from unity (for a noninteracting $\alpha$-particle gas at zero temperature) is also expected in low-density $\alpha$ matter.
 
Turning to a more systematic treatment, we proceed in much the same way as Clark and coworkers~\cite{JC80}, referring especially to the most recent study with M.~T.~ Johnson.  Adopting the $\alpha-\alpha$ interaction potential
\begin{equation}
{V_{\alpha\alpha}(r)} = 475\,\, e^{-(0.7 r/{\rm fm})^2} {\rm MeV} - 130\,\, e^{-(0.475 r/{\rm fm})^2}{\rm MeV}
\label{AliBodmer}
\end{equation}
introduced by Ali and Bodmer~\cite{ali66}, we calculate the reduction of the condensate fraction as function of density within what is now a rather standard variational approach.  Alpha matter is described as an extended, uniform Bose system of interacting $\alpha$ particles, {\it disregarding} any change of the internal structure of the $\alpha$ clusters with increasing density. In particular, the dissolution of bound states associated with Pauli blocking (Mott effect) is not taken into account in the present description.
 
The simplest form of trial wave function incorporating the strong spatial correlations implied by the interaction potential (\ref{AliBodmer}) is the familiar Jastrow choice,
\begin{equation}
\Psi(\vec r_1, \dots, \vec r_A) = \prod_{i<j} f(|\vec r_i - \vec r_j|)\,.
\end{equation}
The normalization condition
\begin{equation}
4 \pi \rho_\alpha \int_0^\infty [f^2(r) - 1]\,\, r^2 dr = -1\,,
\label{norm}
\end{equation}
in which $\rho_{\alpha}$ is the number density of $\alpha$-particles, is imposed as a constraint on the variational wave function, in order to promote the convergence of the cluster expansion used to calculate the energy expectation value \cite{clark79}. In the low-density limit, the energy functional [binding energy per $\alpha$ cluster as a functional of the correlation factor $f(r)$] is given by
\begin{equation}
E[f]= 2 \pi \rho_\alpha \int_0^\infty \left\{ \frac{\hbar^2}{m_\alpha}
    \, \left( \frac{\partial f(r) }{\partial r} \right)^2 +f^2(r)
  V_{\alpha}(r) \right\} r^2 dr \,,
\label{eev}
\end{equation}
where $m_\alpha$ is the $\alpha$-particle mass, while the condensate fraction is given by
\begin{equation}
\rho_0 = \exp \left\{-4 \pi \rho_\alpha \int_0^\infty [f(r) - 1]^2\,\, r^2 dr \right\}\,.
\end{equation}
The variational two-body correlation factor $f$ was taken as one of the forms employed by Clark and coworkers \cite{JC80}, namely
\begin{equation}
f(r) = (1-e^{-ar})(1+be^{-ar}+ce^{-2ar})\,.
\end{equation}
At given density $\rho$, the expression for the energy expectation value is minimized with respect to the parameters $a$, $b$, and $c$, subject to the constraint (\ref{norm}).  It is important to note that these approximations, based on truncated cluster expansions, are reliable only at densities low enough so that the length scale associated with decay of $f^2(1)-1$ is sufficiently small compared to the average particle separation, which is inversely proportional to the cubic root of the density \cite{JC80,SMS06,clark79,Ristig}.
 
\begin{figure}[t]
\begin{center}
\includegraphics[width=0.65\hsize]{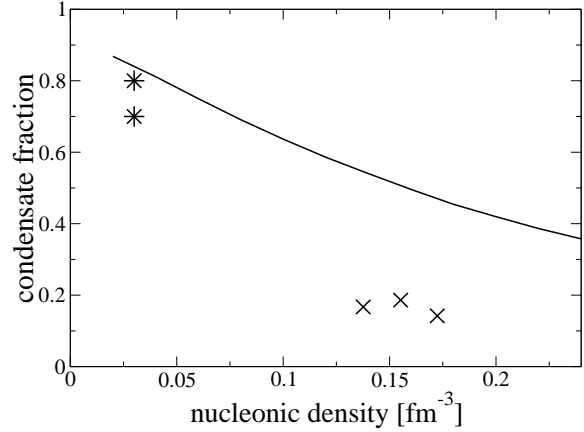}
\caption{Reduction of condensate fraction in $\alpha$ matter with increasing nucleon density. Exploratory calculations (full line) are compared with HNC calculations of Johnson and Clark~\cite{JC80} (crosses). For comparison, we show estimates of the condensate fraction in the $0_2^+$ (Hoyle) state of $^{12}$C, according to Refs.~\cite{yamada05,matsumura04} (stars).}
\label{fig:cond_fraction}
\end{center}
\end{figure}

To give an example, for the nucleon density $4 \rho_\alpha = 0.06$ fm$^{-3}$, a minimum of the energy expectation value (\ref{eev}) was found at $a=0.616$ fm$^{-1}$, $b=1.221$, and $c=-5.306$, with a corresponding energy per $\alpha$ cluster of $-9.763$ MeV and a condensate fraction of 0.750.  The dependence of the condensate fraction on the nucleon density $\rho = 4 \rho_\alpha$ as determined in this exploratory calculation is displayed in Fig.~\ref{fig:cond_fraction}.
 
\begin{figure}[t]
\begin{center}
\includegraphics[width=0.80\hsize]{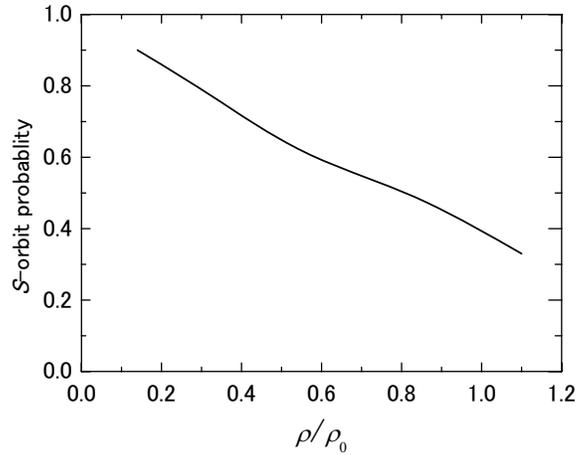}
\caption{Occupation of the $S$ orbital as a function of density using the $3\alpha$ OCM for $^{12}$C~\cite{yamada05}.}
\label{fig:cond_fraction_3alpha}
\end{center}
\end{figure}
 
\begin{figure}[t]
\begin{center}
\includegraphics[width=0.99\hsize]{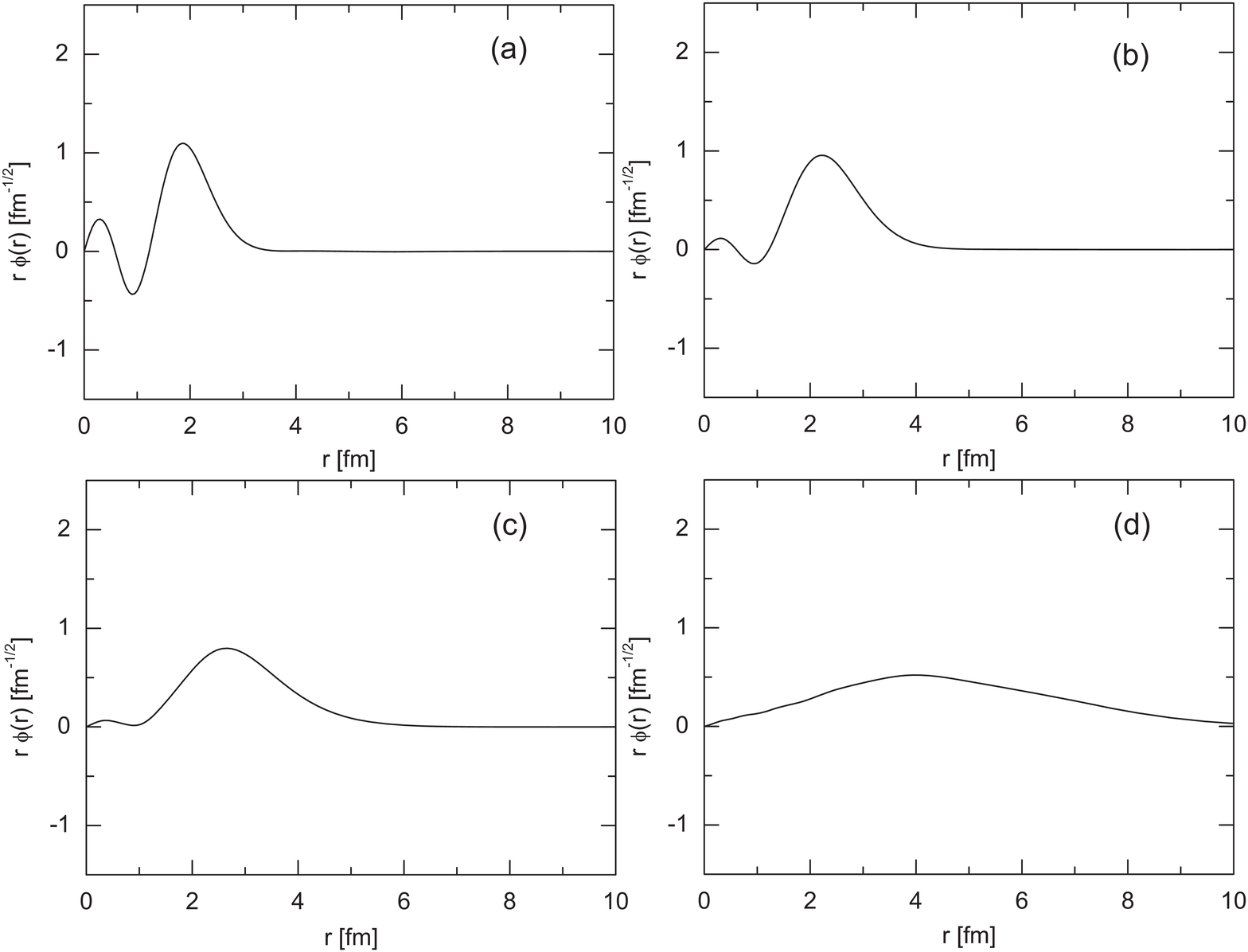}
\caption{Radial behaviors of the $S$ orbit in the $^{12}$C$(0^+)$ state with (a) $R = 2.42$ fm $(\rho/\rho_0\sim1.18)$, (b) $R = 2.70$ fm $(\rho/\rho_0\sim0.85)$, (c) $R = 3.11$ fm $(\rho/\rho_0\sim0.56)$, and (d) $R = 4.84$ fm $(\rho/\rho_0\sim0.15)$, where $R$ denotes the nuclear radius of the $^{12}$C$(0^+)$ state~\cite{yamada05}.}
\label{fig:13a-d}
\end{center}
\end{figure}
 
The reduction of the condensate fraction of $\alpha$ matter to roughly 0.8 as given by our calculation at nucleonic density 0.03 fm$^{-3}$ agrees well with results of Suzuki~ \cite{matsumura04} and Yamada~\cite{yamada05} for $^{12}$C in the Hoyle $0^+_2$ state.  Using many-particle approaches to the ground-state wave function and to the THSR ($0_2^+$) state of $^{12}$C, the occupation of the inferred natural $\alpha$ orbitals is found to be quite different in the two cases. Roughly 1/3 shares (approaching equipartition) are found for the $S$, $D$, and $G$ orbits in the ground ($0_1^+$) state, with $\alpha$-cluster occupations of 1.07, 1.07, and 0.82, respectively (see Sec.~\ref{subsec:4-1-3}).  On the other hand, in the Hoyle ($0^+_2$) state, one sees enhanced occupation (2.38) of the $S$ orbit and reduced occupation (0.29, 0.16, respectively) of the $D$ and $G$ orbits (see also Sec.~\ref{subsec:4-1-3}).  This corresponds to an enhancement of about 70\% compared with equipartition.
 
To get a more extended analysis, OCM calculations have been performed~\cite{yamada05} for studying the density dependence of the $S$-orbit occupancy in the Hoyle state on the different densities $\rho/\rho_0 \sim (R{(0^+_1)}_{\rm exp}/R)^3$, in which the rms radius ($R$) of $^{12}$C is taken as a parameter and $R{(0^+_1)}_{\rm exp} $=2.56 fm. A Pauli-principle respecting OCM basis $\Psi^{\rm OCM}_{0^+}(\nu)$ with a size parameter $\nu$ is used, in which the value of $\nu$ is chosen to reproduce a given rms radius $R$ of $^{12}$C, and the $\alpha$ density matrix $\rho(\vec{r},\vec{r}')$ with respect to $\Psi^{\rm OCM}_{0^+}(\nu)$ is diagonalized to obtain the $S$-orbit occupancy in the $0^+$ wave function. The results are shown in Fig.~\ref{fig:cond_fraction_3alpha}. The $S$-orbit occupancy is $70\sim 80$~\% around $\rho/\rho_0\sim (R{(0^+_1)}_{\rm exp}/R{(0^+_2)}_{\rm THSR})^3 = 0.21$, while it decreases with increasing $\rho/\rho_0$ and amounts to about $30\sim40$ \% in the saturation density region. Figure~\ref{fig:13a-d} shows the radial behaviors of the $S$-orbit with given densities. A smooth transition of the $S$-orbit is observed, with decreasing $\rho/\rho_0$, from a two-node $S$-wave nature $(\rho/\rho_0\sim 1.18)$ in Fig.~\ref{fig:13a-d}(a) to the zero-node $S$-wave one $(\rho/\rho_0\simeq0.15)$ in Fig.~\ref{fig:13a-d}(d)~\cite{yamada05}. The feature of the decrease of the enhanced occupation of the $S$ orbit is in striking correspondence with the density dependence of the condensate fraction calculated for nuclear matter (see Fig.~\ref{fig:cond_fraction}).
 
A more accurate and reliable variational description of $\alpha$ matter can be realized within the hypernetted-chain (HNC) approach to evaluate correlation integrals; this approach~\cite{JC80,clark79} largely overcomes the limitations of the cluster-expansion treatment, including the need for an explicit normalization constraint.  Such an improved approach is certainly required near the saturation density of nuclear matter, where it predicts only a small condensate fraction~\cite{JC80}.  Of course, at high densities the simple Ali-Bodmer interaction~\cite{ali66} ceases to be valid, and it becomes crucial to include the effects of Pauli blocking. Once again, this conclusion reinforces the point of view that we can expect signatures of an $\alpha$ condensate only for dilute nuclei near the threshold of $n \alpha$ decay.
 
%%%%%%%%%%%%%%%%%%%%%%%%%%%%%%%%%%%%%%%%%%%%%%%%%
\subsection{'Gap` equation for quartet order parameter}\label{sebsec:gap_equation}
%%%%%%%%%%%%%%%%%%%%%%%%%%%%%%%%%%%%%%%%%%%%%%%%%
 
For macroscopic $\alpha$ condensation it is, of course, not conceivable to work with a number projected $\alpha$ particle condensate wave function as we did when in finite nuclei only a couple of $\alpha$ particles were present. We rather have to develop an analogous procedure to BCS theory but generalized for quartets. In principle a wave function of the type $|\alpha\rangle = \exp[\sum_{1234}z_{1234}c_1^+c_2^+c_3^+c_4^+]|{\rm vac}\rangle$ would be the ideal generalization of the BCS wave function for the case of quartets. However, unfortunately, it is unknown so far (see, however, Ref.~\cite{jemai10}) how to treat such a complicated many body wave function mathematically in a reasonable way. So, we rather attack the problem from the other end, that is with a Gorkov type of approach, well known from pairing but here extended to the quartet case. Since, naturally, the formalism is complicated, we only will outline the main ideas and refer for details to the literature.
 
\begin{figure}[b]
\begin{center}
\includegraphics[width=60mm]{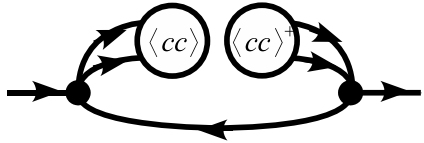}
\caption{\label{fig-bcsmass}Graphic representation of the BCS mass operator in Eq.~(\ref{eq-bcsmass})}
\end{center}
\end{figure}

Actually one part of the problem is written down easily. Let us guide from a particular form of the gap equation in the case of pairing. We have at zero temperature~\cite{RMS,RMS_1}
\begin{equation}
(\varepsilon_1+\varepsilon_2)\kappa_{12}+ (1-n_1 - n_2)\frac{1}{2}\sum_{1'2'} {\bar{V}_{121'2'}} \kappa_{1'2'} = 2\mu\kappa_{12},
\label{eq:gap_eq_two_particles}
\end{equation}
where $\kappa_{12} = \langle c_1c_2\rangle$ is the pairing tensor, $n_i = \langle c_i^+c_i \rangle$ are the BCS occupation numbers, and $\bar{V}_{121'2'}$ denotes the antisymmetrized matrix element of the two-body interaction. The $\varepsilon_i$ are the usual mean field energies. Equation (\ref{eq:gap_eq_two_particles}) is equivalent to the usual gap equation in the case of zero total momentum and opposite spin, i.e. in short hand: $2=\bar 1$ where the bar stands for 'time reversed conjugate'. The extension of (\ref{eq:gap_eq_two_particles}) to the quartet case is formally written down without problem
\begin{eqnarray}
(\varepsilon_{1234} - 4\mu)\kappa_{1234} &=& (1-n_1-n_2)\frac{1}{2}\sum_{1'2'} {{\bar{V}_{121'2'}}} \kappa_{1'2'34} \nonumber\\
&+& (1-n_1-n_3)\frac{1}{2}\sum_{1'3'} {{\bar{V}_{131'3'}}} \kappa_{1'23'4} + {\rm all~permutations}.
\end{eqnarray}
with $\kappa_{1234} = \langle c_1c_2c_3c_4 \rangle $ the quartet order parameter.
This is formally the same equation as in Eq.~(\ref{EWE}) with, however, the Fermi-Dirac occupation numbers replaced by the zero temperature quartet correlated single particle occupation numbers, similar to the BCS case. For the quartet case, the crux lies in the determination of those occupation numbers. Let us again be guided by BCS theory or rather by the equivalent Gorkov approach~\cite{Fetter-Walecka}. In the latter, there are two coupled equations, one for the normal single particle Green's function (GF) and the other for the anomalous GF. Eliminating the one for the anomalous GF in inserting it into the first equation leads to a Dyson equation with a single particle mass operator,
\begin{eqnarray}
M^{\rm BCS}_{1;1'}(\omega) = \sum_{2}\frac{\Delta_{12}\Delta_{1'2}^*}{\omega+\varepsilon_2}~~~{\rm with}~~~{\Delta_{12}=-\frac{1}{2}\sum_{34} {\bar{V}_{12,34}} {\langle c_{4}c_{3}\rangle}}.
\label{eq-bcsmass}
\end{eqnarray}
This can be graphically represented in Fig.~\ref{fig-bcsmass}, where $\langle cc \rangle$ stands for the order parameter $\kappa_{12}$ and the dot for the two body interaction.
 
\begin{figure}[t]
\begin{center}
\includegraphics[width=60mm]{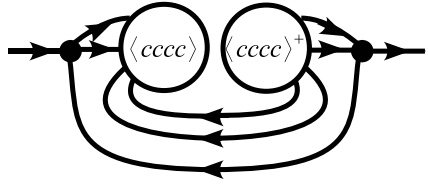}
\caption{\label{fig-alphamassapp}
Graphical representation of the approximate $\alpha$-BEC mass operator $M^{\rm quartet}$ of Eq.~(\ref{eq-Qmassoperator}). }
\end{center}
\end{figure}

The generalization to the quartet case is considerably more complicated but schematically the corresponding mass operator in the single particle Dyson equation can be represented graphically as in Fig.~\ref{fig-alphamassapp}, with the quartet order parameter $\langle cccc\rangle$. Put aside the difficulty to derive a manageable expression for this 'quartet' single-particle mass operator, what immediately strikes is that instead of only one 'backward going line' with $(-{\bf p},-\sigma)$ as in the pairing case, we now have three backwards going lines. As a consequence, the three momenta ${\bf k}_1$, ${\bf k}_2$, ${\bf k}_3$ in these lines are only constrained so that their sum be equal to ${\bf k}_1 + {\bf k}_2 + {\bf k}_3 = - {\bf p}$ and, thus, the remaining freedom has to be summed over. This is in strong contrast to the pairing case where the single backward going line is constrained by momentum conservation to $-{\bf p}$. So, no internal summation occurs in the mass operator belonging to pairing. The consequence of this additional momentum summation in the mass operator for quartetting leads with respect to pairing to a completely different analytic structure of the mass operator in case of quartetting. This is best studied with the so-called three hole level density $g_{3h}(\omega)$ which is related to the imaginary part of the three hole Green's function ${G^{3h}(k_1, k_2,k_3; \omega)} = ({\bar f_1} {\bar f_2} {\bar f_3} + f_1f_2f_3)/(\omega + \varepsilon_{123})$ with $\varepsilon_{123}=\varepsilon_{1}+\varepsilon_{2}+\varepsilon_{3}$ and $\bar f = 1- f$ by
\begin{eqnarray}
g_{3h}(\omega) &=&-\int \frac{d^3k_1}{(2\pi)^3}\frac{d^3k_2}{(2\pi)^3}\frac{d^3k_3}{(2\pi)^3} {\rm Im} G^{(3h)}(k_1,k_2,k_3;\omega+i\eta)
\nonumber \\
&=&\int \frac{d^3k_1}{(2\pi)^3}\frac{d^3k_2}{(2\pi)^3}\frac{d^3k_3}{(2\pi)^3}
\nonumber \\
&&\times
(\bar f_1 \bar f_2 \bar f_3+ f_1 f_2 f_3)
\pi\delta(\omega+\varepsilon_1
                +\varepsilon_2
                +\varepsilon_3).
\label{eq-ld})
\end{eqnarray}
 
\begin{figure}[t]
\begin{center}
\includegraphics[width=65mm]{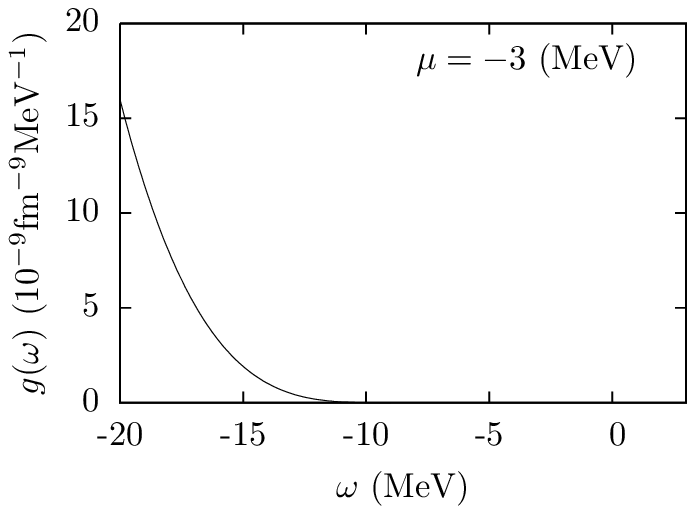}
\includegraphics[width=65mm]{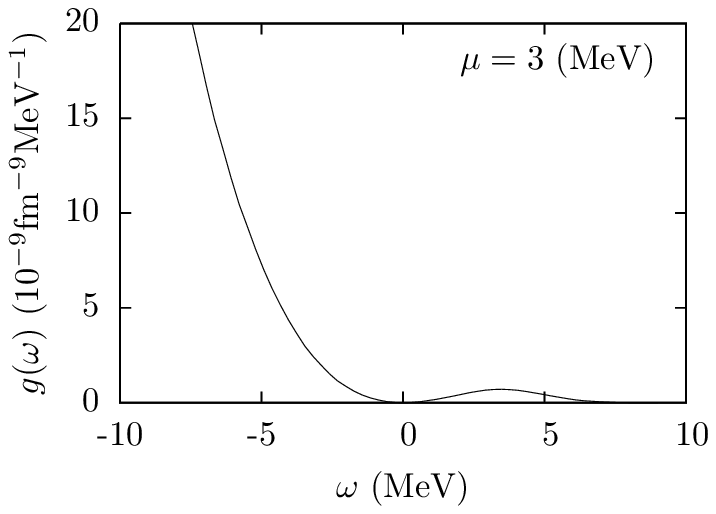}
\includegraphics[width=65mm]{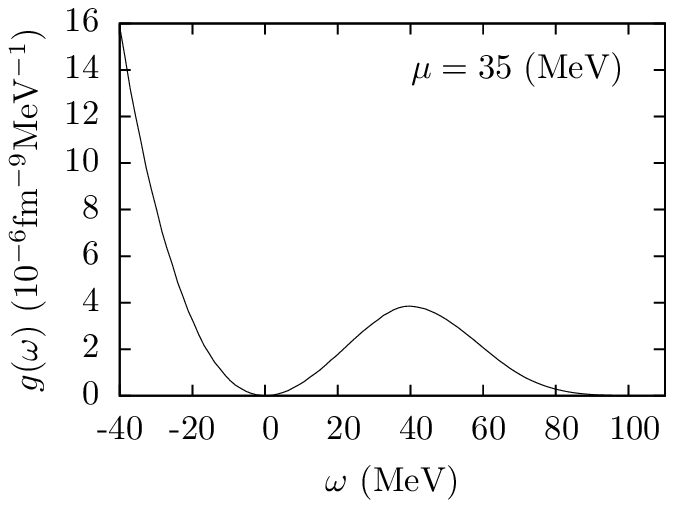}
\caption{\label{fig-ld}
{$3h$} level densities defined in Eq.~(\ref{eq-ld}) for various values of the chemical potential $\mu$ at zero temperature~\cite{sogo10_quartet}. }
\end{center}
\end{figure}
 
\begin{figure*}[t]
\begin{center}
\includegraphics[width=120mm]{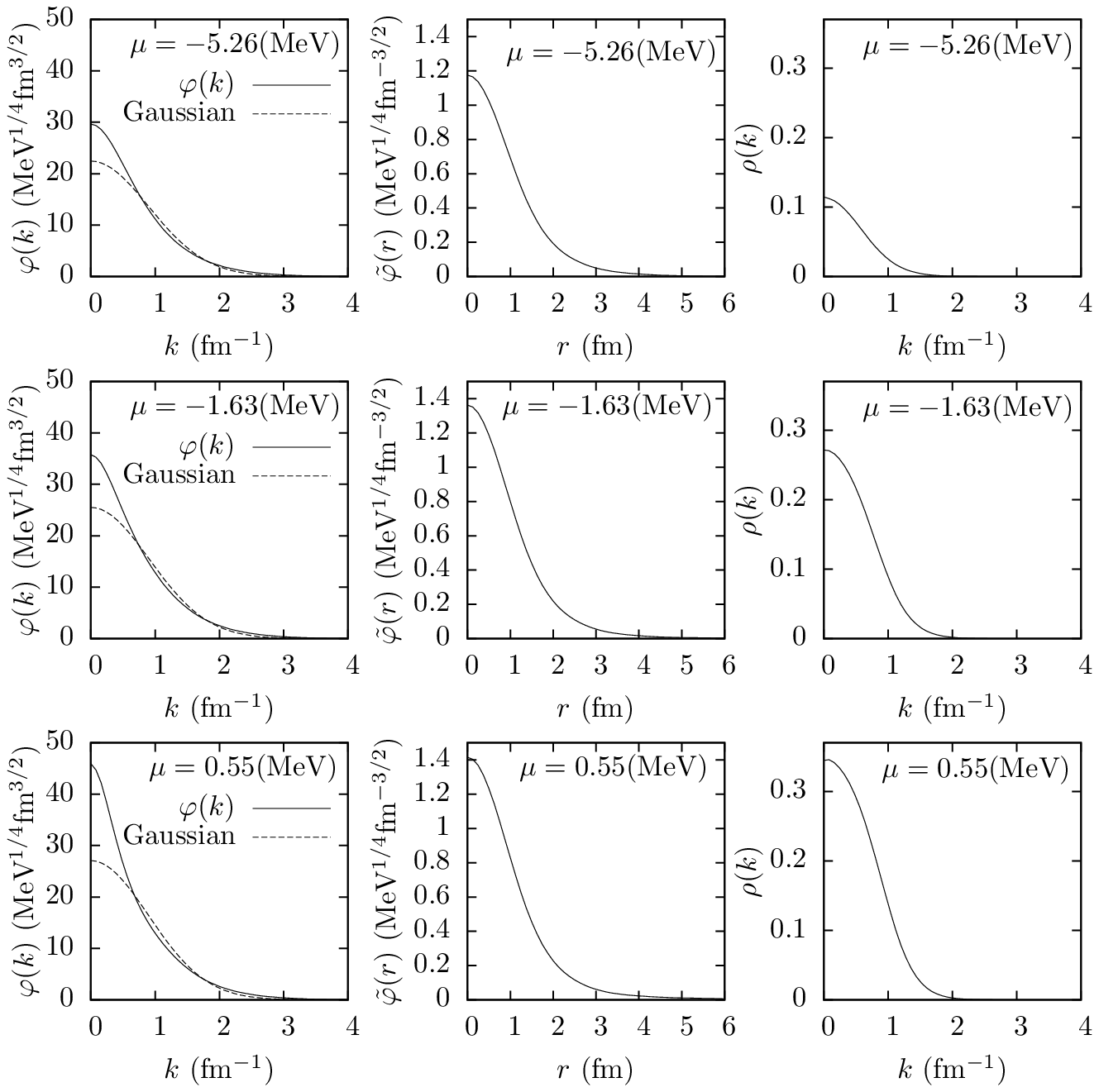}
\caption{\label{fig-spwf_1}
Single particle wave function $\varphi(k)$ in $k$-space (left), for $r$-space $\tilde \varphi(r)$ (middle), and occupation numbers (right) at $\mu=-5.26$ (top), $-1.63$ (middle), and $0.55$ (bottom), with zero temperature. The $r$-space wave function $\tilde \varphi(r)$ is derived from the Fourier transform of $\varphi(k)$ by $\tilde \varphi(r)=\int d^3k e^{i\vec k \cdot \vec r}\varphi(k)/(2\pi)^3$. The dashed line in the left panels correspond to the Gaussian with same norm and r.m.s. momentum as $\varphi(k)$~\cite{sogo10_quartet}}.
\end{center}
\end{figure*}
 
In Fig.~\ref{fig-ld} we show the level density at zero temperature ($f(\omega)=\theta(-\omega)$), where it is calculated with the proton mass $m=938.27$~MeV (natural units)~\cite{sogo10_quartet}. Two cases have to be considered, chemical potential $\mu$ positive or negative. In the latter case we have binding of the quartet. Let us first discuss the case $\mu>0$. We remark that in this case, the $3h$ level density goes through zero at $\omega=0$, i.e., since we are measuring energies with respect to the chemical potential $\mu$, just in the region where the quartet correlations should appear. This is a strong difference with the pairing case where the $1h$ level density, $g_{1h}(\omega)=\int \frac{d^3k}{(2\pi \hbar)^3} (\bar f_k + f_k)\delta(\omega+\varepsilon_k) = \int \frac{d^3k}{(2\pi \hbar)^3}\delta(\omega+\varepsilon_k)$, does not feel any influence from the medium and, therefore, the corresponding level density varies (neglecting the mean field for the sake of the argument) like in free space with the square root  of energy. In particular, this means that the level density is {\it finite} at the Fermi level. This is a dramatic difference with the quartet case and explains why Cooper pairs can strongly overlap whereas for quartets this is impossible as we will see below. We also would like to point out that the $3h$ level density is just the mirror to the $3p$ level density which has been discussed in Ref.~\cite{bhh86}.
 
For the case where $\mu<0$ there is nothing very special, besides the fact that it only is non-vanishing for negative values of $\omega$ and that the upper boundary is given by $\omega = 3\mu$. Therefore, the level density of Eq.~(\ref{eq-ld}) is zero for $\omega>3\mu$.
 
With these preliminary but crucial considerations we now pass to the evaluation of the single-particle mass operator with quartet condensation. Its expression can be shown to be of the following form
\begin{eqnarray}
M^{\rm quartet}_{1;1}(\omega)=\sum_{234}
\frac{\tilde \Delta_{1234}(\bar f_2 \bar f_3 \bar f_4+ f_2 f_3 f_4)
\tilde \Delta_{1234}^*}{\omega+\varepsilon_{234}}
\label{eq-Qmassoperator}
\end{eqnarray}
with
\begin{eqnarray}
\tilde \Delta_{1234}
=
\frac{1}{2} {\bar{V}_{12,1'2'}} \delta_{33'}\delta_{44'}
\langle c_{1'} c_{2'} c_{3'} c_{4'} \rangle.
\label{eq-tildedelta}
\end{eqnarray}
Again, comparing the quartet single-particle mass operator (\ref{eq-Qmassoperator}) with the pairing one (\ref{eq-bcsmass}), we notice the presence of the phase space factors in the former case while in Eq.~(\ref{eq-bcsmass}) they are absent. As already indicated above, this fact implies in the quartet case that only the Bose-Einstein condensation phase is born out whereas a 'BCS phase' (long coherence length) is absent.
The complexity of the calculation in Eq.~(\ref{eq-Qmassoperator}) is  much reduced using for the order parameter ${\langle cccc \rangle}$ our mean field ansatz projected on zero total momentum, as it was already very successfully employed with Eq.~(\ref{eq3}),
\begin{eqnarray}
{\langle c_1c_2c_3c_4 \rangle} &\rightarrow& \phi_{\vec k_1 \vec k_2,\vec k_3 \vec k_4}\chi_0, \nonumber \\
\phi_{\vec k_1 \vec k_2,\vec k_3 \vec k_4}
&=&
\varphi(|\vec k_1|)\varphi(|\vec k_2|)\varphi(|\vec k_3|)\varphi(|\vec k_4|)
\nonumber \\
&&\times
(2\pi)^3\delta(\vec k_1+\vec k_2+\vec k_3+\vec k_4),
\label{eq-PHF4bwf}
\end{eqnarray}
where $\chi_0$ is the spin-isospin singlet wave function. It should be pointed out that this product ansatz with four identical $0S$ single particle wave functions is typical for a ground state configuration of the $\alpha$ particle. Excited configurations with wave functions of higher nodal structures may eventually be envisaged for other physical situations. We also would like to mention that the momentum conserving $\delta$ function induces strong correlations among the four particles and (\ref{eq-PHF4bwf}) is, therefore, a rather non trivial variational wave function.
 
For the two-body interaction of ${\bar V}_{12,1'2'}$ in Eq.~(\ref{eq-tildedelta}), we employ the same separable form (59) as done already for the quartet critical temperature.
 
\begin{figure}[t]
\includegraphics[width=60mm]{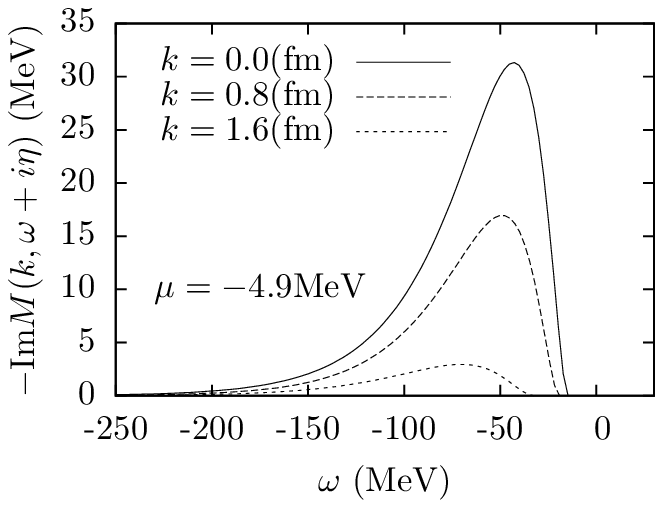}
\includegraphics[width=60mm]{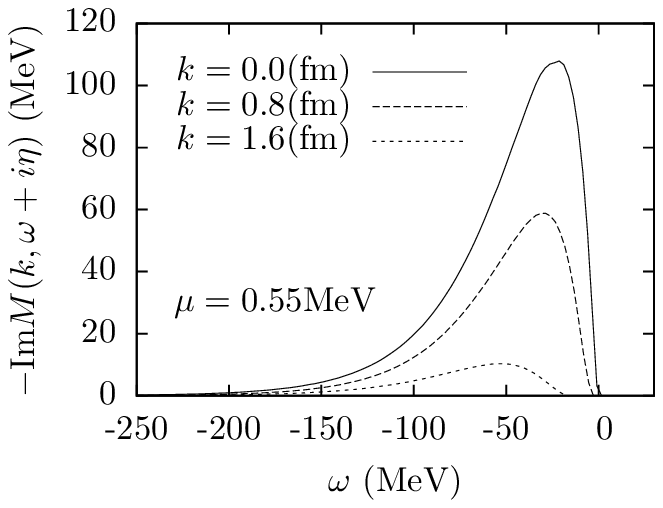}
\caption{\label{fig-immassoperator}
$-$Im$M^{\rm quartet}(k_1,\omega+i\eta)$ in Eq.~(\ref{eq-Qmassoperator}) as a function of $\omega$ for $\mu=-4.9$MeV (left) and for $\mu=0.55$MeV (right)  at  zero temperature~\cite{sogo10_quartet}.}
\end{figure}
 
At first let us mention that in this pilot application of our selfconsistent quartet theory, we only will consider the zero temperature case. As a definite physical example, we will treat the case of nuclear physics with the particularly strongly bound quartet, the $\alpha$ particle. It should be pointed out, however, that if scaled appropriately all energies and lengths can be transformed to other physical systems. For the nuclear case it is convenient to measure energies in Fermi energies $\varepsilon_F = 35$~MeV and lengths in inverse Fermi momentum $k_F^{-1} = 1.35^{-1}$~fm.
 
The single particle wave functions and occupation numbers obtained from the above cycle are shown in Fig.~\ref{fig-spwf_1}. We also insert the Gaussian wave function with same r.m.s. momentum as the single particle wave function in the left figures in Fig.~\ref{fig-spwf_1}. As shown in Fig.~\ref{fig-spwf_1}, the single particle wave function is sharper than a Gaussian.
 
We could not obtain a convergent solution for $\mu>0.55$~MeV. This difficulty is of the same origin as in the case of our calculation of the critical temperature for $\alpha$ particle condensation. In the r.h.s. panels of Fig.~\ref{fig-spwf_1} we also show the corresponding occupation numbers. We see that they are very small. However, they increase for increasing values of the chemical potential. For $\mu = 0.55$~MeV the maximum of the occupation still only attains 0.35 what is far away from the saturation value of one. What really happens for larger values of the chemical potential, is unclear. Surely, as discussed in Sec.~\ref{subsec:four_particle_condensates} the situation for the quartet case is completely different from the standard pairing case. This is due to the fact, as already mentioned, that the 3h level density goes through zero at $\omega=0$, i.e. just at the place where the quartet correlation should build up for positive values of $\mu$. Due to this fact, the inhibition to go into the positive $\mu$ regime is here even stronger than in the case of the critical temperature~\cite{slr09}.
 
\begin{figure}[t]
\begin{center}
\includegraphics[width=60mm]{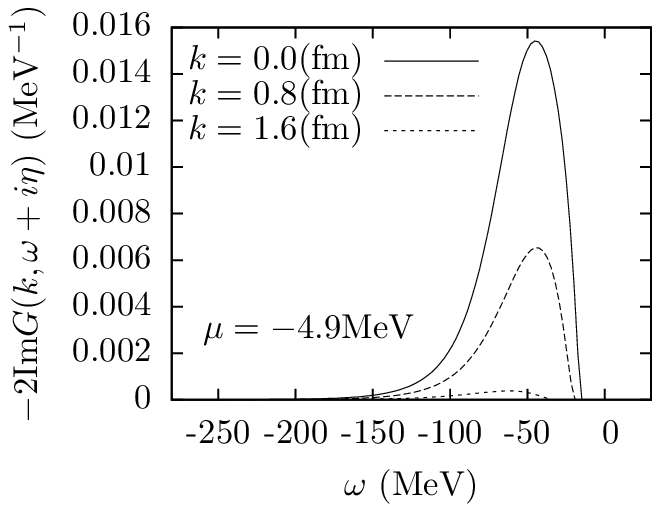}
\includegraphics[width=60mm]{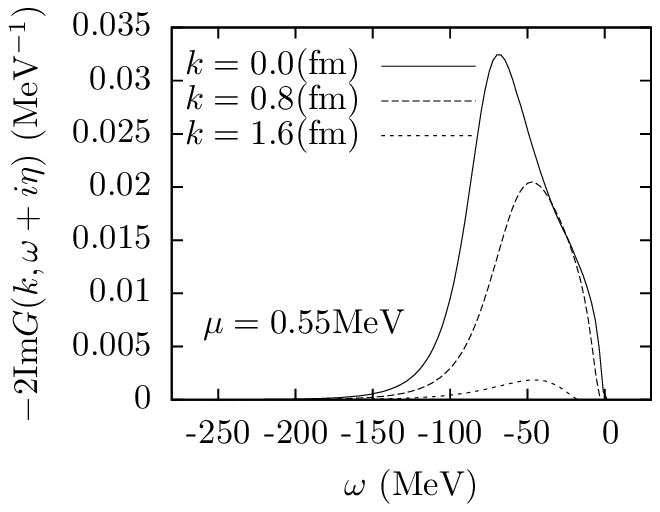}
\caption{
\label{fig-2ImG11}
$-2$Im$G(k,\omega+i\eta)$ in Eq.~(\ref{eq-Qmassoperator}) as function of $\omega$ for $\mu=-4.9$~MeV (top) and for $\mu=0.55$~MeV (bottom) at a zero temperature~\cite{sogo10_quartet}.}
\end{center}
\end{figure}
 
The situation in the quartet case is also in so far much different, as the $3h$ Green's function produces a considerable imaginary part of the mass operator.
Figure~\ref{fig-immassoperator} shows the imaginary part of the approximate quartet mass operator of Eq.~(\ref{eq-Qmassoperator}) for $\mu<0$ and $\mu>0$. These large values of the damping rate imply a strong violation of the quasiparticle picture. In Fig.~\ref{fig-2ImG11} we show the spectral function of the single particle GF. Contrary to the pairing case with its sharp quasiparticle pole, we here only find a very broad distribution, implying that the quasiparticle picture is completely destroyed. How to formulate a theory which goes continuously from the quartet case into the pairing case, is an open question. One solution could be to start right from the beginning with an in medium four body equation which contains a superfluid phase. When the quartet phase disappears, the superfluid phase may remain. Such investigations shall be done in the future.

%%%%%%%%%%%%%%%%%%%%%%%%%%%%%%%%%%%%%%%%%%%%%%%
\section{Summary and conclusions}\label{sec:6}
%%%%%%%%%%%%%%%%%%%%%%%%%%%%%%%%%%%%%%%%%%%%%%%
 
We discussed $\alpha$ condensation in nuclear systems. One remarkable manifestation is the Hoyle state ($0_2^+$) in $^{12}$C at 7.65 MeV with a gas-like structure of three $\alpha$-particles, trapped by a shallow self-consistent mean field of wide extension, in which the c.m. motion of the $\alpha$ particles occurs dominantly in the lowest $0S$ orbit. We found that a simple wave function of the $\alpha$-condensate type, called the THSR wave function, describes very nicely the structure of the Hoyle state and reproduces the inelastic form factor of ${^{12}}{\rm C}(e,e')$ and others quantities. The condensate feature of the Hoyle state was confirmed by the calculation of the bosonic occupation numbers in diagonalizing the bosonic density matrix. It was shown that the occupation of the $0S$ state of the $\alpha$-particles is over 70~\% for the Hoyle state, and the remaining component ($30~\%$) comes from residual correlations, mostly of the Pauli type, among the $\alpha$ particles. In spite of the very different number of particles and other important differences, the situation has some analogy with the case of cold bosonic atoms.
 
We conjectured that the $\alpha$-particle condensates also exist in heavier self-conjugate nuclei. Theoretical calculations of the OCM type indicate that the $0^{+}_{6}$ state at 15.1 MeV in $^{16}$O is a strong candidate. So far we do not dispose of sufficient experimental data to confirm its nature. Experiments are under way and being analyzed. This analogue of the Hoyle state in $^{16}$O has many similarities with the original one in $^{12}$C: it lies a couple of hundred keV above the $4\alpha$ disintegration threshold. It is quite strongly excited by $(e,e')$. Its width is 160 keV. This is much larger than for the Hoyle state in $^{12}$C but with respect to its energy it is still unusually small.  The large width stems from a position higher up in Coulomb barrier and also the Coulomb barrier itself has become slightly lower. The situation in $^{16}$O with respect to alpha clustering is considerably more complicated than in $^{12}$C.  Results from the $4\alpha$ OCM calculations showed that 2nd up to 5-th $0^+$ states in $^{16}$O have $\alpha$+$^{12}$C structures. Only the 6-th $0^+$ state is the analogue to Hoyle state. We also discussed the results of the THSR wave function for $^{16}$O.
 
As for the heavier $\alpha$-particle condensates, we found first that they are predicted to be slightly above their $n\alpha$ threshold in the $A=4n$ nuclei but below the Coulomb barrier, and second the phenomenon will terminate at about eight to ten $\alpha$'s as the confining Coulomb barrier fades away. However, the concept of $\alpha$ condensation in nuclei can be generalized to non self-conjugated nuclei ($A\not = 4n$). Since the nuclear $\alpha$-particle condensation is described dominantly as a product state of $\alpha$ particles occupying the lowest $0S$ orbit, the counterpart in $A\not = 4n$ nuclei should still be presented as a product state of the constituent clusters in the $0S$ state. For instance, we can conjecture product states composed of $\alpha$'s, a few neutrons and/or $s$-wave clusters ($d$, $t$, $^3$He) such as $(0S)^{2}_{\alpha}(0S)_{t}$ in $^{11}$B and $(0S)^3_{\alpha}(0S)_{n}$ in $^{13}$C etc. Indeed, our OCM calculations indicate that they appear slightly above their three- and four-cluster disintegrated thresholds, $2\alpha+t$ for $^{11}$B and $3\alpha+n$ for $^{13}$C, as positive-parity states with $J^{\pi}=1/2^{+}$.  These results encourage us to conjecture that cluster-gas-like states described by antisymmetrized product wave functions of constituent clusters, all in the $0S$ level, can exist in general in excited states of low density in light nuclei.

We dwell on the fact that concepts developed for infinite nuclear matter are of value also to interpret properties in finite nuclei and to construct useful approximations. As examples, we refer to pairing, two and more body correlations, and one body occupation numbers. Pairing definitely also is a useful concept for many finite nuclei, in spite of the fact that nuclei are by far not macroscopic objects. For example, the strong reduction of measured moments of inertia of such nuclei compared with the classical values are explained as a consequence of superfluidity~\cite{bohr,Ring_Schuck,baizot_ripka}. In this sense, we discussed  nuclear $\alpha$-particle condensation as the analogue to the number-projected BCS wave function, replacing Cooper pairs by $\alpha$ particles. A real macroscopic phase of condensed $\alpha$'s may be formed during the cooling process of compact stars~\cite{ShapT}, where one predicts the presence of $\alpha$-particle condensates~\cite{sto98}. On the other hand, a possibility of quartetting with cold atoms in which fermions are trapped in four different magnetic substates also have been discussed~\cite{sg99,Lecheminant,km05}. Theoretical and experimental works in this direction will also be useful and helpful to investigate the low-density bosonic $\alpha$-particle gas states in nuclei.

In conclusion, the idea of $\alpha$-particle condensation in nuclei is novel. A completely new nuclear phase in which $\alpha$ particles move like in a gas as quasi-elementary constituents is surely intriguing. In order to bring deeper insights into the role of clustering and quantum condensates in the systems of strongly interacting fermions, it is hoped that more $\alpha$-particle states in nuclei and/or many $\alpha$'s around a nuclear core, including cluster-gas-like states composed of $\alpha$'s, $t$'s and $n$'s etc., will be observed in the near future.

%%%%%%%%%%%%%%%%%%%%%%%%%%%%%%%%%%%%%%%%%%%%%%%
\section*{Acknowledgments}
%%%%%%%%%%%%%%%%%%%%%%%%%%%%%%%%%%%%%%%%%%%%%%%
 
The authors would like to thanks to B.~Borderie, M.~Freer, Y.~Hatanaka, K.~Ikeda, M.~Itoh, T.~Kawabata, K.~Kat${\rm{\bar{o}}}$, W. von Oertzen, M. F. Rivet, T. Sogo, and T.~Wakasa for useful discussions and comments.
 
%%%%%%%%%%%%%%%

%%%%%%%%%%%%%%%
 

\begin{thebibliography}{999}%
\bibitem{wildermuth77}
K. Wildermuth and Y. C. Tang, {\it A Unified Theory of the Nucleus} (Vieweg, Braunschweig, 1977).
\bibitem{ptp_supple_68}
K. Ikeda, H. Horiuchi, S. Saito, Y. Fujiwara, M. Kamimura, K. K. Kat${\rm{\bar{o}}}$, Y. Suzuki, E. Uegaki, H. Furutani, H. Kanada, T. Kaneko, S. Nagata, H. Nishioka, S.  Okabe, T. Sakuda, M. Seya, Y. Abe, Y. Kond${\rm{\bar{o}}}$, T. Matsuse, and A. Tohsaki-Suzuki  Prog. Theor. Phys. Suppl. {\bf 68} (1980).
\bibitem{oertzen06}
W. von Oertzen, M. Freer, and Y. Kanada-Enyo, Phys. Rep. {\bf 432}, 43 (2006).
\bibitem{hiyama09}
E.~Hiyama and T.~Yamada, Prog. Part. Nucl. Phys. {\bf 63}, 339 (2009).
\bibitem{thsr}
A.~Tohsaki, H.~Horiuchi, P.~Schuck and G.~R\"opke, Phys. Rev. Lett. {\bf 87}, 192501 (2001).
\bibitem{hiura72}
J. Hiura and R. Tamagaki, Prog. Theor. Phys. Supple. {\bf 52}, 25 (1972).
\bibitem{qmc}
R. B. Wiringa, S. C. Pieper, J. Carlson, and V. R. Pandharipande, Phys. Rev. C {\bf 62}, 014001 (2000).
\bibitem{uegaki}
E. Uegaki, S. Okabe, Y. Abe, and H. Tanaka, Prog. Theor. Phys. {\bf 57}, 1262 (1977); E. Uegaki, Y. Abe, S. Okabe, and H. Tanaka, Prog. Theor. Phys. {\bf 59}, 1031 (1978); {\bf 62}, 1621 (1979).
\bibitem{kamimura}
Y. Fukushima and M. Kamimura, {\it{Proc. Int. Conf. on Nuclear Structure}}, Tokyo, 1977, ed. T. Marumori (J. Phys. Soc. Jpn. Suppl. {\bf 44}, 225 (1978)); M. Kamimura, Nucl. Phys. A {\bf 351}, 456 (1981).
\bibitem{funaki03}
Y. Funaki, A. Tohsaki, H. Horiuchi, P. Schuck and G. R\"opke, Phys. Rev. C {\bf 67}, 051306(R) (2003).
\bibitem{chernykh07}
M. Chernykh, H. Feldmeier, T. Neff, P. von Neumann-Cosel, and A. Richter, Phys. Rev. Lett. {\bf 98}, 032501 (2007); T. Neff, talk at YIPQS international molecule workshop on "Alpha- and Dineutron-Correlation in Nuclear Many-Body Systems" Program, Oct.6-24, 2008, Kyoto, Japan. 
\bibitem{hoyle}
F. Hoyle, Astrophys. J. Suppl. {\bf 1}, 121 (1954).
\bibitem{fowler}
C. W. Cook, W. A. Fowler, C. C. Lauritsen, T. B. Lauritsen, Phys. Rev. {\bf 107}, 508 (1957). 
\bibitem{ajzenberg86}
F.~Ajzenberg-Selove, Nucl. Phys. A {\bf 506}, 1 (1990).
\bibitem{schuck07}
P. Schuck, Y. Funaki, H. Horiuchi, G. R\"opke, A. Thosaki, and T. Yamada, Prog. Part. Nucl. Phys. {\bf 59}, 285 (2007).
\bibitem{nupecc}
Y.~Funaki, H. Horiuchi, G. R\"opke, P. Schuck, A. Tohsaki and T. Yamada, {\it Nucl. Phys. News}, {\bf 17}(04), 11 (2007).
\bibitem{funaki09}
Y. Funaki, H. Horiuchi, W. von Oertzen, G. Ropke, P. Schuck, A. Tohsaki, and T. Yamada, Phys. Rev. C {\bf 80}, 064326 (2009).
\bibitem{brenner}
Y. Funaki, T. Yamada, H. Horiuchi, G. R\"opke, P. Schuck, and A. Tohsaki,  {\it Cluster Structure of Atomic Nuclei}, ed. by M. Brenner (Research Signpost/Transworld Research Network, 2010, ISBN: 978-81-308-0403-3), p.1.
\bibitem{Ring_Schuck}
P. Ring, and P. Schuck, {\it The Nuclear Many-Body Problem} (Springer-Verlag, Berlin, 1980).
\bibitem{matsumura04}
H. Matsumura and Y. Suzuki, Nucl. Phys. A {\bf 739}, 238 (2004).
\bibitem{yamada05}
T. Yamada and P. Schuck, Eur. Phys. J. A. {\bf 26}, 185 (2005).
\bibitem{funaki10}
Y. Funaki, T. Yamada, A. Tohsaki, H. Horiuchi, G. R\"opke, and P. Schuck, Phys. Rev. C {\bf 82}, 024312 (2010).
\bibitem{funaki05}
Y. Funaki, A. Tohsaki, H. Horiuchi, P. Schuck, and G. R\"opke, Eur. Phys. J. A. {\bf 24}, 321 (2005).
\bibitem{funaki06}
Y. Funaki, H. Horiuchi, and A. Tohsaki, Prog. Theor. Phys. {\bf 115}, 115 (2006).
\bibitem{kurokawa05}
C. Kurokawa and K. Kat${\rm{\bar{o}}}$, Phys. Rev.  {\bf C 71}, 021301 (2005); Nucl. Phys. A {\bf 792}, 87 (2007).
\bibitem{nocore}
P. Navr\'atil, J. P. Vary, and B. R. Barrett, Phys. Rev. Lett. {\bf 84}, 5728 (2000); P. Navr\'atil, J. P. Vary, and B. R. Barrett, Phys. Rev. C {\bf 62}, 054311 (2000); B. R. Barrett, B. Mihaila, S. C. Pieper, and R. B. Wiringa, Nucl. Phys. News, {\bf 13}, 17 (2003). 
\bibitem{navratil09}
P.~Navr\'atil, S.~Quaglioni, I.~Stetcu, and B.~R.~Barrett, J.~Phys.~G {\bf 36}, 083101 (2009).
\bibitem{itoh04}
M. Itoh et al., Nucl. Phys. A {\bf 738}, 268 (2004).
\bibitem{freer05}
M. Freer et al., Phys. Rev. C {\bf 71}, 047305 (2005); {\bf 76}, 034320 (2007). 
\bibitem{freer09}
M. Freer et al., Phys. Rev. C {\bf 80}, 041303(R) (2009).
\bibitem{kokalova05}
Tz. Kokalova et al., Eur. Phys. J. A {\bf 23}, 19 (2005); Phys. Rev. Lett. {\bf 96}, 192502 (2006).
\bibitem{ohkubo}
S. Ohkubo and Y. Hirabayashi, Phys. Rev. C {\bf 70}, 041602(R) (2004); {\bf 75}, 044609 (2007); Phys. Lett. B {\bf 684}, 127 (2010).
\bibitem{takashina}
M. Takashina and Y. Sakuragi, Phys. Rev. C {\bf 74}, 054606 (2006); M. Takashina, Phys. Rev. C {\bf 78}, 014602 (2008).
\bibitem{enyo}
Y. Kanada-Enyo, Prog. Theor. Phys. {\bf 117}, 655 (2007).
\bibitem{wakasa}
T.~Wakasa {\it et al.}, Phys. Lett. B {\bf 653}, 173 (2007).
\bibitem{Bordeier} 
Ad. R. Raduta et al., talk at {\it 2nd Workshop on "State of the Art in Nuclear Cluster Physics"}, Universite Libre de Bruxelles, May 25-28, 2010.
\bibitem{khoa11}
D. T. Khoa, D. C. Cuonga, and Y. Kanada-En'yo, Phys. Lett. B {\bf 695}, 469 (2011).
\bibitem{funaki08}
Y. Funaki, T. Yamada, H. Horiuchi, G. R\"opke, P. Schuck, and A. Tohsaki, Phys. Rev. Lett. {\bf 101}, 082502 (2008).
\bibitem{saito68}
S. Saito, Prog. Theor. Phys. {\bf 40} (1968); {\bf 41}, 705 (1969); Prog. Theor. Phys. Supple. {\bf 62}, 11 (1977).
\bibitem{tohsaki_nara}
A. Tohsaki, H. Horiuchi, P. Schuck and G. R\"opke, Nucl. Phys. A {\bf 738}, 259 (2004).
\bibitem{yamada04}
T. Yamada and P. Schuck,  Phys. Rev. C {\bf 69}, 024309 (2004).
\bibitem{kawabata07} 
T.~Kawabata et al., Phys. Lett. B {\bf 646}, 6 (2007).
\bibitem{enyo07}
Y. Kanada-En'yo, Phys. Rev. C {\bf 75}, 024302 (2007).
\bibitem{yamada10}
T. Yamada and Y. Funaki, Phys. Rev. C {\bf 82}, 064315 (2010).
\bibitem{kawabata08} 
T. Kawabata et al., Int. J. Mod. Phys. E {\bf 17}, 2071 (2008).
\bibitem{yamada08_IJMPE}
T.~Yamada and Y.~Funaki, Int.~J.~Mod.~Phys.~E {\bf 17}, 2101 (2008).
\bibitem{yoshida09}
T. Yoshida, N. Itagaki, and T. Otsuka, Phys. Rev. C {\bf 79}, 034308 (2009).
\bibitem{roepke98}
G. R\"opke, A. Schnell, P. Schuck, and P. Nozi\`eres, Phys. Rev. Lett. {\bf 80}, 3177 (1998).
\bibitem{beyer00}
M. Beyer, S.A. Sofianos, C. Kurths, G. R\"opke, and P. Schuck, Phys. Lett. B {\bf 488}, 247 (2000).
\bibitem{slr09}
T.~Sogo, R.~Lazauskas, G.~R\"opke, and P.~Schuck, Phys. Rev. C {\bf 79}, 051301(R) (2009).
\bibitem{sogo10}
T. Sogo, G. R\"opke, and P. Schuck, Phys. Rev. C {\bf 82}, 034322 (2010). 
\bibitem{sogo10_quartet}
T.~Sogo, G.~R\"opke, and P. Schuck, Phys. Rev. C {\bf 81}, 064310 (2010).
\bibitem{quartet}
B. Doucot, J. Vidal, Phys. Rev. Lett. {\bf 88}, 227005 (2002); S. Capponi, G. Roux, P. Lecheminant, P. Azaria, E. Boulat, S.R. White, Phys. Rev. A {\bf 77}, 013624 (2008); P. Lecheminant, E. Boulat, and P. Azaria, Phys. Rev. Lett. {\bf 95}, 240402 (2005).
\bibitem{olk08}T.~B.~Ottenstein, T.~Lompe, M.~Kohnen, A.~N.~Wenz, and S.~Jochim, Phys. Rev. Lett. {\bf 101}, 203202 (2008); J.~H.~Huckans, J.~R.~Williams, E.~L.~Hazlett, R.~W.~Stites, and K.~M.~O'Hara, Phys.~Rev.~Lett. {\bf 102}, 165302 (2009).
\bibitem{wheeler37}
J. A. Wheeler, Phys. Rev. {\bf 52}, 1083 (1937); {\bf 52}, 1107 (1937).
\bibitem{ptp_supple_62}
K. Ikeda, R. Tamagaki, S. Saito, H. Horiuchi, A. Tohsaki-Suzuki, and M. Kamimura, Prog. Theor. Phys. Supple. {\bf 62} (1977).
\bibitem{brink}
D. M. Brink, {\it in Proc. International School of Physics ``Enrico Fermi'', Course {\bf 36}} (Academic Press, New York, London, 1966) p. 247.
\bibitem{margenau}
H. Margenau, Phys. Rev. {\bf 59}, 37 (1941).
\bibitem{bayman58}
B. F. Bayman and A. Bohr, Nucl. Phys. {\bf 9}, 596 (1958/59).
\bibitem{yamada_ptp_08}
T. Yamada, Y. Funaki, H. Horiuchi, K. Ikeda, and A. Tohsaki, Prog. Theor. Phys. {\bf 120}, 1139 (2008).
\bibitem{hill53}
D. L. Hill and J. A. Wheeler, Phys. Rev. {\bf 89}, 1102 (1953).
\bibitem{griffin57}
J. J. Griffin and J. A. Wheeler, Phys. Rev. {\bf 108}, 311 (1957).
\bibitem{horiuchi74}
H. Horiuchi, Prog. Theor. Phys.  {\bf 51}, 1266 (1974); {\bf{53}}, 447 (1975).
\bibitem{Suz76}
Y. Suzuki, Prog. Theor. Phys. {\bf 55}, 1751 (1976); {\bf 56}, 111 (1976).
\bibitem{fukatsu89}
K. Fukatsu, K. Kat${\rm{\bar{o}}}$, and H. Tanaka, Prog. Theor. Phys. {\bf 81}, 738 (1989).
\bibitem{Kat92}
K.~Fukatsu and K. Kat${\rm{\bar{o}}}$, {\it Prog. Theor. Phys.} {\bf 87}, 151 (1992).
\bibitem{horiuchi77}
H. Horiuchi, Prog. Theor. Phys. {\bf 58}, 204 (1977); Prog. Theor. Phys. Supple. No.~62, 90 (1977).
\bibitem{kamimura88}
M.~Kamimura, Phys. Rev. A {\bf 38}, 621 (1988).
\bibitem{hiyama03}
E.~Hiyama, Y.~Kino, and M.~Kamimura, Prog. Part. Nucl. Phys. {\bf 51}, 223 (2003).
\bibitem{hiyama97}
E.~Hiyama, M.~Kamimura, T.~Motoba, T.~Yamada, and Y.~Yamamoto, Prog.~Theor.~Phys.~{\bf 97}, 881 (1997).
\bibitem{kukulin}
V. I. Kukulin, V. M. Krasnopol'sky, V. T. Voronchev, P. B. Sazonov, Nucl. Phys. A {\bf 417}, 128 (1984).
\bibitem{suzuki_02}
Y. Suzuki and M. Takahashi, Phys. Rev. C {\bf 65}, 064318, (2002).
\bibitem{suzuki_08}
Y. Suzuki, W. Horiuchi, M. Orabi, K. Arai, Few-Body Systems {\bf 42}, 33 (2008).
\bibitem{yamada08_obdm}
T. Yamada, Y. Funaki, H. Horiuchi, G. R\"opke, P. Schuck, and A. Tohsaki,  Phys. Rev. A {\bf 78}, 035603 (2008).
\bibitem{yamada09_obdm}
T. Yamada, Y. Funaki, H. Horiuchi, G. R\"opke, P. Schuck, and A. Tohsaki,  Phys. Rev. C {\bf 79}, 054314 (2009).
\bibitem{pethick00}
C. J. Pethick and L. P. Pitaevskii, Phys. Rev. A {\bf 62}, 033609 (2000).
\bibitem{pita}
L. P. Pitaevskii, private communication (2009).
\bibitem{44Ti_ohkubo}
T. Yamaya, K. Katori, M. Fujiwara, S. Kato and S. Ohkubo, Prog. Theor. Phys. Suppl. {\bf 132}, 73 (1998); F. Michel, S. Ohkubo and G. Reidemeister, Prog. Theor. Phys. Suppl. {\bf 132}, 7 (1998).
\bibitem{44Ti_horiuchi}
T. Wada and H. Horiuchi, Phys. Rev. C {\bf 38}, 2063 (1988).
\bibitem{itagaki04}
N. Itagaki, T. Otsuka, K. Ikeda and S. Okabe, Phys. Rev. Lett. {\bf 92}, 142501 (2004).
\bibitem{funaki_8be}
Y. Funaki, H. Horiuchi, A. Tohsaki, P. Schuck and G. R\"opke, Prog. Theor. Phys. {\bf 108}, 297 (2002).
\bibitem{volkov65}
A. B. Volkov, Nucl. Phys. A {\bf 74}, 33 (1965).
\bibitem{dalfovo99}
F. Dalfovo, S. Giorgini, L. P. Pitaevskii and S. Stringari, Rev. Mod. Phys. {\bf 71}, 463 (1999).
\bibitem{zinner07} 
N. T. Zinner and A. S. Jensen, Phys. Rev. C {\bf 78}, 041306(R) (2008).
\bibitem{baye2}
M. Libert-Heinemann, D. Baye, P. -H. Heenen, Nucl. Phys. A {\bf 339}, 429 (1980).
\bibitem{Hor68}
H.~Horiuchi and K. Ikeda,  Prog. Theor. Phys. {\bf 40}, 277 (1968).
\bibitem{mhn}
A. Hasegawa and S. Nagata, Prog. Theor. Phys. {\bf 45}, 1786 (1971); F. Tanabe, A. Tohsaki and R. Tamagaki, ibid. {\bf 53}, 677 (1975).
\bibitem{ropke2}
Y. Funaki, H. Horiuchi, G. R\"opke, P. Schuck, A. Tohsaki and T. Yamada, Phys. Rev. {\bf C 77}, 064312 (2008).
\bibitem{r-matrix}
A. M. Lane and R. G. Thomas, {\it Rev. Mod. Phys.} {\bf 30}, 257 (1958).
\bibitem{tohsaki_F1}
A. Tohsaki, Phys. Rev. C {\bf 49}, 1814 (1994).
\bibitem{gross}
L. P. Pitaevskii, Zh. Eksp. Teor. Fiz. {\bf 40}, 646 (1961) [Sov. Phys. JETP {\bf 13}, 451 (1961)]; E. P. Gross, Nuovo Cimento {\bf 20}, 454 (1961); J. Math. Phys. {\bf 4}, 195 (1963).
\bibitem{gogny}
D. Gogny, {\it Proceedings of the International Conference on Nuclear Selfconsistent Fields}, (Trieste, 1975), eds. G. Ripka and M. Porneuf, Noth Holland, Amsterdam, 1975. 
\bibitem{ali66}
S. Ali and A. R. Bodmer, Nucl. Phys. A {\bf 80}, 99 (1966).
\bibitem{Oertzen} 
W.~von~Oertzen, {\it Cluster in Nuclei}, Lecture Notes in Physics {\bf 818}, 109 (2010), ed. C.~Beck, Springer, Heidelberg, 2010.
\bibitem{ajzenberg_A=7}
F.~Ajzenberg-Selove, Nucl. Phys. A {\bf 490}, 1 (1988).
\bibitem{aguilar71}
J.~Aguilar and J.M.~Combes, Commun. Math. Phys. {\bf 22}, 269 (1971); E.~Balslev and J.M.~Combes, Commun. {\it ibid.} {\bf 22}, 280 (1971); B.~Simon, {\it ibid.} {\bf 27}, 1 (1972).
\bibitem{kuruppa88}
A.~T.~Kruppa, R.~G.~Lovas, and B.~Gyarmati, Phys. Rev. C {\bf 37}, 383 (1988).
\bibitem{kuruppa90}
A.~T.~Kruppa and K.~Kat\=o,  Prog.~Theor.~Phys.~{\bf 84}, 1145 (1990).
\bibitem{aoyama06}
S.~Aoyama, T.~Myo, K.~Kat\=o, and K.~Ikeda, Prog.~Theor.~Phys.~{\bf 116}, 1 (2006).
\bibitem{soic04}
N.~Soi\'c et al., Nucl. Phys. A {\bf 742}, 271 (2004).
\bibitem{curtis05}
N.~Curtis et al., Phys. Rev. C {\bf 72}, 044320 (2005).
\bibitem{charity08}
R.~J.~Charity et al., Phys. Rev. C {\bf 78}, 054307 (2008).
\bibitem{yamada11}
T. Yamada and Y. Funaki, private communication.


\bibitem{KKER}
W. D. Kraeft, D. Kremp, W. Ebeling, and G. R\"opke, {\it Quantum Statistics of Charged Particle Systems}, Berlin, Akademie-Verlag 1986.
\bibitem{RMS}
G. R\"opke, L. M\"unchow, and H. Schulz, Nucl. Phys. A {\bf 379}, 536 (1982).
\bibitem{RMS_1}
G. R\"opke, M. Schmidt, L. M\"unchow, and H. Schulz, Nucl. Phys. A {\bf 399}, 587 (1983).
\bibitem{akaishi69}
Y. Akaishi and H. Band${\rm{\bar{o}}}$, Prog. Theor. Phys. {\bf 41}, 1594 (1969). 
\bibitem{brink73}
D. N. Brink and J. J. Castro, Nucl. Phys. A {\bf 216}, 109 (1973).
\bibitem{tohsaki89}
A. Tohaski, Prog. Theor. Phys. {\bf 81}, 370 (1989); {\bf 88}, 1119 (1992); {\bf 90}, 871 (1993).
\bibitem{tohsaki96}
A. Tohsaki, Phys. Rev. Lett. {\bf 76}, 3518 (1996).
\bibitem{takemoto04}
H. Takemoto, M. Fukushima, S. Chiba, H. Horiuchi, Y. Akaishi, and A. Tohsaki, Phys. Rev. C {\bf 69}, 035802 (2004).
\bibitem{MT}
R.~A.~Malfliet and J.~A.~Tjon, Nucl. Phys. {\bf A127}, 161 (1969); G.~L.~Payne, J.~L.~Friar and B.~F.~Gibson, Phys. Rev. C {\bf 26}, (1982) 1385.
\bibitem{km05}
H.~Kamei and K.~Miyake, J. Phys. Soc. Jpn. {\bf 74}, 1911 (2005).
\bibitem{schuck08}
P.~Schuck, {\it State of the Art in Nuclear Cluster Physics}, Strasbourg, May 2008, in Int. J. Mod. Phys. E {\bf 17}, 2136 (2008).
\bibitem{sg99}
A.~S.~Stepanenko, J.~M.~F.~Gunn, arXiv: cond-mat/9901317.
\bibitem{ns85}
P.~Nozi\'eres and S.~Schmitt-Rink, J.~Low~Temp.~Phys. {\bf 59}, 195 (1985).
\bibitem{m06} 
M.~Matsuo, Phys. Rev. C {\bf 73}, 044309 (2006).
\bibitem{afr93}
T.~Alm, B.~L.~Friman, G.~R\"opke, and H.~Schulz, Nucl. Phys. A {\bf 551}, 45 (1993).
\bibitem{lns01}
U.~Lombardo, P.~Nozi\'eres. P.~Schuck, H.-J.~Schulze, and A. Sedrakian, Phys. Rev. C {\bf 64}, 064314 (2001). 
\bibitem{ls91}
J.~M.~Lattimer and F.~D.~Swesty, Nucl. Phys. A {\bf 535}, 331 (1991).
\bibitem{sto98}
M.~Shen, H.~Toki, K.~Oyamatsu, and K.~Sumiyoshi, Prog. Theor. Phys. {\bf 100}, 1013 (1998).


\bibitem{funaki06_epja}
Y. Funaki, A. Tohsaki, H. Horiuchi, P. Schuck, and G. Roepke, Eur. Phys. J. A  {\bf 28}, 259 (2006).
\bibitem{JC80}
M. T. Johnson and  J. W. Clark, Kinam {\bf 2}, 3 (1980) (PDF available at Faculty web page of J. W. Clark at http://wuphys.wustl.edu); see also J. W. Clark and T. P. Wang,  Ann. Phys. (N.Y.) {\bf 40}, 127 (1966) and G.~P.~Mueller and J. W. Clark, Nucl. Phys. A {\bf 155}, 561 (1970).
\bibitem{SMS06}
A. Sedrakian, H. M\"uther, and P. Schuck, Nucl. Phys. A {\bf 766}, 97 (2006). 
\bibitem{PO}
O. Penrose and L. Onsager, Phys. Rev. {\bf 140}, 576 (1956). 
\bibitem{clark79}
J. W. Clark, Prog. Nucl. Part. Phys. {\bf 2}, 89 (1979).
\bibitem{Ristig}
R. Pentf\"order, T. Lindenau, and M. L. Ristig, J. Low Temp. Phys. {\bf  108}, 245 (1997).


\bibitem{jemai10}
M. Jemai and P. Schuck, arXiv:1011.5106.
\bibitem{Fetter-Walecka}
A.~L.~Fetter and J.~D.~Walecka, {\it Quantum Theory of Many-Particle Systems}, (Dover, New York, 2003). 
\bibitem{bhh86}
A.~H.~Blin, R.~W.~Hasse, B.~Hiller, P.~Schuck, and C.~Yannouleas, Nucl. Phys. A {\bf 456}, 109 (1986).


\bibitem{bohr} 
A. Bohr and B. R.Mottelson, {\it Nuclear Structure} (Benjamin, New York, 1975), Vol. 2.
\bibitem{baizot_ripka}
J.-P. Blaizot and G. Ripka, {\it Quantum Theory of Finite Systems} (MIT, Cambridge, MA, 1986).
\bibitem{ShapT} 
S. L. Shapiro and S. A. Teukolsky,  {\em Black holes, white Dwarfs and Neutron Stars: The Physics of Compact Objects} (Wiley, N.Y., 1983); D. Pines, R. Tamagaki and S. Tsuruta (eds.), {\em Neutron Stars} (Addison-Wesley, N.Y., 1992).
\bibitem{Lecheminant} 
B. Doucot, J. Vidal, Phys. Rev. Lett. {\bf 88}, 227005 (2002); S. Capponi, G. Roux, P. Lecheminant, P. Azaria, E. Boulat, S.R. White, Phys. Rev. A {\bf 77}, 013624 (2008).


\end{thebibliography}
\end{document}